\documentclass[a4paper]{article}

\usepackage[utf8]{inputenc}
\usepackage[T1]{fontenc}
\usepackage{lmodern}
\usepackage{fullpage}
\usepackage{microtype}

\usepackage[usenames,dvipsnames,table]{xcolor}

\usepackage{amsmath,amsthm,amssymb}
\usepackage{wasysym} % has \gemini used in reduction

\usepackage{xspace}

\usepackage{subfig}
\usepackage{graphicx}

\usepackage{xifthen}
\usepackage{framed}

\usepackage{tikz} 
\usetikzlibrary{matrix}
\usetikzlibrary{arrows,shapes,positioning}
\usetikzlibrary{decorations.pathreplacing}
\usetikzlibrary{calc}

\usepackage{todonotes}

\usepackage{booktabs} % nice tables

\usepackage{tabularx}
\usepackage{multirow}

 \usepackage[pdftex, plainpages = false, pdfpagelabels, 
                 bookmarks=false,
                 bookmarksopen = true,
                 bookmarksnumbered = true,
                 breaklinks = true,
                 linktocpage,
                 pagebackref,
                 colorlinks = true,  
                 linkcolor = blue,
                 urlcolor  = blue,
                 citecolor = blue,
                 anchorcolor = green,
                 hyperindex = true,
                 hyperfigures
                 ]{hyperref} 

\input{macros}

\author{
  Christophe Crespelle\thanks{Department of Informatics, University of Bergen, Norway -- \texttt{christophe.crespelle@uib.no}}
  \and Pål Grønås Drange\thanks{\texttt{paal.drange@gmail.com}}
  \and Fedor V.~Fomin\thanks{Department of Informatics, University of Bergen, Norway -- \texttt{fedor.fomin@uib.no}}
  \and Petr A.~Golovach\thanks{Department of Informatics, University of Bergen, Norway -- \texttt{petr.golovach@uib.no}}
}

\title{A survey of parameterized algorithms and the complexity of edge modification\thanks{This work has received funding from the European Union's Horizon 2020 research and innovation programme under the Marie Sklodowska-Curie grant agreement No 749022 and is supported by the Research Council of Norway via the project “MULTIVAL”.}}

\begin{document}
\maketitle

\begin{abstract}
The survey   provides an overview of the developing area of parameterized algorithms for graph modification problems. We concentrate on edge modification problems, where the task is to change a small number of adjacencies in a graph in order to satisfy some required property. 
%And we give a lot of open problems! 
\end{abstract}

\bigskip

\noindent\fbox{\parbox{\linewidth-2\fboxrule-2\fboxsep}{\underline{IMPORTANT NOTICE:}\\
This survey is still in a tentative version. If you find some mistakes or missing results, please contact us as soon as possible so that we can correct before final publication.}}

\medskip

\tableofcontents

\bigskip\noindent
%
%Some of the fundamental problems in algorithmic graph theory are the different
%notions of graph modification.  In a graph modification problem, we are given a
%graph target, specified by some property~$\Pi$, and asked to modify a given
%input graph to achieve the property.
 
 %\todo[inline]{Update conference to journal verisions. Interval completion .}
 
 \section{Introduction}
 A  variety of algorithmic graph problems can be formulated as problems of modifying a graph such that the resulting graph  satisfies some desired properties.  In particular, in the past 30 years, graph modification problems served as a strong inspiration for developing
  new  approaches  in 
 {parameterized algorithms and complexity}. 
 In this survey we are concerned with   a  specific  type of graph modification problems, namely edge modification problems. Even for this special version of graph modification problem there is a plethora of algorithmic results in the literature. We focus on new  developments in the area of parameterized algorithms and complexity for edge modification  problems including  kernelization, subexponential algorithms, and algorithms for finding various  cuts and connectivity augmentations, as well as  achieving various vertex-degree constrains. 
We   also provide  open problems for further research.

 One of the classic results about graph modification problems  is the work of Lewis and Yannakakis \cite{lewis1980nodedeletion}, that  provides necessary and sufficient conditions   (assuming  ${\sf P} \neq{\NP}$) of  polynomial time solvability of vertex-removal problems for hereditary properties. However, when it concerns edge-removal problems, no such dichotomy is known.  Since the work of Yannakakis ~\cite{yannakakis1981edgedeletion}, a great deal of work was devoted to establish 
 which edge modification problems are in  ${\sf P}$ and what are 
 $ {\sf NP}$-complete.
There already exists surveys on these topics
\cite{burzyn2006np,mancini2008graph,natanzon2001complexity} that the interested reader can look up.

%\medskip
The edge modification problems discussed in this survey fall mainly in  one of the categories depending on the operations we allow; \emph{adding edges}, \emph{deleting edges}, and the combination of both, which we call \emph{editing edges}.
 Formally, let $\mathcal{G}$  be a graph class.
 In the  \probPedad problem, the task is to decide   whether a given graph $G$ can be transformed into a graph in
$\mathcal{G}$
by \emph{adding} at most $k$ edges.
We use the following notation. For a set $F$ of pairs of $V(G)$, we denote by  $G + F$ the graph obtained from $G$ by making all pairs from $F$ adjacent.  Then we formally define \probPedad as follows:

\defparproblem%
{\probPedad}%
{$k$}%
{Graph $G$ and integer $k$}%
{Decide whether there exists a set $F \subseteq [V(G)]^2$ of size at most $k$
  such that $G + F$ is in $\mathcal{G}$.}

For example, when  $\mathcal{G}$ is the class of chordal graphs, then this is the \probChordComp problem, that is the problem  of
adding at most $k$ edges to make an input graph chordal, i.e., containing no induced cycle of length more than three. If $\mathcal{G}$ is the class of $2$-edge connected graphs, then this is the $\textsc{2-Edge-Connectivity Augmentation}$ problem.
One natural question to ask is why it is that case that \probChordComp is \npci~\cite{yannakakis1981computing}, whereas $\textsc{2-Edge-Connectivity Augmentation}$ (for unweighted graphs) is solvable in polynomial time~\cite{EswaranT76}.%\todo{?}

In the \probPedel problem, the task is to decide whether a given graph $G$ can
be transformed into a graph in $\mathcal{G}$ by \emph{deleting} at most $k$ edges.
We use the notation $G - F$, where $F \subseteq E(G)$, to denote the graph with the vertex set $V(G)$ and edge set
$E(G)\setminus F$.
Then we define \probPedel as follows:

\defparproblem%
{\probPedel}%
{$k$}%
{Graph $G$ and integer $k$}%
{Decide whether there exists a set $F \subseteq E(G)$ of size at most $k$
  such that $G - F$ is in $\mathcal{G}$.}

For example, when $\mathcal{G}$ is the class of acyclic graphs, then \probPedel is trivially solvable in polynomial time (finding a minimum spanning tree). When $\mathcal{G}$
is the class of bipartite graphs, the problem is known as \probOCT\footnote{The problem is strictly speaking called \pname{Edge Bipartization} but is computationally equivalent to \probOCT.} and is \npci~\cite{yannakakis1981edgedeletion}.

%Examples of the latter are deleting edges to obtain an acyclic graph, to obtain
%a bipartite graph, deleting or adding edges---as seen above---to obtain cluster
%graphs, or adding edges to obtain a chordal graph.  Each of these examples have
%important practical and theoretical applications.  

Finally, in the \probPeded problem the task is to decide whether a given graph
$G$ can be transformed into a graph $G + F_{+} - F_{-}$ in $\mathcal{G}$ using
at most $|F_{+}| + |F_{-}| = k$ edges.
For a set $F$ of pairs of $V(G)$, we denote by $G \symdiff F$ the graph with vertex set $V(G)$, and whose edge set is the symmetric difference of~$E(G)$ and~$F$.
We define \probPeded as follows:

\defparproblem%
{\probPeded}%
{$k$}%
{Graph $G$ and integer $k$}%
{Decide whether there exists a set $F \subseteq [V(G)]^2$ of size at most~$k$
  such that $G \symdiff F$ is in $\mathcal{G}$.}

When  $\mathcal{G}$ is the graph class of disjoint unions of complete graphs, i.e.,  cluster graphs, then
this is the problem known as \probCedit or \textsc{Correlation Clustering}, the problem of deleting and adding at most~$k$ edges in a graph~$G$ such that every connected component of the obtained graph is a clique.
This problem is known to be \npci~\cite{shamir2004cluster}.  On the other hand, the \probSplitED problem, the problem  of editing to a split graph (we postpone the definition of this graph till the next section) is solvable in polynomial time~\cite{hammer1981splittance}.

In this survey we intentionally tried to avoid discussions of vertex-modification problems; a survey including both would likely result in a full text-book. Many parameterized and kernelization algorithms for vertex modification problems including \textsc{\probVC}, \textsc{\probFVS}, \textsc{\probOCT} and many others can be found in the books by Cygan et al.~\cite{cygan2015parameterized} and Fomin et al.~\cite{kernelizationbook19}.
We also decided not to discuss parameterized complexity of contraction problems since contraction operation decreases the number of vertices in a graph, and is therefore in some sense closer in spirit to vertex removal problems.  For further reading on contraction problems we refer to existing surveys~\cite{GolovachHP13,guillemot2013faster,guo2015obtaining,HeggernesHLP13}.

\paragraph{Cai's notation.}
Leizhen Cai in  \cite{Cai03a} introduced a notation for   graph modification problems which is widely used in the literature.   Let $\mathcal{G}$ be a  class of graphs, then  $\mathcal{G}-ke$ (respectively $\mathcal{G}+ke$)  is the class of those graphs that can be obtained from a member of $\mathcal{G}$ by deleting at most $k$ edges (respectively adding at most $k$ edges). 
We also can use $\mathcal{G}\pm ke$ for the class of graphs that can be obtained from a member of $\mathcal{G}$ by changing at most $k$ adjacencies. 
With Cai's notation,  the  \probPedad problem is the problem to decide whether graph $G$ is in  $\mathcal{G}-ke$,   \probPedel is to decide whether $G\in \mathcal{G}+ke$, and \probPeded is to decide whether 
 $G\in \mathcal{G}\pm ke$. Similarly, Cai's notation are also used for vertex-modification problems $\mathcal{G}-kv$ and $\mathcal{G}+kv$.

\paragraph{Parameterized complexity}
In most modification problems, and in many naturally occurring problems, we are
interested in finding the \emph{smallest} possible solution---we are looking for
a solution of size at most some prescribed number~$k$.
In \emph{parameterized complexity}, we are taking this value into account in
the analysis of the running time.
We are here looking for algorithms that solve problems in time $f(k) \cdot
\poly(n)$, where~$f$ can be any computable function with input~$k$, called the
\emph{parameter}, and~$n$ is the size of the input, usually measured in the
number of vertices in the input graph.  However,~$\poly$ is restricted to be a
fixed polynomial function.  A problem admitting such an algorithm is said to be
\emph{fixed-parameter tractable}.
This means, informally and vaguely, that for fixed sized solutions, the
problem is in some sense still tractable.
%
%In some contrast to the chaos we have for edge modification problems with
%respect to their \cclass{P}~vs.~\NP{} classification, much can be said about
%their parameterized complexity.
%
 Parameterized complexity offers a more fine-grained analysis than what the
\cclass{P}~vs.~\NP{} classification does.
%
%
%
%
%\begin{definition}[Polynomial kernel]
%  A kernelization algorithm, or simply a \emph{kernel}, for a parameterized
%  problem $Q$ is an algorithm that, given an instance $(I, k)$ of $Q$, in
%  polynomial time returns an equivalent instance $(I', k')$ of $Q$.  Moreover,
%  $|I'|,k' \leq \poly(k)$.
%\end{definition}
%
%

%{\bf Parameterized complexity.} 
\medskip
More formally, 
a \emph{parameterized problem} is a language~$Q\subseteq \Sigma^*\times\mathbb{N}$ where~$\Sigma^*$ is the set of strings over a finite alphabet~$\Sigma$. Respectively, an input  of~$Q$ is a pair~$(I,k)$ where~$I\subseteq \Sigma^*$ and~$k\in\mathbb{N}$;~$k$ is the \emph{parameter} of  the problem. 

A parameterized problem~$Q$ is \emph{fixed-parameter tractable} (\fpt ) if it can be decided whether $(I,k)\in Q$ in  $f(k)\cdot|I|^{\Oh(1)}$ time for some function~$f$ that depends on the parameter~$k$ only. Respectively, the parameterized complexity class \fpt  is composed by  fixed-parameter tractable problems.

Parameterized complexity theory also provides tools to rule-out the existence of \fpt  algorithms under plausible complexity-theoretic assumptions. For this,  a hierarchy of parameterized complexity classes
\[
  \fpt \subseteq \WOne\subseteq  \WTwo\subseteq  \cdots\subseteq \XP
\]
was introduced by Downey and Fellows~\cite{DowneyFellows1992b}, and it was conjectured that the inclusions are proper. The basic way to show that it is unlikely that a parameterized problem admit an \fpt  algorithm is to show that it is~$\WOne$ or~$\WTwo$-hard.
We refer to the many books on the subject~\cite{cygan2015parameterized,DowneyFbook13,Grohe07logic,Niedermeierbook06} for a proper introduction to parameterized algorithms and complexity.

\paragraph{Kernelization}
 A \emph{data reduction rule}, or simply, reduction rule, for a parameterized problem~$Q$ is  a function~$\phi \colon \Sigma^{\ast} \times \mathbb{N} \rightarrow \Sigma^{\ast} \times \mathbb{N}$ that maps an instance~$(I, k)$ of~$Q$ to an 
equivalent instance~$(I', k')$ of~$Q$ such that $\phi$ is computable in time polynomial in~$|I|$ and~$k$. We say that two instances of~$Q$ are \emph{equivalent} if the following holds: $(I, k) \in Q$ if and only if ~$(I',k') \in Q$. We refer to  this property of the reduction rule $\phi$, that it translates an instance to an equivalent one,  as to the \emph{safeness}  of the reduction rule.

Informally, \emph{kernelization} is a {preprocessing algorithm} that consecutively applies various data reduction rules in order to shrink the instance size as much as possible. A preprocessing algorithm takes as input an instance $(I,k)\in\Sigma^{*}\times\mathbb{N}$ of $Q$, runs in polynomial in~$|I|$ and~$k$ time, and returns an equivalent instance $(I',k')$ of $Q$. The quality of a preprocessing algorithm  $\mathcal{A}$ is measured by the size of the output. More precisely,  the \emph{output size} of a preprocessing algorithm $\mathcal{A}$ is a function $\text{size}_{\mathcal{A}}\colon \mathbb{N}\to \mathbb{N}\cup \{\infty\}$ defined as follows:
\[
  \text{size}_{\mathcal{A}}(k) = \sup \{ |I'|+k'\ \mid (I',k')=\mathcal{A}(I,k),\ I\in \Sigma^{*}\}.
\]
A \emph{kernelization algorithm}, or simply a \emph{kernel}, for a parameterized problem~$Q$ is a preprocessing algorithm~$\mathcal{A}$ that, given an instance $(I,k)$ of $Q$, works in polynomial in~$|I|$ and~$k$ time and returns an equivalent instance $(I',k')$ of $Q$ such that $\text{size}_{\mathcal{A}}(k)\leq g(k)$ for some computable function $g\colon \mathbb{N}\to \mathbb{N}$. It is said that~$g(\cdot)$ is the \emph{size} of a kernel.
If~$g(\cdot)$ is a polynomial function, then we say that~$Q$ admits a \emph{polynomial kernel}.

It is well-known that every \fpt problem admits a kernel but, up to some reasonable complexity assumptions, there are \FPT problems that have no polynomial kernels.
In particular, we are using the composition technique introduced by Bodlaender et al.~\cite{BodlaenderDFH09} to show that a parameterized problem does not admit a polynomial kernel unless \PH.
For further references on kernelization we refer to the recent book on the subject~\cite{kernelizationbook19}.

\paragraph{ETH}
The Exponential Time Hypothesis (ETH)  is a widely-believed conjecture of Impagliazzo,   Paturi,   and Zane~\cite{ImpagliazzoPZ01} informally stating that  \tsat has no
algorithm subexponential  in the number of variables. 
It is known that this conjecture
implies that $\FPT\neq  \WOne$, hence it can
be also used to give conditional evidence that certain problems are
not fixed-parameter tractable. More importantly, ETH allows us to
prove quantitative results of various forms. 
In particular, in this survey we mention a number of results ruling out the possibility of solving certain edge modification problems by subexponential parameterized algorithms.   
 
The formal statement of ETH is the following. For $q\geq 3$, let $\delta_q$ be the infinimum of the set of constants $c$ for which there exists an algorithm solving \qsat in time $\Oh(2^{cn})$.  Then  
  ETH is that 
$\delta_3>0$.
We refer to the book of Cygan et al.  \cite{cygan2015parameterized} for more information on ETH and its applications in parameterized algorithms.

\paragraph{Outline of the survey.}
The remaining part of this survey is organized as follows. Section~\ref{sec:hereditary} reviews results about edge modification problems toward hereditary graph classes. Section~\ref{sec:cuts} deals with modification problems related to connectivity, cuts and clustering\footnote{Note that the clustering approaches that consist in modifying the input graph into some hereditary graph class, such as \emph{cluster graphs} for example, are treated in Section~\ref{sec:graph-classes-finite}.}. In Section~\ref{sec:deg} we list results where the aim of the modification problem is to make the input graph satisfy some constraints on the degrees of the vertices. Finally, Section~\ref{sec:misc} reports on variants that do not fit strictly in the scope of the previous sections but are closely related to the questions considered in this survey.

%\todo[inline]{Add  The remaining part of this survey is organized as follows. }

% \input{introduction}

  %!TEX root =survey.tex
\section{Hereditary graph classes}\label{sec:hereditary}

In this section, we review results on edge modifications where the target class of graphs is \emph{hereditary}.   A graph class~$\curs{G}$ is hereditary when for any graph $G \in \curs{G}$, every induced subgraph of~$G$ also belongs to the class. Equivalently, this means that deleting any vertex of a graph in $\curs{G}$ also yields a graph in $\curs{G}$. Restricting ourselves to hereditary graph classes is not a sharp limitation. Although not all classes of graphs are hereditary, most classically studied graph classes are. One reason for this is that heredity is a rather natural property to require from a graph class as soon as belonging to the class is meant to be a characteristic of simplicity for a graph. In this case, it is natural to ask that a subpart of a simple object is also simple.
To illustrate how ubiquitous hereditary graphs classes are, we can count forests, bipartite, planar, distance hereditary, chordal and interval, perfect, comparability, permutation, cluster, cographs, trivially perfect, split, threshold, chain graphs, graphs of bounded treewidth, graphs of bounded degree, to mention just some of them.
Delete a vertex in a graph from any of these classes, and the resulting graph remains in that class. The classical surveys about graph classes are the books of Golumbic~\cite{Golumbic80} and Brandstädt, Le, and Spinrad~\cite{brandstadt1999graph}.

There are also a few notable examples of classes of graphs that are not hereditary, for instance the class of regular graphs, connected graphs, or more generally the class of graphs with at most a certain number of connected components, as
well as graphs with some certain specified connectivity or degree constraints, and sparseness and density requirements. These classes are treated in Sections~\ref{sec:cuts} and~\ref{sec:deg}.

Let $\mathcal{H} = \{H_1, H_2, H_3, \ldots \}$ be a (possibly infinite) set of graphs, we say that a graph $G$ is $\mathcal{H}$-free if for every graph $H \in \mathcal{H}$, $H$ is not an induced subgraph of $G$. The class of graphs~$\mathcal{G}_\mathcal{H}$ is the class of all $\mathcal{H}$-free graphs. We say that~$\mathcal{G}_\mathcal{H}$ is \emph{characterized} by $\mathcal{H}$. When $\mathcal{H}$ is a singleton $\{H\}$, we will simply write $H$-free, and $\mathcal{G}_H$.
It is worth to note that all classes that are defined by forbidden induced subgraphs are hereditary,  and that conversely, all hereditary classes of graphs can be defined by a (possibly infinite) set of forbidden induced subgraphs: those minimal graphs (for the induced subgraph ordering) that do not belong to the class. Therefore, the edge modification problems considered here can be formulated as modifying the edge set of the input graph in order to get rid of each obstacle (i.e., forbidden induced subgraph), either by adding an edge or deleting an edge.
% , without giving rise to new obstacles.
%
As in the rest of the survey, the parameter we consider is the number $k$ of modifications that are allowed. The complexities of these problems span a very broad range. For example, \pname{Split edge Editing} is solvable in polynomial time~\cite{hammer1981splittance} and \pname{Split edge Completion} is \npci~\cite{natanzon2001complexity}, \pname{Planar edge Deletion} is \FPT~\cite{kawarabayashi2007computing} and \pname{Wheel-free edge Deletion} is \cclass{W[2]}-hard~\cite{lokshtanov2008wheelfree}, \pname{$P_4$-free edge Deletion} admits a polynomial kernel~\cite{guillemot2013nonexistence} and \pname{$P_5$-free edge Deletion} does not~\cite{cai2012polynomial}, \pname{Chordal Completion} admits a subexponential time algorithm~\cite{fomin2013subexponential} while \pname{Cograph edge Completion} does not~\cite{drange2015exploring,DP17}.

This section is organized as follows. In the first two subsections, we discuss results on \FPT algorithms and polynomial kernels for hereditary graph classes that are characterized by a finite number of forbidden induced subgraphs (Section~\ref{sec:graph-classes-finite}) and for those characterized by an infinite number of forbidden induced subgraphs (Section~\ref{sec:graph-classes-inf}). The reason for this distinction is the existence of a general result~\cite{cai1996fixedparameter} that guarantees the existence of an \FPT algorithm for any edge modification problem where the target class is characterized by a finite number of forbidden subgraphs. Therefore, for these classes most of the efforts focused on the existence of polynomial kernels. All the results on subexponential parameterized algorithms, both for finitely and non-finitely characterizable classes are listed in Section~\ref{sec:graph-classes-SUBEXP}. Finally, Section~\ref{sec:graph-classes-related} list some results that deal with restricted input graphs or with target classes that are non-hereditary variants of some hereditary classes.

%\todo{Is there another survey?}
For more on polynomial kernels with respect to the aforementioned graph classes, one may consult the  survey on the kernelization complexity by Liu, Wang and Guo~\cite{liu2014overview} and the master thesis of Cai~\cite{cai2012polynomial}.

% % IS THIS PARAGRAPH INTERESTING ENOUGH?
%
% This class of hereditary graphs, the \emph{wheel-free graphs}, is polynomial
% time recognizable by the following algorithm: A graph $G = (V,E)$ has an
% induced wheel if there is a vertex $v \in V$ such that $N(v)$ has a
% cycle~\cite{lokshtanov2008wheelfree}.  Checking for each vertex that the
% neighborhood induces a forest can clearly be done in polynomial time.
% % 
% It follows that the complement class, co-wheel-free graphs has the
% corresponding completion problem \pname{Co-Wheel-free Completion} \wtwo-hard
% as well.

% \input{graph-classes-classical} % how much NP-completeness do we want?

%%%%%%%%%%%%%%%%%%%%%%%%%%%%%%%%%%%%%%%%%%%%%%%%%%%%%%%%%%%%%%%%%%%%%%%%%%%%%%
%%%%%%%%%%%%%%%%%%%%%%%%%%%%%%%%%%%%%%%%%%%%%%%%%%%%%%%%%%%%%%%%%%%%%%%%%%%%%%
%%%%%%%%%%%%%%%%%%%%%%%%%%%%%%%%%%%%%%%%%%%%%%%%%%%%%%%%%%%%%%%%%%%%%%%%%%%%%%
\subsection{Classes characterized by a finite number of minimal forbidden subgraphs}\label{sec:graph-classes-finite}
%%%%%%%%%%%%%%%%%%%%%%%%%%%%%%%%%%%%%%%%%%%%%%%%%%%%%%%%%%%%%%%%%%%%%%%%%%%%%%
%%%%%%%%%%%%%%%%%%%%%%%%%%%%%%%%%%%%%%%%%%%%%%%%%%%%%%%%%%%%%%%%%%%%%%%%%%%%%%
%%%%%%%%%%%%%%%%%%%%%%%%%%%%%%%%%%%%%%%%%%%%%%%%%%%%%%%%%%%%%%%%%%%%%%%%%%%%%%

%%%%%%%%%%%%%%%%%%%%%%%%%%%%%%%%%%%%%%%%%%%%%%%%%%%%%%%%%%%%%%%%%%%%%%%%%%%%%%
\begin{table}
%\normalsize
\small
\centering
\setlength\tabcolsep{0cm}

%\begin{center}
\hspace*{-1.5cm}
\begin{tabular}{| >{\centering}m{2.8cm} || >{\centering}m{2.6cm} | >{\centering}m{2.6cm} || >{\centering}m{2.6cm} | >{\centering}m{2.6cm} || >{\centering}m{2.6cm} | >{\centering}m{2.6cm} |}
\hline
graph class & \multicolumn{2}{c||}{completion} & \multicolumn{2}{c||}{deletion} & \multicolumn{2}{c|}{editing} \tabularnewline
\hline
 & KERNEL & \multicolumn{1}{c||}{\begin{tabular}{>{\centering}m{2.6cm}} TIME \tabularnewline\hline SUBEXP \end{tabular}} & KERNEL & \multicolumn{1}{c||}{\begin{tabular}{>{\centering}m{2.6cm}} TIME \tabularnewline\hline SUBEXP \end{tabular}} & KERNEL & \multicolumn{1}{c|}{\begin{tabular}{>{\centering}m{2.6cm}}TIME \tabularnewline\hline SUBEXP \end{tabular}} \tabularnewline
\hline
\hline

%graph class & KER COMP & \multicolumn{1}{c||}{\begin{tabular}{>{\centering}m{2.6cm}} \FPT COMP \tabularnewline\hline LOW COMP \end{tabular}} & KER DEL & \multicolumn{1}{c||}{\begin{tabular}{>{\centering}m{2.6cm}} \FPT DEL \tabularnewline\hline LOW DEL \end{tabular}} & KER EDIT & \multicolumn{1}{c|}{\begin{tabular}{>{\centering}m{2.6cm}} \FPT EDIT \tabularnewline\hline LOW EDIT \end{tabular}} \tabularnewline \hline

line & OPEN & \multicolumn{1}{c||}{\begin{tabular}{>{\centering}m{2.6cm}}  \FPT by~\cite{cai1996fixedparameter}
 \tabularnewline\hline OPEN  \end{tabular}} &OPEN  & \multicolumn{1}{c||}{\begin{tabular}{>{\centering}m{2.6cm}} \FPT by~\cite{cai1996fixedparameter} \tabularnewline\hline  OPEN  \end{tabular}} & OPEN & \multicolumn{1}{c|}{\begin{tabular}{>{\centering}m{2.6cm}} \FPT by~\cite{cai1996fixedparameter} \tabularnewline\hline OPEN  \end{tabular}} \tabularnewline \hline

$s$-Plex Cluster & - & \multicolumn{1}{c||}{\begin{tabular}{>{\centering}m{2.6cm}} - \tabularnewline\hline - \end{tabular}} & - & \multicolumn{1}{c||}{\begin{tabular}{>{\centering}m{2.6cm}} - \tabularnewline\hline - \end{tabular}} & $s^2k$~\cite{guo2010more} & \multicolumn{1}{c|}{\begin{tabular}{>{\centering}m{2.6cm}} $(2s+\sqrt{s})^k$~\cite{guo2010more} \tabularnewline\hline \cellcolor{red} NOSUB~\cite{komusiewicz2012cluster}  \end{tabular}} \tabularnewline
\hline
\hline

$\set{K_3,2K_2,C_5}$\\ chain & \multicolumn{2}{c||}{as deletion} & $k^2$~\cite{bessy2013polynomial,drange2015threshold} & \multicolumn{1}{c||}{\begin{tabular}{>{\centering}m{2.6cm}} \cellcolor{green} SUBEXP \tabularnewline\hline $2^{\sqrt{k}\log k}$~\cite{drange2015exploring} \end{tabular}} & $k^2$~\cite{drange2015threshold} & \multicolumn{1}{c|}{\begin{tabular}{>{\centering}m{2.6cm}} \cellcolor{green} SUBEXP \tabularnewline\hline $2^{\sqrt{k}\log k}$~\cite{drange2015threshold} \end{tabular}} \tabularnewline
\hline

$\set{K_3,C_4,P_4}$\\ Starforest & \multicolumn{2}{c||}{P} & - & \multicolumn{1}{c||}{\begin{tabular}{>{\centering}m{2.6cm}} \FPT by~\cite{cai1996fixedparameter}   \tabularnewline\hline \cellcolor{red} NOSUB~\cite{DRS+15} \end{tabular}} & \multicolumn{2}{c|}{ as  deletion} \tabularnewline
\hline

$\set{2K_2,C_4,P_4}$\\ threshold \textbf{*} & $k^2$~\cite{drange2015threshold} & \multicolumn{1}{c||}{\begin{tabular}{>{\centering}m{2.6cm}} \cellcolor{green} SUBEXP \tabularnewline\hline $2^{\sqrt{k}\log k}$~\cite{drange2015exploring} NO~$2^{k^{1/4}}$~\cite{bliznets2016lower}\end{tabular}} & $k^2$~\cite{drange2015threshold} & \multicolumn{1}{c||}{\begin{tabular}{>{\centering}m{2.6cm}} \cellcolor{green} SUBEXP \tabularnewline\hline $2^{\sqrt{k}\log k}$~\cite{drange2015exploring} NO~$2^{k^{1/4}}$~\cite{bliznets2016lower} \end{tabular}} & $k^2$~\cite{drange2015threshold} & \multicolumn{1}{c|}{\begin{tabular}{>{\centering}m{2.6cm}} \cellcolor{green} SUBEXP \tabularnewline\hline $2^{\sqrt{k}\log k}$~\cite{drange2015threshold} \end{tabular}} \tabularnewline
\hline

$\set{2K_2,C_4,C_5}$\\ split \textbf{*} & $k^2$~\cite{ghosh2015faster} & \multicolumn{1}{c||}{\begin{tabular}{>{\centering}m{2.6cm}} \cellcolor{green} SUBEXP \tabularnewline\hline $2^{\Oh(\sqrt k)}$~\cite[Exercise~5.17]{cygan2015parameterized} \end{tabular}} & $k^2$~\cite{ghosh2015faster} & \multicolumn{1}{c||}{\begin{tabular}{>{\centering}m{2.6cm}} \cellcolor{green} SUBEXP \tabularnewline\hline $2^{\Oh(\sqrt k)}$~\cite[Exercise~5.17]{cygan2015parameterized} \end{tabular}} & \multicolumn{2}{c|}{P~\cite{hammer1981splittance}} \tabularnewline
\hline
\hline

%$\curs{H}\supseteq \set{K_{1,s}, K_t}$\\ fixed $s>1,t>2$ & - & \multicolumn{1}{c||}{\begin{tabular}{>{\centering}m{2.6cm}} - \tabularnewline\hline - \end{tabular}} & KER~\cite{ASS17} & \multicolumn{1}{c||}{\begin{tabular}{>{\centering}m{2.6cm}} - \tabularnewline\hline - \end{tabular}} & - & \multicolumn{1}{c|}{\begin{tabular}{>{\centering}m{2.6cm}} - \tabularnewline\hline - \end{tabular}} \tabularnewline
%\hline

$\set{P_3,2K_2}$\\ clique + isol. vert. & \multicolumn{2}{c||}{P} & $2k$ [folkl.] & \multicolumn{1}{c||}{\begin{tabular}{>{\centering}m{2.6cm}} \cellcolor{green} SUBEXP \tabularnewline\hline $1.6355^{\sqrt{k \ln k}}$~\cite{damaschke2014editing} \end{tabular}} & $2k$ [folkl.] & \multicolumn{1}{c|}{\begin{tabular}{>{\centering}m{2.6cm}} \cellcolor{green} SUBEXP \tabularnewline\hline $2^{\sqrt{k \ln k}}$~\cite{damaschke2014editing} \end{tabular}} \tabularnewline \hline

$\set{C_4,P_4}$\\ trivially perfect & $k^3$~\cite{guo2007problem} & \multicolumn{1}{c||}{\begin{tabular}{>{\centering}m{2.6cm}} \cellcolor{green} SUBEXP \tabularnewline\hline $2^{\sqrt{k}\log k}$~\cite{drange2015exploring} NO~$2^{k^{1/4}}$~\cite{bliznets2016lower} \end{tabular}} & $k^7$~\cite{DP17} & \multicolumn{1}{c||}{\begin{tabular}{>{\centering}m{2.6cm}} $2.42^k$~\cite{LWY+15} \tabularnewline\hline \cellcolor{red} NOSUB~\cite{drange2015exploring} \end{tabular}} & $k^7$~\cite{DP17} & \multicolumn{1}{c|}{\begin{tabular}{>{\centering}m{2.6cm}} - \tabularnewline\hline \cellcolor{red} NOSUB~\cite{DP17} \end{tabular}} \tabularnewline
\hline

$\set{claw, diamond}$ & OPEN & \multicolumn{1}{c||}{\begin{tabular}{>{\centering}m{2.6cm}} \FPT by~\cite{cai1996fixedparameter} \tabularnewline\hline OPEN \end{tabular}} & $k^{\cO(1)}$~\cite{CPP+17} & \multicolumn{1}{c||}{\begin{tabular}{>{\centering}m{2.6cm}} OPEN \tabularnewline\hline \cellcolor{red} NOSUB~\cite{CPP+17} \end{tabular}} & OPEN & \multicolumn{1}{c|}{\begin{tabular}{>{\centering}m{2.6cm}} \FPT by~\cite{cai1996fixedparameter}  \tabularnewline\hline - \end{tabular}} \tabularnewline
\hline

$\set{2K_2,C_4}$\\ pseudosplit \textbf{*} & - & \multicolumn{1}{c||}{\begin{tabular}{>{\centering}m{2.6cm}} \cellcolor{green} SUBEXP \tabularnewline\hline $2^{\Oh(\sqrt k)}$~\cite{cygan2015parameterized, drange2015exploring} \end{tabular}} &- & \multicolumn{1}{c||}{\begin{tabular}{>{\centering}m{2.6cm}} \cellcolor{green} SUBEXP \tabularnewline\hline $2^{\Oh(\sqrt k)}$~\cite{cygan2015parameterized, drange2015exploring} \end{tabular}} & \multicolumn{2}{c|}{P~\cite{drange2015thesis}} \tabularnewline
\hline
\hline

$\set{P_3}$\\ cluster & \multicolumn{2}{c||}{P} & $k^3$~\cite{GGH+05} & \multicolumn{1}{c||}{\begin{tabular}{>{\centering}m{2.6cm}} $1.41^k$~\cite{BD11} \tabularnewline\hline \cellcolor{red} NOSUB~\cite{komusiewicz2012cluster} \end{tabular}} & $2k$~\cite{chen2012kernel,CC12} & \multicolumn{1}{c|}{\begin{tabular}{>{\centering}m{2.6cm}} $1.76^k$~\cite{BD11} \tabularnewline\hline \cellcolor{red} NOSUB~\cite{komusiewicz2012cluster} \end{tabular}} \tabularnewline
\hline

$\set{K_3}$ & \multicolumn{2}{c||}{P} & $6k$~\cite{brugmann2009generating} & \multicolumn{1}{c||}{\begin{tabular}{>{\centering}m{2.6cm}}  \FPT by~\cite{cai1996fixedparameter} \tabularnewline\hline \cellcolor{red} NOSUB~~\cite{ASS17b}  \end{tabular}} & \multicolumn{2}{c|}{as deletion} \tabularnewline
\hline
\hline

$\set{P_4}$\\ cograph \textbf{*} & $k^3$~\cite{guillemot2013nonexistence} & \multicolumn{1}{c||}{\begin{tabular}{>{\centering}m{2.6cm}} $2.56^k$~\cite{NG12} \tabularnewline\hline \cellcolor{red} NOSUB~\cite{drange2015exploring,DP17} \end{tabular}}  & $k^3$~\cite{guillemot2013nonexistence} & \multicolumn{1}{c||}{\begin{tabular}{>{\centering}m{2.6cm}} $2.56^k$~\cite{NG12} \tabularnewline\hline \cellcolor{red} NOSUB~\cite{drange2015exploring,DP17} \end{tabular}}  & $k^3$~\cite{guillemot2013nonexistence} & \multicolumn{1}{c|}{\begin{tabular}{>{\centering}m{2.6cm}} $4.61^k$~\cite{LWG+12} \tabularnewline\hline \cellcolor{red} NOSUB~\cite{DP17} \end{tabular}} \tabularnewline
\hline

$\set{paw}$ & $k^3$~\cite{eiben2019polynomial} & \multicolumn{1}{c||}{\begin{tabular}{>{\centering}m{2.6cm}} \FPT by~\cite{cai1996fixedparameter} \tabularnewline\hline \cellcolor{red} NOSUB~\cite{ASS17b} \end{tabular}} &  $k^3$~\cite{eiben2019polynomial} & \multicolumn{1}{c||}{\begin{tabular}{>{\centering}m{2.6cm}} \FPT by~\cite{cai1996fixedparameter} \tabularnewline\hline \cellcolor{red} NOSUB~\cite{ASS17b} \end{tabular}} & $k^6$~\cite{eiben2019polynomial} & \multicolumn{1}{c|}{\begin{tabular}{>{\centering}m{2.6cm}} \FPT by~\cite{cai1996fixedparameter}  \tabularnewline\hline \cellcolor{red} NOSUB~\cite{ASS17b} \end{tabular}} \tabularnewline
\hline

$\set{diamond}$ & \multicolumn{2}{c||}{P} & $k^3$~\cite{sandeep2015parameterized,CRS+18} & \multicolumn{1}{c||}{\begin{tabular}{>{\centering}m{2.6cm}} \FPT by~\cite{cai1996fixedparameter} \tabularnewline\hline \cellcolor{red} NOSUB~\cite{sandeep2015parameterized,ASS17b} \end{tabular}} & $k^8$~\cite{CRS+18} & \multicolumn{1}{c|}{\begin{tabular}{>{\centering}m{2.6cm}} \FPT by~\cite{cai1996fixedparameter}  \tabularnewline\hline \cellcolor{red} NOSUB~\cite{ASS17b} \end{tabular}} \tabularnewline
\hline

$\set{claw}$ & OPEN & \multicolumn{1}{c||}{\begin{tabular}{>{\centering}m{2.6cm}} \FPT by~\cite{cai1996fixedparameter}  \tabularnewline\hline \cellcolor{red} NOSUB~\cite{ASS17b} \end{tabular}} & OPEN & \multicolumn{1}{c||}{\begin{tabular}{>{\centering}m{2.6cm}} \FPT by~\cite{cai1996fixedparameter}  \tabularnewline\hline \cellcolor{red} NOSUB~\cite{ASS17b} \end{tabular}} & OPEN& \multicolumn{1}{c|}{\begin{tabular}{>{\centering}m{2.6cm}} \FPT by~\cite{cai1996fixedparameter} \tabularnewline\hline \cellcolor{red} NOSUB~\cite{ASS17b} \end{tabular}} \tabularnewline
\hline

%$\set{C_4}$ & NOKER~\cite{guillemot2013nonexistence} & \multicolumn{1}{c||}{\begin{tabular}{>{\centering}m{2.6cm}} - \tabularnewline\hline \cellcolor{red} NOSUB~\cite{drange2015exploring} \end{tabular}} & NOKER~\cite{guillemot2013nonexistence} & \multicolumn{1}{c||}{\begin{tabular}{>{\centering}m{2.6cm}} - \tabularnewline\hline \cellcolor{red} NOSUB~\cite{drange2015exploring} \end{tabular}} & NOKER~\cite{guillemot2013nonexistence}& \multicolumn{1}{c|}{\begin{tabular}{>{\centering}m{2.6cm}} - \tabularnewline\hline \cellcolor{red} NOSUB~\cite{ASS17b} \end{tabular}} \tabularnewline
%\hline

$\set{K_4}$\\ & \multicolumn{2}{c||}{P} & $k^3$~ \cite{Tsur19} & \multicolumn{1}{c||}{\begin{tabular}{>{\centering}m{2.6cm}}  \FPT by~\cite{cai1996fixedparameter}   \tabularnewline\hline \cellcolor{red} NOSUB~\cite{ASS17b} \end{tabular}} & \multicolumn{2}{c|}{as deletion} \tabularnewline
\hline
\hline

%\set{\KW} & - & \multicolumn{1}{c||}{\begin{tabular}{>{\centering}m{2.6cm}} - \tabularnewline\hline - \end{tabular}} & NOKER~\cite{kratsch2013two} & \multicolumn{1}{c||}{\begin{tabular}{>{\centering}m{2.6cm}} - \tabularnewline\hline - \end{tabular}} & NOKER~\cite{kratsch2013two} & \multicolumn{1}{c|}{\begin{tabular}{>{\centering}m{2.6cm}} - \tabularnewline\hline - \end{tabular}} \tabularnewline
%\hline
%\hline

$\set{P_\ell}$\\ fixed $\ell > 4$ & NOKER~\cite{cai2015incompressibility}  & \multicolumn{1}{c||}{\begin{tabular}{>{\centering}m{2.6cm}} \FPT by~\cite{cai1996fixedparameter} \tabularnewline\hline  \cellcolor{red} NOSUB~\cite{ASS17b}  \end{tabular}} & NOKER~\cite{cai2012polynomial} & \multicolumn{1}{c||}{\begin{tabular}{>{\centering}m{2.6cm}} \FPT by~\cite{cai1996fixedparameter} \tabularnewline\hline  \cellcolor{red} NOSUB~\cite{ASS17b} \end{tabular}} & NOKER~\cite{cai2015incompressibility} & \multicolumn{1}{c|}{\begin{tabular}{>{\centering}m{2.6cm}} \FPT by~\cite{cai1996fixedparameter}  \tabularnewline\hline  \cellcolor{red} NOSUB~\cite{ASS17b}  \end{tabular}} \tabularnewline
\hline

$\set{C_\ell}$\\ fixed $\ell > 3$ & NOKER~\cite{cai2015incompressibility}  & \multicolumn{1}{c||}{\begin{tabular}{>{\centering}m{2.6cm}} \FPT by~\cite{cai1996fixedparameter}  \tabularnewline\hline  \cellcolor{red} NOSUB~\cite{ASS17b}  \end{tabular}} & NOKER~\cite{cai2015incompressibility} & \multicolumn{1}{c||}{\begin{tabular}{>{\centering}m{2.6cm}} \FPT by~\cite{cai1996fixedparameter}  \tabularnewline\hline  \cellcolor{red} NOSUB~\cite{ASS17b}  \end{tabular}} &NOKER~\cite{cai2015incompressibility}  & \multicolumn{1}{c|}{\begin{tabular}{>{\centering}m{2.6cm}} \FPT by~\cite{cai1996fixedparameter}  \tabularnewline\hline  \cellcolor{red} NOSUB~\cite{ASS17b}  \end{tabular}} \tabularnewline
\hline

\end{tabular}
%\end{center}
%\vspace*{-1em}
\medskip
\caption{Parameterized complexity of edge modification problems into hereditary graph classes characterized by a finite number of forbidden induced subgraphs. We provide the best known exponential bound in terms of the parameter $k$ and omit the polynomial part. 
 ``\FPT by~\cite{cai1996fixedparameter}'' means that the problem is  \FPT by the result of Cai and that we are not aware of any more efficient solution. NOKER means that there is no polynomial kernel, while NOSUB means that there is no parameterized subexponential algorithm (of course up to some complexity assumption).  OPEN means that the complexity is widely open, while - means that probably open but most likely nobody looked at this question. P means the problem is solvable in polynomial time.
A star next to the name of the class marks auto-complementary classes, for which any result for one of the completion problem or deletion problem automatically gives the same result for the other problem. \label{tab:finite}}
\end{table}
%%%%%%%%%%%%%%%%%%%%%%%%%%%%%%%%%%%%%%%%%%%%%%%%%%%%%%%%%%%%%%%%%%%%%%%%%%%%%%
 
Several well known graph classes can be characterized by a finite set of forbidden induced subgraphs. This includes cluster, split, threshold, chain, trivially perfect, cographs, triangle-free, claw-free, line graphs and many more. As mentioned above, there is a  general result by Cai~\cite{cai1996fixedparameter} that had a strong impact on the  study of parameterized complexity of edge modification problems into classes of graphs defined by a finite family of forbidden subgraphs. This algorithmic result can be stated for the  generic   \pname{$\mathcal{G}$ $(k_1,k_2,k_3)$-Editing} problem, which is defined as follows.

\defparproblem%
{$\mathcal{G}$ $(k_1,k_2,k_3)$-Editing}%
{$k = k_1 + k_2 + k_3$}%
{$G = (V,E), k_1, k_2, k_3$}%
{Are there sets $V_{-} \subseteq V$ of size at most $k_1$,
  $E_{-} \subseteq E$ of size at most $k_2$, and
  $E_{+} \subseteq [V]^2$ of size at most $k_3$, such that $G - V_{-} - E_{-} + E_{+}$
  is a graph in $\mathcal{G}$?}

\begin{theorem}[Cai's theorem~\cite{cai1996fixedparameter}]\label{thm:Cai}
Let $\mathcal{G}$ be a graph class characterized by a finite set of forbidden induced subgraphs.  Then \pname{$\mathcal{G}$ $(k_1,k_2,k_3)$-Editing} is solvable in $\Oh({c_1}^kn^{c_2})$ time, where $k = k_1+k_2+k_3$ and~$c_1$ and~$c_2$ depend only on the finite characterization of $\mathcal{G}$.
\end{theorem}

In particular, for the problems we are interested in here, this means that completion ($k_1=0$ and $k_2=0$), deletion ($k_1=0$ and $k_3=0$) and editing ($k_1=0$ and try all couples $k_2,k_3$ such that $k_2+k_3\leq l$) are all \FPT parameterized by the number of modifications allowed ($k_3$ in the completion problem, $k_2$ in the deletion problem and $l$ in the editing problem).
This completely settles the parameterized complexity for many problems (see above for a list of some finitely characterizable graph classes) and has two immediate consequences for the domain:
\begin{enumerate}
\item The only hereditary classes for which deciding whether the edge
  modification problems are \FPT are the classes that cannot be characterized by
  a finite family of forbidden graphs (see Section~\ref{sec:graph-classes-inf}).
\item For classes defined by a finite set of forbidden subgraphs, from the
  perspective of parameterized complexity the questions of interest are
  \begin{enumerate}
  \item Improving the (exponential) dependence of the running time in
    Theorem~\ref{thm:Cai} from the parameter~$k$. Such improvements sometime can
    bring to subexponential parameterized running times, see
    Section~\ref{sec:graph-classes-SUBEXP});
  \item Exploring the possibility of polynomial kernelization (we focus on these
    results in this section).
%  \item Above/below guarantee parameterization.  
%  Parameterizations above and below guarantee were introduced by Mahajan and Raman~\cite{MahajanR99}. The idea is that   some optimization problems  have nontrivial lower and/or upper bounds on their optimum solution size. 
%   parameterized problems interesting algorithmic questions can be asked whether there are combinatorial and algorithmic bounds  
%  
%   There are not so many results of this
%    nature for edge modification problems and we mention them in several places
%    of this survey.
\end{enumerate}
\end{enumerate}

%It is worth to note that for finitely characterisable hereditary graph classes Cai's theorem implies that not only are editing, deletion and completion fixed-parameter tractable, but also efficiently tractable with a small single-exponential time algorithm. Let us denote $T(n,m)$ the time complexity of an algorithm solving the recognition problem for class $\curs{G}$ with certificate, i.e. an algorithm that provides a minimal forbidden induced subgraph when $G\not\in\curs{G}$. Then, Cai's theorem gives a \FPT algorithm whose complexity can be alternatively expressed as $\Oh(c^k T(n,m))$, where $c$ depends on the finite characterization of class $\curs{G}$ and on the edge modification problem considered: for completion, $c=\max\set{|E(\overline{H})|\ |\ H\in\curs{H}}$, for deletion, $c=\max\set{|E(H)|\ |\ H\in\curs{H}}$ and for editing, $c=N(N-1)/2$, with $N=\max\set{|V(H)|\ |\ H\in\curs{H}}$.

\noindent
Interestingly, the general result of Cai about the existence of \FPT algorithms extends to kernelization for vertex deletion problems. Indeed, in these settings, the task is to hit all the copies of these forbidden subgraphs (so-called \emph{obstacles}) that are originally contained in the graph. Hence, one can construct a simple reduction to the
\pname{$d$-Hitting Set} problem for a constant~$d$ depending on~$\mathcal{G}$, and use the classic~$\Oh(k^d)$ kernel for the latter that is based on the sunflower lemma~\cite{flum2006parameterized, abu-khzam2010kernelization}. Unfortunately, for edge modification problems, this approach fails utterly: every edge addition and deletion can create new obstacles, and thus it is not sufficient to hit only the original ones. For this reason, kernelization of edge modification problems have received a good deal of attention even for finitely characterizable classes.

From 2007, Guo~\cite{guo2007problem} and Gramm et al.~\cite{gramm2009data} provided kernels for several graph modification problems towards graph classes characterized by a finite set of forbidden induced subgraphs, including cluster, split, threshold, chain and trivially perfect graphs.  Several other positive results followed, which led Fellows et al.\ to ask whether all~$\mathcal{H}$-free modification problems for finite $\mathcal{H}$ admit polynomial, and even linear kernels~\cite{fellows2007efficient}.

% \KW GRAPH
\begin{figure}
  \centering
  \begin{tikzpicture}[every node/.style={circle,draw,scale=.2},scale=.4]
    \def \n {6}
    \def \radius {1}
    \def \margin {8} % margin in angles, depends on the radius
    
    \foreach \s in {1,...,\n}
    {
      \node (\s) at ({360/\n * (\s - 1)}:\radius) {};
    }
    \node (c) at (0:0) {};
    \draw (1)--(c)--(2);
    \draw (4)--(c)--(3);
    \draw (5)--(c)--(6);
    \draw (1)--(2);
    \draw (3)--(4);
  \end{tikzpicture}
  \caption{The graph $\KW$}
  \label{fig:kw}
\end{figure}
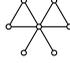

This was refuted by Kratsch and Wahlström~\cite{kratsch2013two} using the framework of Bodlaender et al.\cite{bodlaender2009problems}, who showed that for a certain graph on seven vertices, namely $\KW$ (depicted on Figure~\ref{fig:kw}), none of the problems \pname{$\KW$-free edge Deletion} nor \pname{$\KW$-free edge Editing}, admit polynomial kernels unless \ph.(\ph{} implies that \cclass{PH} is contained in \cclass{$\Sigma^p_3$}. We refer to \cite{cygan2015parameterized} for further discussions.)
This shows that the subtle differences between edge modification and vertex deletion problems have tremendous impact on the kernelization complexity. They conclude by asking whether there is a ``simple'' graph, like a path or a cycle, for which an edge modification problem does not admit a polynomial kernel under similar assumptions.  This question was answered by Guillemot et al.~\cite{guillemot2013nonexistence} who showed that both for the class of~$P_\ell$-free graphs (for~$\ell \geq 7$) and for the class of~$C_\ell$-free graphs (for~$\ell\geq 4$), the edge deletion problems do not have polynomial kernelization algorithms, unless \ph.  They simultaneously gave a cubic kernel for the \pname{Cograph Editing} problem, the problem of editing to a graph without induced paths on four vertices, showing that there is a fundamental difference between $P_4$-free and $P_7$-free graphs when it comes to modification problems.

This led to further developments on polynomial kernelization  for classes characterized by excluding one single graph $H$. The most prominent result in this direction is the one by Cai and Cai~\cite{cai2015incompressibility} who attempted to obtain a complete dichotomy of the kernelization complexity of edge modification problems for classes of~$H$-free graphs, for every graph~$H$. The project has been very successful---the question is settled for all~$3$-connected graphs, all paths and cycles, as well as all but a finite number of trees. They show that when $H$ is 3-connected, \pname{$H$-free edge Deletion} and \pname{Editing} admit no polynomial kernel iff $H$ is not complete; and \pname{$H$-free edge Completion} admits no polynomial kernel iff~$H$ misses at least two edges.
More precisely, the results of Cai and Cai are summarized in the following theorem.

\begin{theorem}[\cite{cai2015incompressibility}]\label{thm:CaiCai}
Let $\mathcal{G}$ be a  hereditary class of graphs characterized by  a graph $H$. Then unless \ph, 
\begin{itemize}
\item
When $H$ is $3$-connected,  \probPedel and  \probPeded admit no polynomial kernel if and only if $H$ is not a complete graph.  \probPedad  admits no polynomial kernel if and only if  $H$ misses at least two edges.
\item When  $H$  is a fixed path or cycle, \probPedel,  \probPeded, and  \probPedad   admit no polynomial kernel if and only if $H$ has at least $4$ edges.
\end{itemize}
\end{theorem}
Moreover, Cai and Cai proved that if $\mathcal{G}$ is characterized by a finite family of forbidden subgraphs $\mathcal{F}$, then  \probPedel admits  no polynomial kernel if all graphs in  $\mathcal{F}$ are $3$-connected and there is a graph $H\in \mathcal{F}$ with fewest edges such that one can add an edge to $H$ to obtain a graph not in $\mathcal{F}$.

 As a consequence of Theorem~\ref{thm:CaiCai}, 
% the question of the existence of a kernel is only open when~$H$ contains an articulation vertex or a disconnecting pair, which is the case of paths and cycles. For~$H$ being a path or a cycle, the three edge modification problems admit polynomial kernels if and only if~$H$ has at most three edges. As an example, one can see in Table~\ref{tab:finite} that edge modification problems into the class of $P_4$-free graphs (also known as cographs) admit polynomial kernels, because a $P_4$ has only 3 edges, while modification problems into the class of $C_4$-free graphs do not, because a $C_4$ has 4 edges. Therefore,
  the existence of a polynomial kernel for any of \pname{$H$-free edge Editing}, \pname{$H$-free edge Deletion}, or \pname{$H$-free edge Completion} problem is in fact a very rare phenomenon. It essentially happens only for very specific graphs~$H$.

Beside this, one can see in Table~\ref{tab:finite} that the question of existence of a kernel for edge modification problems into $H$-free graphs has been answered for all graphs on three vertices ($K_3$ and $P_3$) and for almost all graphs on four vertices. The only case remaining is the claw ($K_{1,3}$), which is unsolved for completion, deletion, and editing.
For $C_4$-free graphs, Guillemot et al.~\cite{guillemot2013nonexistence} showed that none of the three modification problems admit a kernel. On the positive side, they show the existence of a cubic kernel for each of the three modification problems into the class of $P_4$-free graphs (cographs). For the class of cographs, there were also some effort put in obtaining the best possible \FPT algorithm resulting in $2.56^k$ complexity for completion and deletion~\cite{NG12} and $4.61^k$ for editing~\cite{LWG+12}. The case of diamond-free graphs also drew quite a bit of attention. Fellows et al.~\cite{FGK+11} designed a $k^4$ vertex kernel for \pname{Diamond-free edge Deletion}, which was improved to $k^3$ by Sandeep and Sivadasan~\cite{sandeep2015parameterized}. Cao et al.~\cite{CRS+18} also provided a $k^3$ vertex  kernel for the deletion problem, following a different approach, and a $k^8$ kernel for \pname{Diamond-free edge Editing}.
Tsur \cite{Tsur19} gave a $k^{t-1}$ vertex kernel for the $K_t$-free deletion problem.

%%% CLAW, PAW
The question  about the existence of   polynomial kernel for \pname{Claw-free edge Deletion}  highlights how little help a finite characteristic provides.
Cygan et al.  in ~\cite{CPP+17} using modulator techniques (obtaining a specific vertex deletion
modulator $X$) similar to that used for showing kernels for modification to
trivially perfect graphs, threshold and chain graphs~\cite{DP17,
  drange2015threshold} showed that deletion to a subclass of claw-free graphs,
\pname{Claw-Diamond-free edge Deletion}, admits a polynomial
kernel, pinpointing the really hard cases that are left to solve in order to obtain a
polynomial kernel for \pname{Claw-free edge Deletion}.  On the negative side, Cai showed that \pname{$S_{11}$-free edge Deletion} does not have a kernel unless \ph~\cite{cai2012polynomial}. Here, $S_{11}$ is the star on~11 vertices, while the claw is the star on 4 vertices. Moreover, since forbidding induced $S_3 = P_3$ is the characterization for cluster graphs, \pname{$S_3$-free edge Deletion} admits a polynomial kernel, and thus there is a gap in our knowledge for the \pname{$S_t$-free edge Deletion} problems with $4 \leq t \leq 10$.

\begin{open}[\cite{cygan2013open, cai2015incompressibility,
    CPP+17}]
  Does \pname{claw-free edge Deletion} admit a polynomial kernel?
\end{open}
By the well-known characterization of line graphs by Beineke~\cite{MR262097}, a graph  
 is a line graph if and only if it does not contain one of   nine graphs as an induced subgraph. One of these graphs is a claw. 

\begin{open}[\cite{cygan2013open}]
  Does \pname{Line Graph Deletion} admit a polynomial kernel?
\end{open}
Similar questions are open for \pname{Line Graph Completion} and \pname{Line Graph Editing}.

%\begin{open}[\cite{sandeep2015parameterized}]
%  Does \pname{paw-free Deletion} admit a polynomial kernel?
%\end{open}

There has also been some attempt to generalize the approach of Cai and Cai  \cite{cai2015incompressibility} to families of hereditary graphs characterized by not only a single obstruction but a finite number of them. This gave the very nice result contained in the work of Aravind,  Sandeep, and Sivadasan~\cite{ASS17}, but which is valid only for restricted input graphs: if the input graphs have bounded degree and if the graphs in $\curs{F}$ are connected, then the \pname{$\curs{F}$-free edge Deletion} problem admits a polynomial kernel.

%%% CLUSTER

Among the classes of graphs listed in Table~\ref{tab:finite}, one received a particular attention: cluster graphs (see the survey by B{\"o}cker and Baumbach~\cite{BB13} for more on the topic). The reason is that cluster graph modification problems, more precisely deletion and editing, are closely related to the question of community detection, which is central in the domain of complex networks. It is striking to see that despite the simplicity of the structure of cluster graphs (they are disjoint union of cliques), both the editing and deletion problems remain \NP-complete. Completion is trivially polynomial: simply turn each connected component into a clique.
From a kernelization perspective, Gramm et al.~\cite{GGH+05} first showed the existence of a~$k^3$ kernel both for \pname{Cluster Deletion} and \pname{Cluster Editing}. The editing kernel was improved to linear size, namely~$6k$, by Fellows et al. ~\cite{fellows2007efficient} and there were several works putting efforts to further reduce the size of the kernel to~$4k$, by Guo~\cite{Guo09}, and then to~$2k$ by Chen and Meng~\cite{chen2012kernel} and by Cao and Chen~\cite{CC12} independently. The same efforts were put in trying to obtain the best possible complexity for \FPT algorithms solving these modification problems. Gramm et al.~\cite{GGH+05} first obtained a~$2.27^k$ complexity for editing and~$1.77^k$ for deletion, which was improved by B{\"o}cker and Damaschke~\cite{BD11} to~$1.76^k$ and~$1.41^k$ respectively.

Van Bevern, Froese, and Komusiewicz~\cite{DBLP:journals/mst/BevernFK18} looked at parameterized algorithms and kernelization for graph modification problems above packing guarantee. For example, if an input graph~$G$ contains~$\ell$ modification-disjoint induced~$P_3$s (no pair of these~$P_3$s share an edge or non-edge), then in order to be transformed into a cluster, graph~$G$ requires at least~$\ell$ edits.  Then a perhaps more ``honest'' question is whether~$\ell+k$ edits will suffice. For \pname{Cluster Editing}, Li, Pilipczuk, and Sorge~\cite{LiPiliSor19} show that the problem is \NP-complete for~$\ell=0$.

\begin{open}[\cite{DBLP:journals/mst/BevernFK18}]
  Is \pname{Cograph Editing} (editing to a $P_4$-free graph) with~$\ell +k$ edits,  where~$\ell$ is the number of vertex disjoint induced~$P_4$s in the input graph,    \FPT parameterized by~$k$?
\end{open}

Many variants of the problem of cluster editing have been considered in the literature. They are not listed in Table~\ref{tab:finite} and we report them below.
%%% S-PLEX
Guo et al.  generalized the \pname{Cluster Editing} problem to a problem called \pname{$s$-Plex Editing}~\cite{guo2010more}. An~$s$-plex is one way of generalizing the notion of a clique. A set~$S$ is an~$s$-plex in a graph~$G$ if every vertex~$v \in S$ has degree at least~$|S| - s$ in~$G[S]$.  Hence, a clique is a~$1$-plex. A graph~$G$ is then an~$s$-plex cluster if every connected component is an~$s$-plex. They show that the~$s$-plex cluster graphs are characterizable by a finite set of forbidden induced subgraphs and they give an $\Oh(s^2k)$ vertex kernel for the problem as well as an $\Oh(\sqrt{s}^k)$-time \FPT algorithm. It is worth noting that the number of obstructions of these classes depends exponentially on $s$, but each of the obstructions is of  size $\Oh(s)$.

In \cite{FGK+11}, another relaxed version of the cluster editing problem is studied, where a vertex ($s$-vertex-overlap) or an edge ($s$-edge-overlap) is allowed to participate in at most~$s$ clusters, where~$s$ is part of the input. All the corresponding modification problems are shown to be \NP-hard when~$s\geq 1$ ($s\geq 2$ in the case of completion),~$\WOne$-hard when parameterized by~$k$ and \FPT parameterized by~$(s,k)$. The authors of \cite{FGK+11} also give an~$\Oh(k^4)$ kernel for \pname{1-edge-overlap edge Deletion} (which is exactly \pname{diamond-free edge Deletion}) and an~$\Oh(k^3)$ kernel for \pname{2-vertex-overlap edge Deletion}.

Other results about different approaches to clustering problems are given in Section~\ref{sec:cuts}.

%%% CYCLE TRANSVERSAL
\newcommand{\ct}[1]{\pname{{#1}-cycle Transversal}}%
Xia and Zhang~\cite{xia2012kernelization} studied the problems \ct{$s$} and
\ct{$\leq s$}.  In these problems, the task is to find a set of edges~$S
\subseteq E(G)$ of a given graph~$G$ of size at most some given budget~$k$, such
that every (not necessarily induced) cycle of length (at most)~$s$ in~$G$ has an
edge in~$S$.  For~$s = 3$ these problems become \pname{Triangle-Free edge Deletion},
which is known to admit a linear kernel~\cite{brugmann2009generating}.
Xia and Zhang show that \ct{$\leq s$} is \NP-complete, even on planar graphs of
maximum degree seven, for any~$s \geq 3$.
They give a~$6k^2$ vertex kernel for both \ct{$4$} and \ct{$\leq 4$}, implying that
the \pname{$\{C_3, C_4\}$-free edge Deletion} problem admits a $6k^2$ kernel. The problems were already known to admit $\Oh(k^{s-1})$ vertex kernels by a reduction to \pname{Hitting Set}~\cite{xia2012kernelization, abu-khzam2010kernelization}.

%\todo[inline]{Define the problems mentioned below!}
%%% THRESHOLD
Due to the structure of the two classes, the modification problems into threshold graphs and chain graphs are closely related. Guo~\cite{guo2007problem} gave a cubic vertex kernel for \pname{Threshold Completion} and \pname{Threshold Deletion} (the class is auto-complementary) and Bessy et al.~\cite{bessy2013polynomial} gave a quadratic kernel for \pname{Chain Deletion}. (The characterizations of all these graph classes in the form of forbidden subgraphs is given in Table~\ref{tab:finite}.) Until recently, it was unknown whether \pname{Threshold Editing} and \pname{Chain Editing} were \NP-hard or not.  This was shown 
by Drange et al.~\cite{drange2015threshold}, who   obtained  quadratic kernels for all three modification problems towards threshold graphs and chain graphs. Furthermore, \pname{Chain Deletion} was shown to be solvable in $2.57^k \polyn$ time by Liu et al.~\cite{LWY+15}.
For split graphs completion and deletion  (graphs excluding $\set{2K_2,C_4,C_5}$), Guo~\cite{guo2007problem} initially gave a~$k^4$ kernel which was later improved to~$k^2$~\cite{ghosh2015faster}.

In the same article, Guo~\cite{guo2007problem} also provided a~$k^3$ kernel for \pname{Trivially Perfect edge Completion} and polynomial~$k^7$ kernels have been obtained for the deletion and editing versions of the problem by Drange and Pilipczuk~\cite{DP17}. Nastos and Gao~\cite{NG12} designed a $ 2.45^k \polyn $ time \FPT algorithm for \pname{Trivially Perfect Deletion} which was later improved to $ 2.42^k \polyn $ by Liu et al.~\cite{LWY+15}.

%A graph is pseudosplit if it does not contain a $2K_2$ nor a $C_4$ as an induced
%subgraph (it is the $\{2K_2, C_4\}$-free graphs). The following problem is open:
%\begin{open}[\cite{liu2014overview}]
%  Does \pname{Pseudosplit Completion} admit a quadratic or linear kernel?
%\end{open}
%
%

%%%%%%%%%%%%%%%%%%%%%%%%%%%%%%%%%%%%%%%%%%%%%%%%%%%%%%%%%%%%%%%%%%%%%%%%%%%%%%
%%%%%%%%%%%%%%%%%%%%%%%%%%%%%%%%%%%%%%%%%%%%%%%%%%%%%%%%%%%%%%%%%%%%%%%%%%%%%%
%%%%%%%%%%%%%%%%%%%%%%%%%%%%%%%%%%%%%%%%%%%%%%%%%%%%%%%%%%%%%%%%%%%%%%%%%%%%%%
\subsection{Classes characterized by an infinite number of minimal forbidden subgraphs}\label{sec:graph-classes-inf}
%%%%%%%%%%%%%%%%%%%%%%%%%%%%%%%%%%%%%%%%%%%%%%%%%%%%%%%%%%%%%%%%%%%%%%%%%%%%%%
%%%%%%%%%%%%%%%%%%%%%%%%%%%%%%%%%%%%%%%%%%%%%%%%%%%%%%%%%%%%%%%%%%%%%%%%%%%%%%
%%%%%%%%%%%%%%%%%%%%%%%%%%%%%%%%%%%%%%%%%%%%%%%%%%%%%%%%%%%%%%%%%%%%%%%%%%%%%%

%%%%%%%%%%%%%%%%%%%%%%%%%%%%%%%%%%%%%%%%%%%%%%%%%%%%%%%%%%%%%%%%%%%%%%%%%%%%%%
\begin{table}
\normalsize
\centering
\setlength\tabcolsep{0cm}

%\begin{center}
\hspace*{-1.2cm}\begin{tabular}{| >{\centering}m{2cm} || >{\centering}m{2.6cm} | >{\centering}m{2.6cm} || >{\centering}m{2.6cm} | >{\centering}m{2.6cm} || >{\centering}m{2.6cm} | >{\centering}m{2.6cm} |}
\hline
graph class & \multicolumn{2}{c||}{completion} & \multicolumn{2}{c||}{deletion} & \multicolumn{2}{c|}{editing} \tabularnewline
\hline
 & KERNEL & \multicolumn{1}{c||}{\begin{tabular}{>{\centering}m{2.6cm}} \FPT \tabularnewline\hline SUBEXP \end{tabular}} & KERNEL & \multicolumn{1}{c||}{\begin{tabular}{>{\centering}m{2.6cm}} \FPT \tabularnewline\hline SUBEXP \end{tabular}} & KERNEL & \multicolumn{1}{c|}{\begin{tabular}{>{\centering}m{2.6cm}} \FPT \tabularnewline\hline SUBEXP \end{tabular}} \tabularnewline
\hline
\hline

%graph class & KER COMP & \multicolumn{1}{c||}{\begin{tabular}{>{\centering}m{2.6cm}} \FPT COMP \tabularnewline\hline LOW COMP \end{tabular}} & KER DEL & \multicolumn{1}{c||}{\begin{tabular}{>{\centering}m{2.6cm}} \FPT DEL \tabularnewline\hline LOW DEL \end{tabular}} & KER EDIT & \multicolumn{1}{c|}{\begin{tabular}{>{\centering}m{2.6cm}} \FPT EDIT \tabularnewline\hline LOW EDIT \end{tabular}} \tabularnewline \hline

Linear forest & \multicolumn{2}{c||}{P} & $9k$~\cite{feng2014randomized} & \multicolumn{1}{c||}{\begin{tabular}{>{\centering}m{2.6cm}} $2.29^k$~\cite{feng2014randomized} randomized \tabularnewline\hline  \cellcolor{red} NOSUB (Hamiltonicity)\end{tabular}} & \multicolumn{2}{c|}{as deletion} \tabularnewline
\hline

Distance-hereditary & OPEN & \multicolumn{1}{c||}{\begin{tabular}{>{\centering}m{2.6cm}} \FPT (from~\cite{courcelle2001fixed}) \tabularnewline\hline - \end{tabular}} & OPEN & \multicolumn{1}{c||}{\begin{tabular}{>{\centering}m{2.6cm}} \FPT (from~\cite{courcelle2001fixed}) \tabularnewline\hline \cellcolor{red} NOSUB~\cite{drange2015exploring,DP17} \end{tabular}} &OPEN & \multicolumn{1}{c|}{\begin{tabular}{>{\centering}m{2.6cm}} \FPT (from~\cite{courcelle2001fixed}) \tabularnewline\hline \cellcolor{red} NOSUB~\cite{drange2015exploring,DP17} \end{tabular}} \tabularnewline
\hline

Planar & \multicolumn{2}{c||}{P} &OPEN & \multicolumn{1}{c||}{\begin{tabular}{>{\centering}m{2.6cm}} \FPT~\cite{kawarabayashi2007computing}
    (minor closed~\cite{robertson1994graph}) \tabularnewline\hline OPEN \end{tabular}} & \multicolumn{2}{c|}{as deletion} \tabularnewline
\hline

$H$-minor-free & \multicolumn{2}{c||}{P} &OPEN & \multicolumn{1}{c||}{\begin{tabular}{>{\centering}m{2.6cm}}\FPT minor closed~\cite{robertson1994graph}  \tabularnewline\hline OPEN \end{tabular}} & \multicolumn{2}{c|}{as deletion} \tabularnewline
\hline

Bipartite & \multicolumn{2}{c||}{P} & $k^{3}$~\cite{kratsch2014compression}$^\dagger$ randomized & \multicolumn{1}{c||}{\begin{tabular}{>{\centering}m{2.6cm}} $2^k$~\cite{guo2006compressionbased}  $1.977^k$~\cite{PilipczukPW19} \tabularnewline\hline \cellcolor{red} NOSUB (folk.) \end{tabular}} & \multicolumn{2}{c|}{as deletion} \tabularnewline
\hline
\hline

3-leaf power & $k^3$~\cite{bessy2010polynomial} & \multicolumn{1}{c||}{\begin{tabular}{>{\centering}m{2.6cm}} \FPT~\cite{DGH+06} \tabularnewline\hline OPEN \end{tabular}} & $k^3$~\cite{bessy2010polynomial} & \multicolumn{1}{c||}{\begin{tabular}{>{\centering}m{2.6cm}} \FPT~\cite{DGH+06} \tabularnewline\hline \cellcolor{red} NOSUB (Clustering) \end{tabular}} & $k^3$~\cite{bessy2010polynomial} & \multicolumn{1}{c|}{\begin{tabular}{>{\centering}m{2.6cm}} \FPT~\cite{DGH+06} \tabularnewline\hline \cellcolor{red} NOSUB (Clustering) \end{tabular}} \tabularnewline
\hline

$4$-leaf power & OPEN & \multicolumn{1}{c||}{\begin{tabular}{>{\centering}m{2.6cm}} \FPT~\cite{dom2005extending, dom2008closest} \tabularnewline\hline - \end{tabular}} & OPEN & \multicolumn{1}{c||}{\begin{tabular}{>{\centering}m{2.6cm}} \FPT~\cite{dom2005extending, dom2008closest} \tabularnewline\hline -\end{tabular}} & OPEN & \multicolumn{1}{c|}{\begin{tabular}{>{\centering}m{2.6cm}} \FPT~\cite{dom2005extending, dom2008closest} \tabularnewline\hline -\end{tabular}} \tabularnewline
\hline

proper interval & $k^3$~\cite{bessy2013polynomial} & \multicolumn{1}{c||}{\begin{tabular}{>{\centering}m{2.6cm}} \cellcolor{green} SUBEXP \tabularnewline\hline $2^{\Oh(k^{2/3})\log k}$~\cite{bliznets2015subexponential} NO~$2^{k^{1/4}}$~\cite{bliznets2016lower} \end{tabular}} & OPEN & \multicolumn{1}{c||}{\begin{tabular}{>{\centering}m{2.6cm}} \FPT~\cite{Cao17}  \tabularnewline\hline OPEN \end{tabular}} & OPEN & \multicolumn{1}{c|}{\begin{tabular}{>{\centering}m{2.6cm}} \FPT~\cite{Cao17}  \tabularnewline\hline OPEN \end{tabular}} \tabularnewline
\hline

interval & OPEN & \multicolumn{1}{c||}{\begin{tabular}{>{\centering}m{2.6cm}} \cellcolor{green} SUBEXP \tabularnewline\hline $2^{\sqrt{k}\log k}$~\cite{bliznets2014interval} NO~$2^{k^{1/4}}$~\cite{bliznets2016lower} \end{tabular}} & OPEN & \multicolumn{1}{c||}{\begin{tabular}{>{\centering}m{2.6cm}} $2^{\Oh(k)\log k}$~\cite{cao2016linear} \tabularnewline\hline OPEN \end{tabular}} & OPEN & \multicolumn{1}{c|}{\begin{tabular}{>{\centering}m{2.6cm}} OPEN \tabularnewline\hline OPEN \end{tabular}} \tabularnewline
\hline
    
strongly chordal & OPEN & \multicolumn{1}{c||}{\begin{tabular}{>{\centering}m{2.6cm}} $64^k$~\cite{kaplan1999tractability} \tabularnewline\hline OPEN \end{tabular}} & OPEN & \multicolumn{1}{c||}{\begin{tabular}{>{\centering}m{2.6cm}} OPEN \tabularnewline\hline OPEN \end{tabular}} & OPEN & \multicolumn{1}{c|}{\begin{tabular}{>{\centering}m{2.6cm}} OPEN \tabularnewline\hline OPEN \end{tabular}} \tabularnewline
\hline

chordal & $k^2$~\cite{natanzon2000polynomial} & \multicolumn{1}{c||}{\begin{tabular}{>{\centering}m{2.6cm}} \cellcolor{green} SUBEXP \tabularnewline\hline $2^{\sqrt{k}\log k}$~\cite{fomin2013subexponential} NO~$2^{\sqrt{k}}$~\cite{CS17} \end{tabular}} & OPEN & \multicolumn{1}{c||}{\begin{tabular}{>{\centering}m{2.6cm}} $2^{\Oh(k\log k)}$~\cite{CM16} \tabularnewline\hline OPEN \end{tabular}} & OPEN & \multicolumn{1}{c|}{\begin{tabular}{>{\centering}m{2.6cm}} $2^{\Oh(k\log k)}$~\cite{CM16} \tabularnewline\hline OPEN \end{tabular}} \tabularnewline
\hline
%\hline

%Wheel-free & \multicolumn{2}{c||}{P} & - & \multicolumn{1}{c||}{\begin{tabular}{>{\centering}m{2.6cm}} \wtwo~\cite{lokshtanov2008wheelfree} \tabularnewline\hline - \end{tabular}} & - & \multicolumn{1}{c|}{\begin{tabular}{>{\centering}m{2.6cm}} - \tabularnewline\hline - \end{tabular}} \tabularnewline
%\hline

\end{tabular}
%\end{center}
%\vspace*{-1em}
\medskip
\caption{Parameterized complexity of edge modification problems into hereditary graph classes whose number of minimal forbidden induced subgraph is infinite.  $^\dagger$~means a \corp{} kernel, and the size is the number of bits in the representation up to a polylogarithmic factor. 
NOKER means that there is no polynomial kernel, while NOSUB means that there is no parameterized subexponential algorithm (of course up to some complexity assumption).  OPEN means that the complexity is widely open, while - means that probably open but most likely nobody looked at this question. P means the problem solvable in polynomial time.
For linear forest the subexponential lower bound follows from from reduction from Hamiltonicity. For $3$-leaf deletion and editing, the lower bound follows from the lower bounds for clustering.  \label{tab:inf}}
\end{table}
%%%%%%%%%%%%%%%%%%%%%%%%%%%%%%%%%%%%%%%%%%%%%%%%%%%%%%%%%%%%%%%%%%%%%%%%%%%%%%
%\marginpar{TODO PETR: are the kernels for distance hereditary graphs really open?}

%%% CHORDAL AND SUBCLASSES
Although many studied graph classes are finitely characterizable, there are important examples that are outside this regime, such as chordal graphs (defined as graphs with no induced cycle of length at least four) or interval graphs (chordal graphs without asteroidal triples) for example. Therefore, Cai's theorem does not directly cover modification problems into chordal or interval graphs. However, consider the problem \pname{Chordal Completion}, which constituted a seminal case study for parameterized complexity of edge modification problems.
%
%\todo{Cite FV? \cite{fomin2013subexponential}}
%
Given an input instance~$(G,k)$ of \pname{Chordal Completion}, we may observe that if~$G$ has an induced cycle of length more than~$k+3$, then~$(G,k)$ is a \noinstance~\cite{cai1996fixedparameter}. Therefore, even though chordal graphs do not have a finite characterization, the set of obstacles can be bounded by a function of~$k$:  $(G,k)$ is a \yesinstance of \pname{Chordal Completion} if and only if~$(G,k)$ is a \yesinstance of \pname{$\mathcal{H}$-free edge Completion} for $\mathcal{H} = \{ C_4, C_5, \dots , C_{k+4}\}$ and the $\mathcal{H}$-free graph output is chordal.

Thanks to this fundamental property, Kaplan, Shamir, and Tarjan~\cite{kaplan1999tractability} showed as early as 1994
that \pname{Chordal Completion} (usually called \pname{Minimum Fill-In}) can be solved in $ 16^k \cdot \polyn $ time and admits a polynomial kernel with~$\Oh(k^3)$ vertices. In 1996, Cai improved their result on \pname{Chordal Completion} by giving an \FPT algorithm for the problem running in time $\Oh(4^k \cdot(n+m))$~\cite{cai1996fixedparameter} and in 2000, the analysis of the kernelization algorithm of~\cite{kaplan1999tractability} was improved by Natanzon, Shamir, and Sharan~\cite{natanzon2000polynomial} to show that it actually produces a kernel of size $\Oh(k^2)$. For deletion and editing, no polynomial kernel is known.

\begin{open}[]
  Do \ChordD{} and \ChordE{} admit polynomial kernels?
\end{open}
The related problem of deleting at most $k$ vertices to obtain a chordal graph admits a polynomial kernel \cite{AgrawalLMSZ19,JansenP18}.

A  general version of \pname{Chordal Editing} was shown to be \FPT by Cao and Marx~\cite{CM16}.  In fact, they showed that \pname{$\mathcal{G}$ $(k_1,k_2,k_3)$-Editing} (see Section~\ref{sec:graph-classes-finite} for the definition of the problem), with~$\mathcal{G}$ being the class of chordal graphs, is \FPT parameterized by $k=k_1+k_2+k_3$. That is, vertex deletion, edge deletion~\footnote{The existence of an \FPT algorithm for \pname{Chordal Deletion} had been already established by Marx~\cite{Mar10}.}, edge completion as well as edge editing to chordal graphs are all \FPT as a result. Then, the result of Cao and Marx can be seen as an extension of Cai's theorem to graph classes without finite characterizations. 
%\todo[inline]{Stopped here}

It could be interesting to see if there are natural ways of extending Cai's theorem to include this result. Answering that question, one needs to take into account that \pname{Wheel-free edge Completion} is \wtwo-hard, so any such characterization should exclude this class.

\begin{open}
Are there natural extensions of Cai's theorem to include also chordal graphs?
\end{open}

Kaplan et al.~\cite{kaplan1999tractability} also provided \FPT-like algorithms for completion into subclasses of chordal graphs, namely \pname{Strongly Chordal Completion} and \pname{Proper Interval Completion}, in $\Oh(64^k \polyn)$ time and $\Oh(16^k \polyn)$ respectively, and they asked for a similar result for \pname{Interval Completion}. This was not solved until almost ten years later, when Villanger, Heggernes, Paul, and Telle~\cite{villanger2009interval} showed that \pname{Interval Completion} was indeed fixed-parameter tractable. The complexity of the best \FPT algorithm available for the problem was later lowered to $\Oh(6^k \polyn)$ time by Cao in~\cite{Cao13,cao2016linear}. In \cite{villanger2009interval}, the authors conclude by asking specifically for a polynomial kernel for \pname{Interval Completion}.
This question was raised again in the work of Bliznets, Fomin, Pilipczuk, and Pilipczuk~\cite{bliznets2014interval} and became one notoriously hard problem in the domain, but the existence of kernels both for the subclass of proper interval graphs and for the superclass of chordal graphs makes the question particularly appealing.

\begin{open}[\cite{bessy2013polynomial, bliznets2014interval}]
  Does \pname{Interval Completion} admit a polynomial kernel?
\end{open}

\begin{open} 
  Does \pname{Interval Deletion} admit a polynomial kernel?
\end{open}
Note that the problem of deleting at most $k$ vertices to obtain an interval graph admits a  polynomial kernel \cite{AgrawalM0Z19}.

Cao \cite{cao2015interval} gave an algorithm of running time $k^{\Oh(k)} \Oh(n+m) $ for \pname{Interval Deletion}. The existence of single-exponential algorithm for this problem is open. 
\begin{open} 
 Could  \pname{Interval Deletion} be solved  in time $2^{\Oh(k)} \polyn$? 
\end{open}

\begin{open}[\cite{cao2015interval}]
  Is \pname{Interval Editing} fixed-parameterized tractable?
\end{open}

For the subclass of proper interval graphs, the \FPT running time of ~\cite{kaplan1999tractability} was improved to $\Oh(4^k \polyn)$ by Liu et al.~\cite{LWX+15}. Bessy and Perez~\cite{bessy2013polynomial} gave a polynomial kernel for \pname{Proper Interval Completion} with~$\Oh(k^3)$ vertices and Cao recently showed that \pname{Proper Interval Deletion} is \FPT, namely solvable in $\Oh(2^{\Oh(k\log k)} (n+m))$ time~\cite{cao2016linear}.

\begin{open}[\cite{bessy2013polynomial}]
Do \pname{Proper Interval Deletion} and \pname{Proper Interval Editing} admit polynomial kernels?
\end{open}

We observe that the question above is actually open for most of the subclasses of chordal graphs shown in Table~\ref{tab:inf} (except 3-leaf powers). Finding a kernel for one of these classes or proving that there is no is a question of high interest.

%%% P-LEAF POWER
Another subclass of chordal graphs that received quite a bit of attention in the parameterized framework is the class of $p$-leaf power graphs. Motivated by the problem of reconstructing evolutionary history, Nishimura et al.~\cite{nishimura2002graph} defined $p$-leaf powers as follows. Let $T$ be a tree and $L_T$ be the leaves of $T$. The $p$-leaf power of $T$ is the graph $G = (L_T, E)$ where $uv \in E$ if and only if $\dist_T(u,v) \leq p$. It follows that the $1$-leaf power graphs are the independent sets and the $2$-leaf power graphs are the cluster graphs, i.e. the $P_3$-free graphs. The editing, deletion and completion problems towards $p$-leaf power graphs are \NP-hard for every $p \geq 3$.

All three modification problems into the class of \emph{$3$-leaf-power} graphs, which are also chordal bull-dart-gem-free graphs, were shown to be \FPT by Dom et. al~\cite{DGH+06} and Bessy, Paul, and Perez~\cite{bessy2010polynomial} later showed that these three problems also admit linear time cubic vertex kernels.
The $4$-leaf power modification problems were all shown to be fixed-parameter tractable in two
articles by Dom, Guo, Hüffner, and Niedermeier~\cite{dom2005extending,dom2008closest}.
For $5$-leaf power graphs, there is a linear time recognition algorithm, which leaves the obvious open question below. The question is actually open for all $p\geq 5$, but there is currently no polynomial recognition algorithm known for $p\geq 6$.

\begin{open}[\cite{dom2008closest}]
Is \pname{5-Leaf Power Editing} (also known as \pname{Closest 5-Leaf Power}) \FPT?
\end{open}

%%% DISTANCE HEREDITARY
One large graph class for which there is no result in the parameterized framework is the class of perfect graphs. It might therefore seem reasonable to start working with modification towards some of its subclasses as a first step in gaining insight into modification towards perfect graph themselves. One of their interesting subclasses is distance-hereditary graphs. A connected graph~$G$ is a distance-hereditary graph if and only if for every two vertices~$v$ and~$u$ in~$G$, and every connected induced subgraph~$G'$ of~$G$, containing~$v$ and~$u$, $\dist_G(v,u) = \dist_{G'}(v,u)$. The class is obviously hereditary and it is exactly the house, hole (induced cycle of length at least five), domino, gem-free graphs, or so-called HHDG-free graphs~\cite{brandstadt1999graph}. For distance hereditary graphs, the existence of \FPT algorithms for edge modification problems is granted from \cite{courcelle2001fixed} as for any graph class~$\mathcal{G}$ with bounded rank-width and for which membership is definable in the variant of Monadic Second Order Logic without edge set quantifiers. 
Nevertheless, it would be interesting to improve the complexity of \FPT algorithms resulting from the general theorem mentioned above and the question of the existence of polynomial kernels for the three modification problems into distance hereditary graphs is still open.
%\marginpar{TODO PETR: are the kernels for distance hereditary graphs really open?}

%%% PLANAR
Another problem (or class of problems) that admits fixed-parameter tractable algorithms as a result of general tools is the problem of \pname{Planar Deletion}. Here the task is to delete at most $k$ edges to obtain a planar graph. Since the class of graphs $Planar +ke$ is minor-closed and thus  by the fundamental result of  
Robertson and Seymour \cite{RobertsonS04}  is characterized by a finite set of forbidden  minors,
 minor testing algorithm by Robertson and Seymour from the
graph minors project~\cite{robertson1994graph} implies that the problem is non-uniformly \FPT. 
\pname{Planar Deletion} was shown by Kawarabayashi and Reed to admit a linear
time \FPT algorithm~\cite{kawarabayashi2007computing}.
%
%However, as for other minor closed families, the parameterized
%tractability of the edge deletion problem is a consequence of their finite
%characterization in terms of forbidden minors.
%
Using the algorithm by Adler, Grohe, and Kreutzer~\cite{adler2008computing}
combined with the minor testing algorithm by Robertson and Seymour, we obtain uniform \FPT algorithms
for \pname{$\mathcal{H}$-minor free Deletion}. But the existence of polynomial kernels for these problems is open.
\begin{open} Does 
\pname{Planar Deletion} admit a polynomial kernel?
\end{open}

\begin{open} Does 
\pname{$\mathcal{H}$-minor free Deletion} admit a polynomial kernel?
\end{open}
%The question is  open even for very special cases 
 For the related problem of deleting at most $k$ vertices to obtain  an ${\mathcal{F}}$-minor free graph, a non-uniform polynomial kernel is known when family $\mathcal{F}$ contains at least one  planar graph \cite{FominLMS12}.  

% Jansen, Lokshtanov, and Saurabh~\cite{jansen2014nearoptimal} expect that their algorithm for \pname{$k$-Vertex Planarization} can be applied to also obtain a near-optimal algorithm for \pname{Planar Deletion}.\todo{cite what?}

%%% BIPARTITE
In 2004, \pname{Odd Cycle Transversal} (which is \pname{Bipartite Vertex Deletion}) and its edge version, called \pname{Edge Bipartization} (which is \pname{Bipartite Deletion}), were shown to be solvable in time $3^k \polyn$ by Reed, Smith and Vetta~\cite{reed2004finding}, inventing the now well-known technique \emph{iterative compression}.
\pname{Edge Bipartization} was shown later to be solvable in time~$2^k \cdot\polyn$~\cite{guo2006compressionbased}.
Iterative compression has proven to be a very successful technique. One challenge is to get it to work naturally with edge modification problems.  The technique has been extremely helpful for many vertex deletion problems, but few edge modification problems. In the case of \pname{Edge Bipartization}, one reason for the success of iterative compression is the close relation between the edge version and the vertex version of the problem; there is a parameter-preserving reduction from \pname{Odd Cycle Transversal} to \pname{Edge Bipartization}~\cite{wernicke2003algorithmic}.

Kratsch and Wahlström~\cite{kratsch2014compression} proved that there exists a
randomized compression such that \pname{Edge Bipartization} as well as the
vertex version, \pname{Odd Cycle Transversal} admits a~$k^{4.5}$ \corp{} kernel.
Here, \corp{} allows false positives in the sense that if an instance is a
no-instance, then the compressed instance is a no-instance with probability at
least~$1/2$, while any yes-instance will be compressed to a yes-instance.
Here, we may boost the success probability by running the algorithm polynomially
in~$k$ many times (not polynomial in~$n$ as that would defeat the purpose of a
kernelization procedure), and the output instance will then be the ``and'' over
all the compressed instances. Nevertheless, the question is still open in deterministic settings.

\begin{open}[\cite{drange2015thesis}]
  Does \pname{Edge Bipartization} admit a deterministic polynomial kernel?
\end{open}

%%% LINEAR FOREST
Finally, let us mention the class of \emph{linear forests}, which are the graphs whose connected components are paths. Though the class is pretty simple, it does not admit a finite number of forbidden subgraphs. Feng, Zhou, and Li showed that \pname{Linear Forest Deletion} admits a polynomial kernel with $9k$ vertices~\cite{feng2014randomized}. They also provide an $\Oh(2.29^k \polyn)$ time randomized \FPT algorithm for solving the problem.

%%% NOT YET RESULTS

There are many important hereditary classes for which the parameterized complexity of the edge modification problems is still unknown. Among them are comparability, co-comparability and permutation graphs, which are subclasses of perfect graphs, as well as circular-arc and circle graphs. Obtaining positive or negative results for any of these classes would be of high interest. In particular, the following questions were already asked in the literature.

\begin{open}[\cite{drange2015thesis}]
  Are any of \pname{Comparability Completion}, \pname{Comparability Deletion} and \pname{Permutation Completion} in \FPT?
\end{open}

\begin{open}[\cite{hof2013proper}]
  Is \pname{Proper Circular Arc Deletion} in \FPT?
\end{open}

\begin{open}[\cite{villanger2009interval, heggernes2013parameterized}]
  Is \pname{Perfect Deletion} in \FPT?
\end{open}

%%%%%%%%%%%%%%%%%%%%%%%%%%%%%%%%%%%%%%%%%%%%%%%%%%%%%%%%%%%%%%%%%%%%%%%%%%%%%%
%%%%%%%%%%%%%%%%%%%%%%%%%%%%%%%%%%%%%%%%%%%%%%%%%%%%%%%%%%%%%%%%%%%%%%%%%%%%%%
%%%%%%%%%%%%%%%%%%%%%%%%%%%%%%%%%%%%%%%%%%%%%%%%%%%%%%%%%%%%%%%%%%%%%%%%%%%%%%
\subsection{Subexponential time algorithms}\label{sec:graph-classes-SUBEXP}
%%%%%%%%%%%%%%%%%%%%%%%%%%%%%%%%%%%%%%%%%%%%%%%%%%%%%%%%%%%%%%%%%%%%%%%%%%%%%%
%%%%%%%%%%%%%%%%%%%%%%%%%%%%%%%%%%%%%%%%%%%%%%%%%%%%%%%%%%%%%%%%%%%%%%%%%%%%%%
%%%%%%%%%%%%%%%%%%%%%%%%%%%%%%%%%%%%%%%%%%%%%%%%%%%%%%%%%%%%%%%%%%%%%%%%%%%%%%

As usual in algorithms, a natural, but difficult, question to ask is about lower bounds. The case of parameterized complexity is not different. Once a problem has been shown to admit a fixed-parameter tractable
algorithm, a natural next question is whether it is possible to improve upon
that algorithm.  This is especially interesting when the algorithms have running
times that are of the order $2^{\Omega(k^2)} \cdot \polyn$, or even
$2^{\Omega(2^k)} \cdot \polyn$.
 
As mentioned above, the modification problems for finite forbidden induced
subgraphs already have nice running times like $6^k \cdot \polyn$ or even $3^k
\cdot \polyn$ and $2^k \cdot \polyn$.  Is it possible to obtain faster
algorithms?  Can we improve from $3^k\polyn$ to, say, $2^k\polyn$ or $1.5^k\polyn$?
Are there reasons to suspect that we cannot get better than $2^k\cdot\polyn$ algorithms?  These questions were at the core of what is known as the \emph{optimality programme}, see for example \cite{marx2012whats}.

Simultaneously with the optimality programme and the development of polynomial kernel theory, some problems were shown to be solvable in \emph{subexponential parameterized time}, i.e., in time $2^{o(k)} \polyn$, or $(1+\epsilon)^k \polyn$ for every $\epsilon > 0$, and there was a strong interest in identifying parameterized problems that admits such subexponential parameterized algorithms. The complexity class of problems admitting such an algorithm is called \cclass{SUBEXP} and was defined by Flum and Grohe in their seminal work on parameterized complexity~\cite{flum2006parameterized}. They noticed that most natural problems do in fact \emph{not} live in this complexity class: the classical \NP-hardness reductions paired with the \emph{Exponential Time hypothesis} (\ETH) of Impagliazzo et al.~\cite{impagliazzo2001which} is enough to show that no $2^{o(k)} \cdot \polyn$ algorithm exists.

%
%Although for many classical parameterized problems like \probVC or \probFVS already known \NP-hardness
%reductions show that the existence of such an algorithm would contradict the
%\emph{exponential time hypothesis} of Impagliazzo, Paturi, and
%Zane~\cite{impagliazzo2001which}.  

%
%
% CHEN's QUESTION
%
%

As the first known subexponential parameterized algorithms were for problems with restricted input graphs, such as planar, or more generally $H$-minor free graphs~\cite{demaine2005subexponential}, Chen posed the following question~\cite{bodlaender2006open}: are there examples of natural problems on graphs, that do not have such a topologically constrained input, and also admit subexponential parameterized algorithms?

Such  a problem was first found by  Alon, Lokshtanov, and Saurabh~\cite{alon2009fast} who designed a new algorithmic technique called \emph{chromatic coding} and used it to solve \pname{Feedback Arc Set on Tournaments} (\pname{FAST}) on tournament graphs in time $2^{\Oh(\sqrt k \log k)} \cdot \polyn$. A tournament graph is a directed graph obtained from a compete undirected graph by choosing an orientation for each edge. Then \pname{FAST}  is the problem of identifying at most $k$ arcs  in the given tournament whose deletion transforms the tournament into an acyclic graph.  
 \pname{FAST} is also known to admit a quadratic vertex kernel~\cite{dom2010fixedparameter} which was improved to a linear kernel by Bessy et al.~\cite{bessy2011kernels}.

%However, even though tournament graphs are not topologically constrained, they still constitute a restricted input in the sense that they are a very particular case of directed graphs. Hence, Chen's question was therefore not considered fully
%answered---are there problems which are in \cclass{SUBEXP} on general input graphs? The question was settled in full when 

Fomin and Villanger~\cite{fomin2013subexponential} gave an algorithm for \pname{Chordal Completion} (\pname{Minimum Fill-In}) using ideas from the techniques developed for minimal triangulations and treewidth computations. Numerous $2^{\Oh(k)} \polyn$ algorithms were known~\cite{cai1996fixedparameter, kaplan1999tractability, bodlaender2011faster} for \pname{Chordal Completion}, but  Fomin and Villanger proved that this problem is solvable in time $\Oh(2^{\Oh(\sqrt k \log k)} + k^2 n m)$.  The additive polynomial factor was due to first preprocessing the graph, thereby obtaining a kernelized instance of polynomial size. The main tools in this algorithm were that of \emph{minimal triangulations and potential maximal cliques}, a framework developed  by Bouchitté and Todinca~\cite{bouchitte2001treewidth, bouchitte2002listing}, see also~\cite{fomin2008exact}.

%A potential maximal clique (given~$G$ and~$k$) is a set of vertices $\Omega \subseteq V(G)$ for which there is a minimal triangulation of size at most~$k$ in which~$\Omega$ is a clique.
%It remains an open question to determine whether the number of potential maximal cliques is always bounded subexponentially in~$k$\todo{or some other $\mathcal{Q}$?}.
%\begin{open}
%  Given a graph $G$ and an integer $k$ as input to the \pname{Chordal
%    Completion} problem, is the number of potential maximal cliques bounded by
%  $2^{o(k)} \polyn$?\todo{what? citation?}
%\end{open}

Following the results of Fomin and Villanger, several new subexponential
parameterized time completion results followed.  Based on the chromatic coding technique of Alon et al.~\cite{alon2009fast}, Ghosh et al.~\cite{ghosh2015faster}
gave an algorithm with the same running time\footnote{The best complexity known for the problem does not need the $\log k$ factor in the exponent, see Exercise~5.17 in~\cite{cygan2015parameterized}.}, $2^{\Oh(\sqrt k \log k)} +
\polyn$, for \pname{Split Completion}, thus also giving an algorithm for the
equivalent problem of \emph{deleting} to a split graph.  A natural question
arose again on the complexity of completing to~$\mathcal{H}$-free graphs: Could
this be subexponential time for all~$\mathcal{H}$? for finite $\mathcal{H}$?
The result by Lokshtanov~\cite{lokshtanov2008wheelfree} again immediately gives
a negative result here, as his result implies that for~$\mathcal{H}$ being the
complement of the wheels, \pname{$\mathcal{H}$-free edge Completion}, that is
\pname{co-wheel-free edge Completion}, is \wtwo-hard.
So for general $\mathcal{H}$, the answer is indeed clearly negative. Therefore, a next
question was to look for simple $\mathcal{H}$.

And while the classes of chordal and split graphs are rather ``simple'', they
certainly are much more complex than the simple cluster %or bicluster
graphs.
Therefore, the problems \pname{Cluster Editing} and \pname{Cluster Deletion}
were natural candidates for subexponential time algorithms%, together with the similar question for bicluster graphs
. From Cai's theorem, we immediately obtain $2^k\polyn$ and $3^k\polyn$
algorithms for \pname{Cluster Deletion} and \pname{Cluster Editing}, respectively.
This question was first answered in the negative by Komusiewicz and Uhlmann
studying this problem on bounded degree graphs~\cite{komusiewicz2012cluster},
and then independently by Fomin et al.~\cite{fomin2014tight}.
Again somewhat surprisingly, we cannot expect algorithms running in time
$2^{o(k)} \polyn$ solving \pname{Cluster Editing}.  Komusiewicz and Uhlmann
gave an elegant reduction proving that both parameterized and exact
subexponential time algorithms are not achievable, unless the exponential time
hypothesis fails~\cite{komusiewicz2012cluster}.
In other words, under the exponential time hypothesis, there is an $r \in (1,2]$
such that none of these problems are solvable in time $r^k \polyn$.

Following the subexponential algorithm for \pname{Chordal Completion} and \pname{Split Completion}, it was shown that \pname{Trivially Perfect Completion}, as well as \pname{Chain Completion}, \pname{Threshold Completion}, and \pname{Pseudosplit Completion} all were solvable in subexponential parameterized time~\cite{drange2015exploring}. They simultaneously give negative results, showing that neither completing to a cograph (and thus also deleting, since the class is auto-complementary), deleting to trivially perfect graphs, nor completing to $C_4$-free or $2K_2$-free graphs are in SUBEXP under
ETH. Later, \pname{Trivially Perfect Editing}~\cite{DP17} and \pname{Starforest\footnote{Starforest are the graphs where each connected component is a star, they are also the triangle-free trivially perfect graphs.} Deletion}~\cite{DRS+15} were also added to the list of problems that are not in SUBEXP under ETH. 

Then followed two results by Bliznets et al.~\cite{bliznets2014interval,
  bliznets2015subexponential}, that \pname{Interval Completion} and
\pname{Proper Interval Completion} both are solvable in subexponential time, $2^\osl\polyn$ and $2^{\Oh(k^{2/3} \log k)} + \polyn$, respectively.

\begin{open}[\cite{bliznets2015subexponential, bodlaender2014graph}]
  Does \pname{Proper Interval Completion} admit an algorithm of running time
  $2^\osl \polyn$?
\end{open}

Drange et al.  gave algorithms for \pname{Threshold
  Editing} and \pname{Chain Editing} running in time $2^{\Oh(\sqrt k \log k)} +
\polyn$, thereby adding these problems to the line of subexponential
parameterized time solvable problems~\cite{drange2015threshold}.
These two graph classes, threshold and chain graphs, are the only classes known
for which all three edge modification problems are \npci{} and solvable in
subexponential parameterized time.
Drange et al. ~\cite{drange2015exploring} showed that
also for \pname{Trivially Perfect Completion}, or \pname{$\{C_4, P_4\}$-free
edge  Completion}, as well as for \pname{Pseudosplit Completion} and
\pname{Threshold Completion}, we have subexponential time algorithms.

Later a problem known as \pname{Clique Editing}, or \pname{Sparse Split Editing}
was introduced as a model for core/periphery
structures~\cite{borgatti2000models}, and for noise
reduction~\cite{damaschke2014editing}.
This problem consists of editing a graph to a disjoint union of a clique and an
independent set, or, \pname{$\{2K_2, P_3\}$-free edge Editing}. The problem was shown to be \NP-hard independently by Damaschke and
Mogren~\cite{damaschke2014editing} and Kovác, Selecéniová, and Steinová~\cite{kovac2014clique} and is solvable in subexponential time. Indeed, a polynomial
kernel is quite trivial after a twin reduction rule, and then the result follows
from guessing a vertex in the clique and a (small) number of other vertices with which its adjacency relationships have to be changed. Damaschke and Mogren showed that several similar problems are solvable in
subexponential parameterized time and they showed that \pname{Clique Deletion}
is solvable in time~$O (1.6355^{\sqrt{k \ln k}}\polyn)$~\cite{damaschke2014editing}.

Using the \pname{$H$-Bag Editing} problem from Damaschke and Mogren~\cite{damaschke2014editing}, Meesum, Misra, and
Saurabh~\cite{meesum2015reducing} show that the \pname{$r$-Rank Reduction
  Editing} problem is solvable in time $2^{\Oh(\sqrt k \log k)} \cdot \polyn$.
In this problem, we are asked to edit the input graph $G$ to a graph $G'$ by
modifying at most $k$ edges so that $\rank(A_{G'})$, the rank of the adjacency
matrix of $G'$ is at most $r$. A similar result is shown in~\cite{meesum2016rank} for the directed case.

There are still some graph classes for which the question of whether the edge modification problems admit a subexponential algorithm is not entirely settled. In particular, the case of triangle free graphs and 3-leaf powers is appealing since there exist some polynomial kernels for these problems.

\begin{open}
Do the following problems admit subexponential parameterized algorithms: \pname{Triangle-free edge Deletion}, \pname{Linear Forest edge Deletion}, \pname{Planar edge Deletion} and \pname{$3$-Leaf Powers edge Completion} (as well as \pname{deletion} and \pname{editing})?
\end{open}

As for the lower computational bounds, 
 many problems are known not to be solvable in time $2^{o(k)} \polyn$, that is they do not admit subexponential parameterized algorithms, under some complexity hypothesis such as $P\neq NP$ or \ETH{} or \noph, see Tables~\ref{tab:finite} and~\ref{tab:inf}. For many problems  the question of obtaining lower bounds on the subexponential complexity is open.  
 
Fomin and Villanger~\cite{fomin2013subexponential} noted that, unless \ETH{}  fails, \pname{Chordal Completion} cannot be solved in time $2^{o(k^{1/6})} \polyn$. Later, Bliznets et
al.~\cite{bliznets2016lower} showed that this can be tightened quite a bit: unless
\ETH{} fails, there is a positive natural number~$c >
1$ such that \pname{Chordal Completion} can not be solved in time $2^{\Oh(k^{1/4} / \log^c k)} \polyn$, and the same lower bound holds for \pname{Interval Completion}, \pname{Proper Interval Completion}, \pname{Trivially Perfect Completion},
\pname{Threshold Completion} (and so \pname{Threshold Deletion} since the class is auto-complementary).
This, however, still leaves a gap for almost all the problems between $k^{1/2}$
and $k^{1/4}$ in the exponent. Is the correct running times for these problems
closer to $2^{\Oh(k^{1/4} / \log^c k)} \polyn$, to $2^{\Oh(k^{1/2})} + \polyn$
or to $2^\osl + \polyn$?
For chordal graphs, we know that the exponent $1/2$ of $k$ is optimal as it was shown by Cao and Sandeep~\cite{CS17} (again up to \ETH{}). Therefore, the only open question is on the optimality of the $2^{\Oh(k^{1/2}\log k)}$. For \pname{Proper Interval Completion} the gap on the exponent of $k$ is larger than for the other problems cited above since we
only know an algorithm running in time $k^{\Oh(k^{2/3})} +
\polyn$~\cite{bliznets2015subexponential}.

There were also some attempts to obtain general results about the (non-)existence of subexponential parameterized algorithms for edge modification problems into $H$-free graph classes~\cite{ASS17b}. These are results of impossibility: when $H$ has at least two edges (resp. non-edges), $H$-free edge deletion (resp. completion) is NP-complete and not in SUBEXP; when $H$ has at least three vertices, $H$-free edge editing is NP-complete and not in SUBEXP. 

\begin{theorem}[\cite{ASS17b}]\label{thm:ASS}
Let $\mathcal{G}$ be a  hereditary class of graphs characterized by    graph $H$. Then unless \ETH{} fails, 
\begin{itemize}
\item If $H$ has less than two edges, then \probPedel is solvable in polynomial time. Otherwise, the problem cannot be solved in time
$2^{o(k)}\cdot \polyn$ unless \ETH{} fails. 
\item If $H$ has less than two non-edges, then  \probPedad is solvable in polynomial time. Otherwise, the problem cannot be solved in time
$2^{o(k)}\cdot \polyn$ unless \ETH{} fails. 
\item 
If $H$ has less than three vertices, then \probPeded is solvable in polynomial time. Otherwise, the problem cannot be solved in time
$2^{o(k)}\cdot \polyn$ unless \ETH{} fails. 
\end{itemize}
\end{theorem}

Even more recently, such kind of results where extended to the question of the existence of approximation algorithms~\cite{BCK+18}: when $H$ is 3-connected and has at least two non-edges, then there does not exist any poly(OPT)-approximation algorithm running in parameterized subexponential time (in OPT), unless ETH fails, for $H$-free edge deletion as well as $H$-free edge completion. Moreover, the same holds for $H$ being a cycle on at least $4$ vertices or a path on at least $5$ vertices. With previous results, this solves all cases of paths and cycles except the cograph edge deletion problem, for which \cite{BCK+18} suggests the existence of a parameterized subexponential approximation algorithm, because of the existence of a kernel to the problem~\cite{guillemot2013nonexistence}.

Among the most interesting open questions in the topic of subexponential parameterized algorithms is to explain why some problems admit such algorithms, which we saw is an exceptional case. In particular, one can ask whether the existence of a polynomial kernel is a prerequisite for a \pname{$\mathcal{H}$-free Modification} problem to admit a subexponential time algorithm. Actually, there are examples of problems that admit subexponential time algorithms and that do not
have polynomial kernels under the assumption of \noph, but these problems are not of the \pname{$\mathcal{H}$-free Modification} type. Indeed, it is easy to come up with problems that trivially or-cross-composes, like the
\pname{or-Minimum Fill-In} which asks whether a graph has a connected component
that can be completed to a chordal graph. This problem cannot have a polynomial
kernel under \noph, but does admit a subexponential time algorithm, simply by
running the algorithm by Fomin and Villanger~\cite{fomin2013subexponential} for
each connected component. However, the \pname{or-Minimum Fill-In} problem is not of the \pname{$\mathcal{H}$-free Modification} type and it turns out that for all such problems that we know to admit a subexponential time
algorithm, we also have polynomial kernels---with the possible exception of
\pname{Interval Completion} for which existence of a polynomial kernel remains open.

\begin{open}
  Does there exist an $\mathcal{H}$ for which \pname{$\mathcal{H}$-free
    Completion} or \pname{$\mathcal{H}$-free Editing} is solvable in time
  $2^{o(k)} \polyn$ but does not admit a polynomial kernel unless \ph?
\end{open}

Finally, another important question to address is about the tools used to show lower bounds. Many results are established on complexity hypotheses that are not as reliable as the assumption that $P\neq NP$, such as, for example, the assumption that \noph. It would be highly desirable to develop the necessary techniques to fond more of these results on the sole hypothesis that $P\neq NP$.

\begin{open}
Which lower bounds can be shown assuming only $P\neq NP$?
\end{open}

%%% Local Variables:
%%% mode: latex
%%% TeX-master: "survey"
%%% End:

%%%%%%%%%%%%%%%%%%%%%%%%%%%%%%%%%%%%%%%%%%%%%%%%%%%%%%%%%%%%%%%%%%%%%%%%%%%%%%
%%%%%%%%%%%%%%%%%%%%%%%%%%%%%%%%%%%%%%%%%%%%%%%%%%%%%%%%%%%%%%%%%%%%%%%%%%%%%%
%%%%%%%%%%%%%%%%%%%%%%%%%%%%%%%%%%%%%%%%%%%%%%%%%%%%%%%%%%%%%%%%%%%%%%%%%%%%%%
\subsection{Related results}\label{sec:graph-classes-related}
%%%%%%%%%%%%%%%%%%%%%%%%%%%%%%%%%%%%%%%%%%%%%%%%%%%%%%%%%%%%%%%%%%%%%%%%%%%%%%

%%% BIPARTITE
In some cases, the input graph is naturally a bipartite graph.
%%% CHAIN COMPLETION
The \pname{Chain Completion} problem is the problem of making a bipartite graph a bipartite chain graph, that is, a bipartite graph with no induced $2K_2$. The problem was first shown to admit a polynomial kernel by Guo~\cite{guo2007problem} when the bipartition is fixed. Fomin and Villanger showed that this version of the problem admits a subexponential time algorithm~\cite{fomin2013subexponential} and Bliznets et al.~\cite{bliznets2016lower} showed that it cannot be solved in time $\Oh(2^{k^{1/4}})$ unless ETH fails. Drange et al.~\cite{drange2015threshold} relaxed the input requirements, showing that the problem still admits a quadratic kernel even when the bipartition is not fixed.

\medskip

%%% FLIP CONSENSUS
In the \pname{Minimum Flip Consensus Tree} problem, we are asked to turn an input bipartite graph into a bipartite graph with same partition that contains no $P_5$ starting from any top vertex, called a \emph{consensus tree}.
This kind of graph, a consensus tree, arises in computational phylogenetics, with the bottom vertices being characters and the top vertices being the taxa.
The problem is solvable in time $c^k \polyn$ by Cai's theorem. Chen~\cite{chen2006minimumflip} proved that it is \npci and gave an $\Oh(6^k n^2)$ \FPT algorithm, which was later improved to $\Oh(4.42^k n)$ by B\"{o}cker, Bui and Truss \cite{BBT12}. Finally, Komusiewicz and Uhlmann~\cite{komusiewicz2014cubicvertex} gave a $\Oh(3.68^k n^3)$ algorithm and a $\Oh(k^3)$ kernel for \pname{Minimum Flip Consensus Tree}.

\medskip

%%% BIPARTITE CLUSTER
Several variants of cluster editing were introduced for the special case of bipartite graphs. The main one, called \BicE{}, aims at obtaining a union of complete bipartite graphs. It admits a $4k$ linear kernel and a \FPT algorithm running in $\Oh(3.24^k+|E|)$~\cite{guo2008improvedbicluster}.
Drange et al.~\cite{DRS+15} considered the extension of \pname{$p$-Cluster Editing} to \BicE{} and the more general \pname{$t$-Partite Cluster Editing}, yielding the problems \pname{$p$-Bicluster Editing} and \pname{$t$-Partite $p$-Cluster Editing}. None of the classical parameterized versions are solvable in subexponential time, but fixing the number $p$ of connected components in the solution, the problems become solvable in subexponential time.
In \cite{DRS+15}, it is shown that a problem called \PStarE{} is solvable in time $\Oh(2^{3\sqrt{pk}} +m+n)$, whereas an algorithm of running time $2^{\Oh(p \sqrt k \log(pk))} + \Oh(m+n)$ is given for \PBicE, as well as \pname{$t$-Partite $p$-Cluster Editing}.

\medskip

%%% PLANAR
Let us also mention that for planar input graphs, Xia and Zhang~\cite{xia2012kernelization} give a linear kernel for \ct{$5$} and \ct{$\leq 5$}, thereby showing that \pname{$\{C_3, C_4, C_5\}$-free Deletion}, or \pname{Girth-$6$ Deletion} admits a linear kernel on planar graphs.

%%% KONIG
A non-hereditary variant of \pname{Edge Bipartization} is edge deletion toward K{\"o}nig graphs. An undirected graph is a K{\"o}nig graph if it admits a vertex cover of size equal to the size of its maximum matching. This class contains all bipartite graphs, but not every K{\"o}nig graph is bipartite; for example, a triangle with a pendant vertex attached to one vertex is a K{\"o}nig graph.
For the following \ProblemName{K{\"o}nig Edge Deletion} problem, it is known that it is at least as hard as 
\ProblemName{Almost 2-SAT}. Its vertex-deletion variant is known to be \FPT~\cite{LokshtanovNR12}.
    \begin{open}[\cite{CyganKPopen13}]

Consider the \ProblemName{K{\"o}nig Edge Deletion} problem where we are to delete at most $k$ edges from the given graph to obtain a K{\"o}nig graph. Is this problem \FPT {} parameterized by $k$? 
 \end{open}

 %!TEX root =survey.tex
 
\section{Connectivity,  cuts, and clustering}\label{sec:cuts}
In this section we consider  problems around edge cuts and connectivity augmentations. 
By cut problems here we mean a wide class of problems where one wants for a given (directed) graph $G$  to identify a minimum-sized set of edges $X$ (edge-cut) such that  in the new graph $G-X$ obtained by deleting $X$ from $G$, some connectivity conditions change.  For example, the condition  can be  that a set of specific terminals becomes separated or that at least one connected component in the new graph is of a certain size.  
Clustering problems  can be seen as a hybrid of connectivity and cuts, where we want to identify highly connected areas of a graph that can be easily cut from each other. 
Most of these problems
are \NP-hard, except 
 several notable
exceptions, like minimum $s-t$ cut or minimum multiway cut in
planar graphs with fixed number of terminals,  Several interesting algorithmic techniques were developed
in order to establish fixed-parameter tractability of various cut problems. 

 The ``dual" set of problems concerns with adding edges in order to augment some connectivity properties of the graph.
 
\subsection{Cuts}

\paragraph{Edge Multiway Cut}
In the \probEMWay problem, we are given 
a graph $G$, a set $T \subseteq V(G)$ of terminal vertices, and an integer $k$. The task is to decide whether 
there exists a set $X$ of at most $k$ edges of $G$ such that every element of $T$
lies in a different connected component of $G-X$.

\defparproblem             %
{\probEMWay}               % name 
{$k$}                      % parameter
{Graph $G$,  $T \subseteq V(G)$ and integer $k$} % input
{Does there exists a set $X$ of at most $k$ edges of $G$ such that every
  element of $T$ lies in a different connected component of $G-X$. }

The related problem is \probVMWay, where one wants to delete at most $k$ vertices to separate terminals. For most of the variants of the cut problems an \FPT-algorithm for edge-deletion version  can be obtained from the vertex-deletion variant. This is why most of the work in the area was concentrated on vertex-deletions. 
  % Thus most of the algorithmic work on cut problems concerns 
  
   For $|T|=2$, \probEMWay is the classical 
  \probMinCut and  is solvable in polynomial time due to  its duality with the maximum flow problem  \cite{MR0079251}. 
  %The  {Ford-Fulkerson algorithm} for
  %finding maximum flows was published in 1956 \cite{MR0079251}. For
  %other, more efficient, algorithms for finding maximum flows, see any
  %standard algorithm textbook such as Cormen et al. 
%  \cite{DBLP:books/daglib/0023376}.  
 However,  as it was shown by Dalhaus et al.~\cite{MR95h:90039},  \probEMWay    is \NP-complete for $|T|=3$. 

In influential paper~\cite{DBLP:journals/tcs/Marx06},  Marx established   fixed-parameter tractability of    \probEMWay and \probVMWay parameterized by $k$. For that  Marx developed  the technique of important separators based on submodular properties of cuts. The technique appeared to be handy for many problems in this area. Algorithms for  \probEMWay with running times $2^k\cdot
  n^{\Oh(1)}$ and  $1.84^k\cdot n^{\Oh(1)}$ were given by 
Xiao~\cite{DBLP:journals/mst/Xiao10} and   Cao, Chen, and Fan  \cite{Cao2014167},  correspondingly.  Chapter~8 of the textbook \cite{cygan2015parameterized} contains an overview of basic techniques around important separators and parameterized algorithms for finding cuts in graphs.

\probEMWay remains \NP-complete on planar graphs but as it was shown by Dalhaus et al. ~\cite{MR95h:90039},  for a fixed number  of
  terminals,   the problem can be solved in time
  $n^{\Oh(|T|)}$   on planar graphs. The running time for planar graphs was improved
  to $2^{\Oh(|T|)}\cdot n^{\Oh(\sqrt{|T|})}$ by Klein and
  Marx~\cite{DBLP:conf/icalp/KleinM12}.

Lokshtanov and Ramanujan~\cite{DBLP:conf/icalp/LokshtanovR12} studied  the version of    \probVMWay  called \ProblemName{Parity Multiway Cut}.
Here the terminal set $T$ consists of  two not necessarily disjoint subsets  $T_o$ and $T_e$.  The objective is to decide whether there exists a $k$-sized vertex (or edge) subset $S$ such that $S$ intersects all odd-lenght paths from $v \in T_o$  to $T -v$ and all even-length paths from  $v \in T_e$  to $T -v$.
The edge-deletion case with $T_o=T_e$ is exactly \probEMWay.  Lokshtanov and Ramanujan proved that both edge- and vertex-deletion versions of \ProblemName{Parity Multiway Cut} are \FPT{} parameterized by $k$. Chandrasekaran and
              Mozaffari in 
\cite{ChandrasekaranM17} studied parity variants of these problem on directed acyclic graphs.

The {random sampling of important separators} technique developed  by Lokshtanov
  and Ramanujan  was   generalized to directed graphs 
by Chitnis et al.~\cite{doi:10.1137/12086217X} who showed the fixed-parameter
  tractability of \probDirEMWay and \probDirVMWay parameterized by the
  size of the solution.

The technique based on important separators was used by 
Chen et al.~\cite{DBLP:journals/jacm/ChenLLOR08} in their \FPT {} algorithm for \probDFVS, and
  \probDFAS, which parameterized complexity was open for a long time. In this problem the task is to decide whether at most $k$ vertices (or correspondingly, arcs) can be removed from a directed graph such that the resulting graph is acyclic. 
The generalization  of the problem, namely     \ProblemName{Directed Subset Feedback Vertex Set}, was studied 
  by Chitnis et al. ~\cite{chitnis2015directed}.  Xiao and   Nagamochi gave an \FPT{} algorithm for   \ProblemName{Subset Feedback Arc Set}
\cite{XiaoN12}. Kratsch et al. in  \cite{KratschLMPW18} define multi-budgeted variant 
of   \probDFAS and some versions of \textsc{Min-Cut} and establish fixed-parameter tractability of these problems.
The existence of a polynomial kernel for   \ProblemName{Directed   Feedback Vertex Set} and for 
  \ProblemName{Directed   Feedback Arc Set} is widely open. 

\begin{open} Do   \probDFVS and   \probDFAS admit a polynomial kernel? 
  \end{open}
  
 Lucchesi and Younger \cite{Lucchesi} proved that on  planar graphs   \probDFAS is solvable in polynomial time.  
\begin{open}Does 
  \probDFVS  admit a polynomial kernel on planar graphs? 
  \end{open}
Whether the running time $k^{\Oh(k)}\polyn$ of Chen et al.~\cite{DBLP:journals/jacm/ChenLLOR08}  for  \probDFVS  is  tight,  is another open question. 
 
\begin{open}
 Could \probDFVS and  \probDFAS be solved in time $2^{O(k)} n^{O(1)}$? Could \probDFVS  be  solved in time $2^{O(k)} n^{O(1)}$ on planar graphs?
  \end{open}

\paragraph{Edge Multicut}
In the related  \probEMCut  problem, 
we are given 
a graph $G$, a set of pairs $(s_i,t_i)_{i=1}^\ell$ of vertices of $G$, and an integer $k$. 
The question is if  there exists a set $X$ of at most $k$ edges of $G$ such that for every $1 \leq i \leq \ell$,
vertices $s_i$ and $t_i$ lie in different connected components
 of $G-X$. The vertex-deletion version of the problem is \probVMCut.

\probEMCut is \NP-hard on trees   \cite{GargVaziraniYannakakis1997}. Guo and
  Niedermeier~\cite{NET:NET20081} obtained a $2^k\cdot
  n^{\Oh(1)}$-time algorithm   for the problem on trees. 
%  The polynomial-time algorithm for
%  solving \probVMCut on trees  was
%  described first by Calinescu, Fernandes, and
%  Reed~\cite{Calinescu2003333}.
 For general graphs, 
  the fixed-parameter tractability of \probEMCut and \probVMCut
  parameterized by the solution size $k$ was a long-standing open question, which was resolved independently by
  Bousquet et al. ~\cite{bousquet2011multicut} and Marx and
  Razgon~\cite{marx-razgon-stoc2011-multicut}. 
   
  On general directed graphs \probEMCut is \FPT  parameterized by $k$ for the special case with two terminal pairs $(s_1,t_1), (s_2,t_2)$  \cite{doi:10.1137/12086217X} and is \WOne-hard for  four terminal pairs \cite{PilipczukW16}. The complexity of the case with three terminal pairs is open.
  
  \begin{open}[\cite{PilipczukW16}] 
  What is the parameterized complexity of \probEMCut on directed graphs when the three pairs of terminal sets  $(s_i,t_i)_{i=1}^3$ parameterized by the cut-size $k$?
\end{open}

  Kratsch et al.~\cite{DBLP:conf/icalp/KratschPPW12}  proved that
  \probEMCut is \FPT parameterized by $k$ and $|T|$ on directed
  acyclic graphs and that it remains \WOne-hard parameterized by $k$ even on DAGs.  Chitnis and Feldmann in \cite{ChitnisF19} study FPT innaproximability of 
 \probEMCut  on directed graphs. The following question about kernelization of \probEMCut is open.
 
  \begin{open}[\cite{CyganKPopen13}]
 Does \probEMCut admit a polynomial kernel on directed acyclic graphs, when parameterized by $k$ and $|T|$? Or when parameterized by $k$ and when the number of terminal pairs is constant? 
 %\marginpar{TODO FEDOR: check the statement of this open problem.}
  \end{open}

Bringmann et al. \cite{bringmann_et_al:LIPIcs:2015:4911} provide a detailed study of the following generalization of  \probEMCut.
In the  \ProblemName{Steiner   Multicut} we are given an undirected graph $G$, a collection $T = \{T_{1},...,T_{t}\}, T_i \subseteq V(G)$, of terminal sets of size at most $p$, and an integer $k$. The task is to decide  whether there is a set $S$ of at most $k$ edges  such that of each set $T_{i}$ at least one pair of terminals is in different connected components of $G- S$.  \probEMCut is the special case for $p=2$.
%Directed Multicut with k = 2 can be solved in time O?(22O(p) )

Parameterized complexity of a variant of the cut problem called
 \textsc{Length-Bounded Edge-Cut} (delete at most $k$ edges such that the resulting graph has no $s-t$ path of length shorter than $\ell$) was studied by 
 Golovach and  Thilikos \cite{GolovachT11}. They showed that \textsc{Length-Bounded Edge-Cut}  is in \FPT for the combined parameter $k+\ell$ .  Fluschnik et al. \cite{FluschnikHNN18} proved that it is unlikely to admit a polynomial kernel in $k+\ell$ even when the input graph is planar. When it concerns  structural parameterized complexity, 
 Dvor{\'{a}}k and  Knop \cite{DBLP:journals/algorithmica/DvorakK18}
  showed that  the problem is \WOne-hard when parameterized by the pathwidth  and is fixed-parameter tractable when parameterized by the treedepth of the input graph.  Bazgan et al. \cite{DBLP:journals/networks/BazganFNNS19}  provided an \XP algorithm for the parameter $\Delta$, the maximum degree of the input graph $G$, and an \FPT algorithm for the feedback edge number.
  Bentert, Heeger, and Knop \cite{DBLP:journals/corr/abs-1910-03409}
prove   \WOne-hardness for the combined parameter pathwidth and  maximum degree $\Delta$ of the input graph. They also prove that  \textsc{Length-Bounded Edge-Cut}   is \WOne-hard for the feedback vertex number.    
  
   Kolman \cite{kolman2017algorithms} showed that \textsc{Length-Bounded Edge-Cut}  is \FPT  on planar graphs when parameterized by $\ell$. Parameterized complexity of the problem with parameter $k$ on planar graphs is open.
   \begin{open}
 What is the parameterized complexity of   \textsc{Length-Bounded Edge-Cut}  when the input graph $G$ is planar and the parameter is the cardinality of the cut $k$?
 \end{open} 

 The problem of  metric repair, which generalizes both  \probEMCut and \textsc{Length-Bounded Edge-Cut},  was  studied in~\cite{FanRB18,GilbertS18}.

\paragraph{Constrained cuts}
Here we collected the results on the problems of the following type: is it possible to delete at most $k$ edges from the graph such that some of the required constraints like on  the size of a connected component or on the number of connected components hold. 
  For a vertex set $X\subseteq V(G)$, we denote by $\partial(X)$ the set of edges between $X$ and $V(G)\setminus X$.

A general framework for defining constrained cuts was suggested by Lokshtanov and Marx in~\cite{lokshtanov-marx-ic-clutering}. 
Let 
  $\mu$ be a function that assigns a non-negative integer to each subset of vertices in the graph.
Following the notation of  Lokshtanov and Marx, we say that a vertex set $X\subseteq V(G)$ is a  \emph{($\mu,p, q$)-cluster}, if 
$|\partial(X)|\leq q$ and $\mu(X)\leq p$. For example, if $\mu(X)$ is the number of non-edges in the subgraph induced by $X$, then 
($\mu,0, q$)-cluster is a clique which can be cut from the graph by at most $k$ edges. 

Then in the  
 \ProblemName{($\mu,p, q$)-Cut} problem, for a given graph $G$ the task is to identify whether $G$ contains a {($\mu,p, q$)-cluster}. In the  
 \ProblemName{Terminal ($\mu,p, q$)-Cut} problem, we are given  graph $G$ and vertex $v$, the task is to decide whether there is a {($\mu,p, q$)-cluster} containing $v$. 
 
 Lokshtanov and Marx proved that  
 \ProblemName{Terminal ($\mu,p, q$)-Cut}  is solvable in time $2^{\Oh(q)}n^{\Oh(1)}$  ($p$ being a  part of the input)   and in time $2^{\Oh(p)}n^{\Oh(1)}$  ($q$ being a  part of the input) for the following important special cases
 \begin{itemize}
 \item $\mu(X)$ is the number of nonedges in the subgraph induced by $X$;
\item $\mu(X)$  is the maximum degree of $\overline{G}[X]$, the complement of the graph induced by $X$; 
 \item $\mu(X)$  is the number of vertices of $X$.
 \end{itemize}
 Let us note that for each of the above cases the  \ProblemName{Terminal ($\mu,p, q$)-Cut}  problem is \NP-complete when both $p$ and $q$ are part of the input \cite{FominGK13}. 
 An \FPT{} algorithm for \ProblemName{Terminal ($\mu,p, q$)-Cut} trivially  implies an \FPT{} algorithm for 
  \ProblemName{($\mu,p, q$)-Cut}; we just try all possible terminal vertices.

    \begin{open}What is the parameterized complexity of the weighted versions of   \ProblemName{($\mu,p, q$)-Cut} when parameterized by $p$ and by $q$ and function $\mu(X)$ being the number of nonedges in the subgraph induced by $X$ and 
 the maximum degree of $\overline{G}[X]$?
\end{open}

    \begin{open}
 What is the parameterized complexity of deciding for given graph $G$ and integers $p$ and $q$, if $G$ contains a set of vertices $X$ such that $|X|=p$ and $|\partial (X)|\leq q$ with parameter $p$ or $q$?
 \end{open}

The related \pname{Bisection} problem,  the problem of separating a graph into two equally large
graphs cutting at most $k$ edges,  was shown to be solvable in
time $2^{\Oh(k^3)} \cdot \polyn$ by Cygan et al.~\cite{CyganLPPS19}.  
The incompressibility of the problem was shown by van Bevern et al.~\cite{bevern2013parameterized}.

  Lokshtanov and Marx  also  show that when $\mu$ is monotone, then the solution for  \ProblemName{Terminal ($\mu,p, q$)-Cut} can be in polynomial time transformed into a solution of the following
  \ProblemName{($\mu$,$p$, $q$)-Partition} problem. In this problem we are given a  graph $G$ and the task is to decide whether there is a partition of the vertex set  $V(G)$ into ($\mu,p, q$)-clusters. In particular, since 
  for every monotone polynomial time computable function $\mu$,  \ProblemName{Terminal ($\mu,p, q$)-Cut} is solvable in time  $n^{\Oh(q)}$ by a brute-force algorithm trying all cuts of size at most $q$, 
  this yields that \ProblemName{($\mu$,$p$, $q$)-Partition} is solvable in time $n^{\Oh(q)}$. 
  
  Kim et al. \cite{kim_et_al:LIPIcs:2015:5573} considered a related problem, under name
  \ProblemName{Min-Max Multiway Cut}, where we are  given a graph $G$, a non-negative integer $\ell$, and a set $T$ of  terminals, the question is whether we can partition the vertices of $G$ into $|T|$ parts such that (a) each part contains one terminal and (b) there are at most $\ell$ edges with only one endpoint in this part. They gave an algorithm solving this problem in time  
  $2^{\Oh((\ell |T|)^2 \log{\ell |T|})} n^4\log{n}$. 

An interesting variant of \probEMCut was introduced by 
Chitnes et al. \cite{ChitnisEM17} under the name \textsc{Chain SAT}.
 In the graph version, the problem can be  phrases as follows. 
The input is a parameter $k$ and a directed graph $ G$ with specified source $s$ and sink $t$. The edges of $G$ are partitioned into sets $E_1,\dots, E_m$, such that each $E_i$ is an edge-set of a path of length at most $L$. 
The goal is to select $k$ sets $E_i$, such that if you delete edges from all selected $E_i$s, then there is no path from $s$ to $t$. That is, the edges of the graph are bundled into paths of length at most $L$ and one pays one to delete edges of one bundle.

    \begin{open}[\cite{ChitnisEM17}]
 Is \textsc{Chain SAT} \FPT parameterized by $L$?
 \end{open} 

Parameterized complexity of the multi-budgeted variant of \probEMCut, where arcs  are colored and the required cut should contain a certain amount of arcs of each color, was investigated by 
Kratsch et al. in \cite{kratsch_et_al:LIPIcs:2019:10219}.

\medskip 
Another problem of cutting a graph is  \pname{Minimum $k$-way Cut of Bounded Size}, where we are given 
graph  $G$ and integers $k$ and $s$.  The task is to decide whether 
  there is a set of at most $s$ edges $X$ such that $G-X$ has at least $k$ connected components.  
 Downey et al.~\cite{downey2003cutting} prove that
 the problem parameterized by $k$ is \wone-hard.
  Kawarabayashi and Thorup~\cite{kawarabayashi2011minimum} show that 
  the
problem is fixed parameter tractable when parameterized by $s$.

A \emph{matching cut} is an edge cut that is a matching. \textsc{Matching Cut} is fixed-parameter tractable parameterized by the size of the solution. This result, as well as tractability of cut problems with other various constraints, follows from the work of Marx et al. \cite{MarxOR13}. Kernelization algorithms for various structural parameterization of  \textsc{Matching Cut}  and its generalization are given in \cite{KomusiewiczKL18} and \cite{GomesS19}.

\subsection{Connectivity}
 
In this subsection we discuss results around connectivity augmentation problems. In such problems the input is a (multi) graph and the objective is to increase edge or vertex connectivity by adding the minimum number (weight) of additional edges, called links. 
 
 This problem was first studied by Eswaran and Tarjan~\cite{EswaranT76} who showed that increasing the edge connectivity of a given graph to 2 by adding a minimum number of links (also called an augmenting set) is polynomial time solvable. Subsequent work of Watanabe and  Nakamura  \cite{Watanabe198796}, Cai and Sun~\cite{cai1989minimum},  and Frank~\cite{Frank92} established that the problem is also polynomial time solvable for any given target value of edge connectivity to be achieved. However, if the set of links is restricted, that is, there are pairs of vertices in the graph which do not constitute a link, or if the links have (non-identical) weights on them, then the problem of computing the minimum size (or weight) augmenting set is  \NP-complete~\cite{EswaranT76}. 
%Therefore, these variants have attracted substantial attention in the field of approximation algorithms.

It is interesting to note that  the vertex version of the problem is substantially less understood  even when the set of links which can be added is unrestricted. Vegh ~\cite{Vegh11} obtain a polynomial time solution for  the special case when the connectivity of the graph is required to increase by 1. The complexity of  the general case is  open. Jackson and   Jord{\'{a}}n \cite{JacksonJ05a} gave a $2^{\Oh(\lambda)}n^{\Oh(1)}$-time algorithm for the problem of finding a minimum number of edges to make a graph $\lambda$-vertex connected. Let us note, that according to the current knowledge,  the problem still can  be solved in polynomial time.

In the {\sc Weighted Minimum-Cost Edge-Connectivity Augmentation by One}, we are given  graph $G$
  which is $\lambda$-edge connected, set of links $\cal L$, integer $k$, weight function $w$ on $\cal L$,  $p\in \mathbb{R}$. 
  The task is to decide whether there is a link set $F\subseteq \cal L$ such that $w(F)\leq p$, $\vert F\vert \leq k$ and $G\cup F$ is $\lambda+1$-edge connected?
  
The first parameterized algorithm for the connectivity augmentation problem was considered by Nagamochi~\cite{Nagamochi200383}, who gave a $2^{\Oh(k\log k)}|V|^{\Oh(1)}$ algorithm for the case when the weights on the links are identical and $\lambda$ is odd.  Guo and Uhlmann~\cite{GuoU10} gave a kernel with $\Oh(k^2)$ vertices and links for the same case. Marx and Vegh~\cite{MarxV13} studied the problem in its full generality and gave a kernel with $\Oh(k)$ vertices, $\Oh(k^3)$ links and weights of $(k^6\log k)$ bit integers. Basavaraju et al. \cite{BasavarajuFGMRS14} gave an algorithm solving {\sc Weighted Minimum-Cost Edge-Connectivity Augmentation by One} in time $9^k n^{\Oh(1)}$.

%Furthermore, the special case of edge connectivity augmentation, where the objective is to increase the edge connectivity of a graph by 1 by adding edges from a \emph{restricted} set,  is also known to be \NP-complete~\cite{FredericksonJ81}. 
 
%In this paper, we will focus on the parameterized complexity aspects of the problems,  {\sc Weighted Edge Connectivity Augmentation By One (w-Aug-One)} and {{\delaug}}. 

Another variant of connectivity concerns problems where one has to delete a set of edges while still keeping some connectivity requirements on the remaining graph. 
   
  Basavaraju et al. \cite{BasavarajuFGMRS14}  study the following  
\pname{Deletion with $\lambda$ connectivity}   problem. In this problem we are given a triple 
 {$(G,{\cal L},k)$ where $G$ is a $\lambda$-edge connected, $\cal L$ is a set of edges, called links, $G+\cal L$ is $(\lambda+1)$-edge connected, and $k$ a positive integer.} {The task is to decide whether there is a set of $k$ links in $\cal L$ whose deletion from $G+\cal L$ maintains $(\lambda+1)$-edge connectivity. }   Basavaraju et al. gave an algorithm solving \pname{Deletion with $\lambda$ connectivity}  in time $2^{\Oh(k)} n^{\Oh(1)}$.

Hüffner et al.~\cite{huffner2015finding}  introduced  the following edge deletion problem. We say that an $n$-vertex graph $G$  is highly connected, if every vertex of $G$ is of degree at least $\lfloor n/2\rfloor +1$.
In the \pname{Seeded Highly Connected Edge Deletion} problem, the input is graph $G$, vertex set $S\subseteq V(G)$ and integers $k$ and $\alpha$. The task is to decide whether there is an edge  set $X\subseteq E(G)$ of at most $\alpha$ edges such that $G-X$ consists only of degree-zero vertices and a $(k + |S|)$-vertex highly connected subgraph containing $S$.  Hüffner et al. obtained a 
 kernel with at most $2\alpha + 4\alpha/k$  vertices and $\binom{2\alpha}{2}+\alpha$ edges computable
in $\Oh(\alpha^2 nm)$ time. They also gave a subexponential algorithm of running time $\Oh( 2^{4\cdot \alpha^{0.75}} +\alpha^2 nm)$.

%
%In this chapter, we investigate the fixed-parameter
%tractability of some of these problems parameterized by the size of the
%solution, that is, the size of the cut that we remove from the graph
%(one could parameterize these problems also by, for example, the number of
%terminals, while leaving the size of the solution unbounded, but 
%such parameterizations are not the focus of this chapter). It turns
%out that small cuts have certain very interesting extremal
%combinatorial aspects that can be exploited in FPT algorithms. The
%notion of important cuts formalizes this extremal property and gives a
%very convenient tool for the treatment of these problems.

%Edge Multiway Cut

%Directed Edge Multiway cut

%
%Steiner Multicut (STACS 2015)
%
%
%DFVS

%Edge Subset Feedback Edge Set (Xiao Nagamochi)

%Basavaraju et al. augmenting to connectivity~\cite{basavaraju2014parameterized}?

Adler et al. \cite{AdlerKT15} introduced the problem of   augmenting a planar graph $G$ with a given set of $k$ pairs of terminals. The task is to augment $G$ with the minimum number of edges  such that 
all edges are added within one face of $G$, the augmented graph is planar and all  terminal-pairs are linked with vertex-disjoint paths. This problem is \FPT{} parameterized by $k$ \cite{AdlerKT15}.

%\paragraph{Open problems in connectivity and cuts}

 %%%%%%%%%%%%%%%%%%
 \subsection{Clustering} One of the simplest variants of clustering is \probCedit or \textsc{Correlation Clustering}, where the task is to  delete/add in total at most $k$ edges from/to graph $G$ such that every connected component of the obtained graph is a clique. Since a clique is a graph containing no induced path $P_3$, \probCedit is also a special case of the problem of editing to a graph class  
 characterised by a finite number of minimal forbidden subgraphs. This is why we discussed this problem in Section~\ref{sec:graph-classes-finite}. However, different variants of clustering do not fit this scheme and we discuss them here.   
% \subsubsection{Non-hereditary variants of hereditary classes}

%I could not check this:% \cite{bessy2010polynomial} gives a polynomial kernel $O(dt)$ for cluster editing when parameterized both by number of clusters $d$ and maximum number of modifs $t$ touching one vertex.

%%% GRID
%Díaz and Thilikos showed that editing towards \pname{Grid Editing}---as well as towards a wide range of graph classes of other highly regular patterns, such as walls, and radars---admits polynomial kernels~\cite{diaz2006fast} on $O(k^4)$ vertices.

%%% CLUSTER
One can generalize the concept of cluster graphs as follows. A graph is a $s$-club cluster if every connected component has diameter at most $s$. These graph classes are not hereditary, as adding a universal vertex will transform  any graph into a $2$-club cluster. Liu, Zhang, and Zhu~\cite{liu2012editing} studied \pname{$2$-Club Cluster Deletion} as well as \pname{$2$-Club Cluster Editing}. They show that both these problems (and the vertex deletion version) are \NP-complete and they give a $2.74^k \cdot\polyn$ time algorithm for \pname{$2$-Club Cluster Deletion}. Whether the problem admits a polynomial kernel is open.

\begin{open}[\cite{liu2012editing}]
  Does \pname{$2$-Club Cluster Deletion} admit a polynomial kernel?
\end{open}

As mentioned earlier, \pname{Cluster Editing} does not admit a subexponential time algorithm~\cite{komusiewicz2012cluster} unless ETH fails. On the other hand, the problem \pname{$p$-Cluster Editing}, where the number of components in the target class is fixed to be exactly~$p$---rather surprisingly---does indeed admit a subexponential parameterized time algorithm. This was shown by Fomin et al.~\cite{fomin2014tight}, who designed an algorithm solving this problem in time $2^{\Oh(\sqrt{pk})} \cdot \polyn$. The \pname{$p$-Cluster Editing} problem, as well as \pname{$p$-Cluster Deletion} was first studied by Shamir, Sharan, and Tsur~\cite{shamir2004cluster}, who showed that even \pname{$2$-Cluster Deletion} is \npci.

Motivated by the result of Fomin et al.~\cite{fomin2014tight}, Misra, Panolan and Saurabh~\cite{misra2013subexponential} studied a similarly constrained version of another clustering problem: \pname{$s$-Club $d$-Cluster Editing}. In this version, the clusters are $s$-club graphs and their number is \emph{at most} $s$. Misra et al. show that in this case, constraining the number of desired clusters does not help in obtaining a subexponential time algorithm: \pname{$s$-Club $d$-Cluster Editing} does not admit a subexponential time fixed parameter algorithm when parameterized by both $s$ and $d$, even for $s = d = 2$.

 H{\"{u}}ffner et al. \cite{HuffnerKLN14} considered parameterized complexity and kernelization of  a clustering variant called \pname{Highly Connected Deletion}. 
In this problem one seeks to delete at most $k$ edges such that, in the resulting graph, each vertex in each connected component is adjacent to at least half of the vertices of this component. 
See also the work of Bliznets and Karpov for further improvements and other variants of this problem~\cite{BliznetsK17}. Golovach and Thilikos \cite{GoplovachT19} studied a related notion of connectivity clustering, where the task is to delete at most $k$ edges to obtain clusters of given size and of given connectivity.

%The  \pname{Cluster Deletion} problem, that is, the problem of deleting at most $k$ edges resulting in $P_3$-free graph is related to the so-called  \pname{Strong Triadic Closure} \cite{DBLP:journals/tcs/KonstantinidisN18}. Here we  aim to label the edges of graph $G$  as strong and weak such that at most $k$ edges are weak and $G$ contains no induced $P_3$ with two strong edges. Parameterized complexity of this problem was studied by Golovach et al.~\cite{DBLP:conf/swat/GolovachHKLP18}
%and  Gr{\"{u}}ttemeier and   Komusiewicz~\cite{DBLP:conf/wg/GruttemeierK18}.\todo{see Section 4.4. We have to decide where we put these things.}
%
%\todo[inline]{ANY open problem about triadic closure?}

Finally, let us mention that for weighted graphs, the \pname{Cluster Editing}  problem also admits a $2k$ kernel with integer   weights~\cite{CC12} and an  \FPT algorithm in $\Oh(1.82^k)$ time~\cite{BBB+09}. The parameterized complexity of the dynamic version of  \pname{Cluster Editing} was studied by Luo et al.~\cite{LuoMNN18}.

%Open problems 

%\marginpar{WARNING: from here, it seems that we start to mix between "modification" and "editing" to refer to the generic problem. It was modification until now.}

%!TEX root = survey.tex

\section{Degree constraints}\label{sec:deg}
In this section, we survey the advances in modifying graphs to have some specified degree constraints possibly together with other properties.   
The degree constraints may be related to degrees of individual vertices or degree sequences. Speaking about other properties, we focus on the connectivity.
Such problems have a long history in the literature as they encompass such classical problems like 
\pname{Perfect Matching}, \pname{$r$-Factor}, \pname{Hamiltonian Path} or \pname{Hamiltonian Cycle}.
Typically, whenever the parametrized complexity of problems of this kind was investigated, the authors also considered vertex deletions besides edge modification operations. Hence, to present the full spectra of the work, we extend our framework in this section to include the results about vertex deletion when appropriate.

\subsection{Modification to satisfy individual degree constraints}\label{sec:inddeg}
The investigation of the parameterized complexity of the problems where the aim is to satisfy some degree restrictions for each vertex were initiated by Thilikos and Moser~\cite{MoserT09} and 
Mathieson and Szeider~\cite{mathieson2012editing}.

In particular, Thilikos and Moser~\cite{MoserT09} considered the \pname{$k$-Almost $r$-Regular Graph} problem, which asks, given a graph $G$ and a non-negative integer $k$, whether $G$ can be made $r$-regular by deleting at most $k$ vertices. They proved that the problem admits a kernel with $\Oh(kr(r+k)^2)$ vertices. Despite the fact that Thilikos and Moser were interested solely in vertex deletions, we discuss this result briefly, because the approach that was used is generic for similar problems. 
Since the deletion of a vertex decreases the degrees of its neighbors by one, a vertex of degree at most $r-1$ or at least $r+k+1$ should be deleted.  Applying this straightforward reduction rule to the input graph, we obtain a graph $G$ of bounded degree. For a set of vertices $X$ of $G$ of degree at least $r+1$, one can observe that its size should be polynomially bounded in $k$ and $r$ for any yes-instance, since $G$ has bounded degree. Further, we have that the vertices of $G-X$ have the same degrees $r$, and for each component $H$ of $G-X$, it holds that if any vertex of $H$ or a neighbor of a vertex of $H$ is deleted, then $H$ should be deleted completely. This observation allows to construct reduction rules for the components of $G-X$. This way a polynomial kernel could be constructed.  Thilikos and Moser complement this results by observing that because  \pname{Cubic Subgraph} is one of the fundamental \NP-complete problems discussed by
Garey and Johnson~\cite{garey1979computers} (in fact, this problem is \NP-complete for very restricted inputs~\cite{Stewart94,Stewart96,Stewart97}), \pname{$k$-Almost $r$-Regular Graph} is \paraNP-complete when parameterized by $d$. 
%They also show that this problem is \WOne-hard when parameterized by $k$ only.

The most general variant of the modification problem to satisfy degree constraints was introduced by Mathieson and Szeider~\cite{mathieson2012editing} (see also the
thesis  of Mathieson~\cite{mathieson:thesis} for more details). For a set of modification operations $S$, they defined

\defproblem{Weighted Degree Constraint Editing $(S)$(WDCE$(S)$)}{A graph $G$, non-negative integers $k$ and $r$, a weight function $\rho\colon V(G)\cup E(G)\rightarrow \mathbb{N}_0$, and a degree list function $\delta\colon V(G)\rightarrow 2^{\{0,\ldots,r\}}$.}{Is it possible to obtain a graph $G'$ from $G$  such that for every $v\in V(G')$, $\sum_{vx\in E(G')}\rho(vx)\in\delta(v)$, using at most $k$ modification operations from $S$?}

\noindent
Here, the aim is to obtain a graph such that the (weighted) degree of every vertex is in a given set defined by the list function. 
They considered \pname{WDCE$(S)$} for various non-empty 
$$S\subseteq \{\textrm{vertex deletion},\textrm{edge deletion},\textrm{edge addition}\}.$$
Mathieson and Szeider also considered the unweighted variant of the problem that we call  \pname{Degree Constraint Editing $(S)$} (\pname{DCE$(S)$}). In this variant the weight function 
$\rho\equiv 1$, but \pname{DCE$(S)$} is not a restricted version of \pname{WDCE$(S)$}. 
For \pname{DCE$(S)$}, vertex and edge deletions and edge additions are defined in the standard way. For the weighted problem, it is more complicated. It is assumed that each modification operation has unit cost. If a vertex $v$ has weight $\rho(v)=1$, then the vertex deletion operation deletes $v$ together with incident edges, but if $\rho(v)>1$, then this operation just reduces its weight by 1. Similarly, the edge deletion operation deletes an edge $e$ of weight 1 and reduces its weight by 1 if $\rho(e)>1$.  Respectively, the edge addition operation can be applied to an existing edge, and it increases the weight of such an edge by 1. 

 Mathieson and Szeider~\cite{mathieson2012editing} proved that \pname{WDCE$(S)$} and \pname{DCE$(S)$} are \WOne-hard when parameterized by $k$ for any non-empty $S$. Moreover, the hardness for \pname{DCE$(S)$} holds even if $\delta(v)=\{r\}$ if $\textrm{vertex deletion}\in S$, that is, for the case where the aim is to obtain an $r$-regular graph. 
It is interesting to observe that for the important case of degree lists of size 1, \pname{WDCE$(S)$} and \pname{DCE$(S)$} can be solved in polynomial time if $\textrm{vertex deletion}\notin S$~\cite{mathieson2012editing} by the reduction to the \pname{Perfect Matching} problem. 

From the positive side, it is proved that \pname{WDCE$(S)$} and \pname{DCE$(S)$} are \FPT when parameterized by $k+r$ for any non-empty $S$. To achieve this result, Mathieson and Szeider showed that for any $k$ and $r$, the problems can be expressed in the first-order logic. Applying the similar to the described above arguments of Thilikos and Moser~\cite{MoserT09}, an instance of \pname{WDCE$(S)$} or \pname{DCE$(S)$} can be reduced to an equivalent instance with a graph of bounded degree. Then the meta-theorem of Frick and Grohe~\cite{FrickG01} gives the result (the same can be obtained without preprocessing by the meta-theorem of Bulian and Dawar~\cite{BulianD15}).  Clearly, this approach only allows to classify \pname{WDCE$(S)$} and \pname{DCE$(S)$} to be in \FPT. For the special case of \pname{DCE$(S)$} with degree lists of size 1, Golovach~\cite{golovach2015editing} used the random separation technique (see the book~\cite{cygan2015parameterized} for an introduction to this technique) to show that the problem can be solved in time $2^{\Oh(kr^2+k\log k)}\cdot \poly(n)$.  It gives rise to the following open problem.

\begin{open}
Is it possible to give efficient \FPT algorithms for  \pname{DCE$(S)$} and/or \pname{WDCE$(S)$} parameterized by $k+r$ for general degree list functions? 
\end{open}

For the case $\textrm{vertex deletion}\in S\subseteq\{\textrm{vertex deletion},\textrm{edge deletion}\}$ and single-element degree lists, Mathieson and Szeider~\cite{mathieson2012editing} showed that \pname{WDCE$(S)$} admits a kernel with  $\Oh(kr(k+r))$ vertices. For general degree lists, they demonstrated a kernel with $\Oh(k^2r^{k+1}+kr^{k+2})$ vertices. These results were complemented by
Froese, Nichterlein and Niedermeier~\cite{FroeseNN16}, who proved that if only edge additions are allowed (i.e, for the completion problem), then \pname{DCE$(S)$} has kernels with $\Oh(kr^2)$ and $\Oh(r^5)$ vertices, that is, it admits a polynomial kernel whose size depends only on $r$.  To obtain the latter result, they prove that the problem can be solved in polynomial time if $k$ is sufficiently large (greater that some polynomial of $r$). The latter result is  based on a clever application of  combinatorial results about existence of $f$-factors. Hence, the following win-win approach can be used: if $k$ is large, then the problem is solved in polynomial time, and if $k$ is bounded by a polynomial of $r$, then the kernelization algorithm for the case where the parameter is  $k+r$ is applied. Froese, Nichterlein and Niedermeier~\cite{FroeseNN16} also give lower bounds by proving that \pname{DCE$(S)$} parameterized by $k+r$ has no polynomial kernel unless \ph{} if $S=\{\textrm{vertex deletion}\}$ or $S=\{\textrm{edge addition}\}$. Another lower bound for this parameterization was given by  Golovach~\cite{golovach2015editing} who proved that 
\pname{DCE$(S)$} with degree lists of size one has no polynomial kernel unless \ph{} if $\{\textrm{vertex deletion},\textrm{edge addition}\}\subseteq S$.

The variant of \pname{DCE$(S)$} with degree lists of size one, where $S\subseteq\{\textrm{vertex deletion},\textrm{edge deletion}\}$ and where we are given separate bounds $k_v$ and $k_e$ for the number of vertex and edge deletions respectively, was considered by  Dabrowski et al.~\cite{DabrowskiGHPT17} on planar graphs. They proved that the problem admits a polynomial kernel when parameterized by $k_v+k_e$.

Golovach~\cite{Golovach17} introduced the degree constrained modification problem with connectivity restrictions. In~\cite{Golovach17} it was called \pname{Edge Editing to a Connected Graph of Given Degrees} but later the other title \pname{Edge Editing to Connected $f$-Degree Graph} was proposed and we use it in the survey. 

\defproblem{Edge Editing to Connected $f$-Degree Graph (EECG)}{A graph $G$, non-negative integers $d$ and $k$, and a function $f\colon V(G)\rightarrow \{0,\ldots,d\}$.}
{Is it possible to obtain a connected graph $G'$ from $G$  with $d_{G'}(v)=f(v)$ for every $v\in V(G')$ by at most $k$ edge deletions and additions?}

\noindent
Recall (see~\cite{mathieson2012editing}) that if the degree lists have size 1,  \pname{DCE$(S)$} is polynomial if only edge deletions and additions are allowed. Contrary to this, \pname{EECG} is \NP-hard even if $f(v)=2$ for all $v\in V(G)$: it is straightforward to see that \pname{EECG} for $f(v)=2$ for $v\in V(G)$ and $k=m-n$ is equivalent to the \pname{Hamiltonian Cycle} problem that is well-known to be \NP-complete~\cite{garey1979computers}.

Golovach~\cite{Golovach17} proved that \pname{EECG} has a kernel with $\Oh(kd^3(k+d)^2)$ vertices. The results is obtained using the generic approach of Thilikos and Moser~\cite{MoserT09}, but due to allowing edge additions and connectivity restrictions, the reduction rules are great deal more complicated and are based on the following structural observations. It can be easily seen that if $v$ is a vertex of $G$ with $d_G(v)=f(v)$, then the number of deleted edges incident to $v$ equals to the number of added edges incident to $v$. Therefore, if $X$ is the set of vertices of $G$ whose degrees are different from the values of $f$, then for any solution, the set of deleted edges $D$ and the set of added edges $A$ compose a graph which can be covered by edge disjoint walks without repeated edges joining vertices of $X$ and closed walks that are alternating in the sense that if an edge of a walk is from $D$, then the next is from $A$ and vice versa. 
Golovach~\cite{Golovach17} also constructed an algorithm running in time $k^{\Oh(k^3)}\cdot \poly(n)$ for the case $f(v)=d$ for each $v\in V(G)$, that is, for the modification to a connected regular graph, but left open the question whether \pname{EECG} if \FPT when parameterized by $k$ only. This question was resolved by Fomin, Golovach, Panolan, and Saurabh~\cite{FominGPS19}. They show that  \pname{EECG} is solvable in $2^{\Oh(k)} \cdot \poly(n)$ time. Fomin et al.~\cite{FominGPS19} use the same structural properties of solutions as Golovach in~\cite{Golovach17}, but the crucial new component is the application of the recently developed matroid representative sets techniques combined with color coding (we refer to~\cite{cygan2015parameterized} for an introduction to these techniques). It is still open whether \pname{EECG} has a polynomial kernel whose size depends on $k$ only. For the special case of planar graphs, Dabrowski et al.~\cite{DabrowskiGHPT17} proved that the problem admits a polynomial kernel even if we additionally permit vertex deletions and parameterize the problem by the number of vertex and edge deletions.

\begin{open}
Does  \pname{EECG} parameterized by $k$ have a polynomial kernel?
\end{open}

For the weighted variant of  \pname{EECG}, Fomin et al.~\cite{FominGPS19} proved that it is \WOne-hard when parameterized by $k+d$.

Recall that in  \pname{DCE$(S)$} we require that each vertex has the degree from a given list but in \pname{DCE$(S)$} these lists have size one. It leads to the following open problem.

\begin{open}\label{prob:degree-list}
Investigate the parameterized complexity of the variant of \pname{EECG} where instead of the degree function $f$ a degree list function $\delta\colon V(G)\rightarrow 2^{\{0,\ldots,r\}}$ is given and   it is asked whether it possible to obtain a connected graph $G'$ from $G$  with $d_{G'}(v)\in \delta(v)$ for every $v\in V(G')$ by at most $k$ edge deletions and additions. 
\end{open}

Notice that if we have a choice of degrees, then the structural properties of solutions used in~\cite{FominGPS19,Golovach17} could not be applied any more. The problem is open even when the degree constraints are intervals of bounded size.  Harraberg in~\cite{Haarberg19} considered the special case where the degree constraints are given by inequalities. More precisely, he considered the \pname{Edge Editing to a Connected Upper (Lower) Bounded Degrees} (\pname{EditUBD} and \pname{EditLBD}, respectively) problems. \pname{EditUBD} asks, given a (multi) graph $G$, a non-negative integer $k$ and a function $f\colon V(G)\rightarrow \mathbb{Z}^+$, whether  it is possible to obtain a connected graph $G'$ from $G$ with $d_{G'}(v)\leq f(v)$ for every $v\in V(G')$ by at most $k$ edge deletions and additions. It is shown in~\cite{Haarberg19} that this problem is \NP-complete, has a kernel with $\Oh(k^3)$ vertices and $\Oh(k^6)$ edges, and can be solved in time $2^{\Oh(k)}\cdot\poly(n)$. In \pname{EditLBD}, it is required that $d_{G'}(v)\geq f(v)$ for every $v\in V(G')$, that is, the upper bounds on the degrees are replaced by lower bounds. Interestingly, \pname{EditLBD} is proved in~\cite{Haarberg19} to be polynomially time solvable. 

%It also would be interesting to consider the variant of \pname{EECG} where vertex deletions are allowed.

All aforementioned problems are stated for undirected graphs. The systematic study of the degree constraint modification problems for directed graphs was recently initiated by Bredereck, Froese, Koseler, Millani, Nichterlein and Niedermeier~\cite{BredereckFKMNN19}. We will return to this paper in the next section where we consider degree sequence restriction, but here we mention only that they considered the 
\pname{Digraph Degree Constraint Completion} problem that could be seen as a variant of \pname{DCE$(S)$}, where for each vertex, a degree list  function that assigns to each vertex a set of pairs of non-negative integers from $\{0,\ldots,r\}$ that specify the desired pairs of values of in- and out-degrees respectively are given and $S=\{\textrm{edge (arc) addition}\}$. Bredereck et al.~\cite{BredereckFKMNN19} show that this problem admits a kernel with $\Oh(r^5)$ vertices. 
 
\begin{open}
Investigate the parameterized complexity of variants of \pname{DCF$(S)$} and \pname{EECG} for directed graphs.
\end{open}

Besides vertex degree constraints, it could be interesting to consider edge degree constraints or combined vertex and edge degree constraints.  In particular, Mathieson~\cite{Mathieson17} considered a number of problems of this type. For a edge weighted graph, the weighted degree of a vertex is defined as the sum of weights of incident edges. Respectively, the weighted sum of an edge is the sum of the vertex degrees of its end-points. Mathieson~\cite{Mathieson17} considered the following problems for edge weighted graphs:
\begin{itemize}
\item\pname{Weighted Edge Degree Constraint Editing}, where for each edge, it is given a list of weighted degrees, and the aim is to obtain a graph, by at most $k$ modification operations, such that every vertex has a  degree from its list.
\item\pname{Weighted Bounded Degree Editing}, where a degree bound for each vertex is given, and the aim is to obtain a graph, by at most $k$ modification operations, such that the weighted degree of a vertex does not exceed its bound.
\item\pname{Weighted Edge Regularity Editing}, where for each vertex, it is given a list of weighted degrees, and for each pair of vertices, it is given a set of feasible sizes of the set of common neighbors, and the aim is to obtain a graph,  by at most $k$ modification operations, such that every vertex has the weighted degree from its list and for every edge, the size of the set of common neighbors is feasible.  
\item\pname{Weighted Strongly Regular Editing}, that is a variant of\pname{ Weighted Edge Regularity Editing}, where additionally a second set of allowed sizes of the set of common neighbors is given for each pair of vertices, and for every pair of nonneighbors of the modified graph,  the size of the set of common neighbors should belong to this set.
\end{itemize}
The allowed modification operations are vertex deletions, edge deletions and edge additions. Mathieson presented
the essentially complete picture of the complexity of these problems parameterized by the number of modification operations $k$ and/or the upper bound of the feasible degrees for various combinations of allowed operations: the cases when the problems are \WOne-hard, \FPT, have polynomial kernels or do not have them up to some conjectures are distinguished. He also investigated special cases, in particular, the unweighted problems (i.e., the problems for unit weights) and the case when the sets of feasible degrees are singletons.  Additionally, the structural parameterization by the treewidth of an input graph was considered.  We do not  discuss the details of these results, because they proved to be similar to the results about \pname{DCS$(S)$} and are obtained by similar techniques.

Another direction of research would be to consider the discussed problems for graph classes. Up to now, a very little work was done in this direction. 
Dabrowski et al.~\cite{DabrowskiGHPT17} considered variants of \pname{DCF$(S)$} and \pname{EECG} for planar graphs. 
More precisely, they considered the problems that asks for a given planar graph $G$, a degree function $f\colon V(G)\rightarrow \mathbb{N}_0$ and two non-negative integers $k_e$ and $k_v$, whether it is possible to obtain a (connected) graph $G'$ from $G$ with $d_{G'}(v)=f(v)$ for $v\in V(G')$ by deleting at most $k_v$ vertices and at most $k_e$ edges. Notice that $G'$ is not required to be planar. They proved that these problems have polynomial kernels when parameterized by $k_v+k_e$. In fact, more general kernalization results were obtained as it is shown that it could be assumed that vertices and edges have costs and the task is to delete at most $k_v$ and $k_e$ to satisfy degree restriction and achieve the minimum total cost of deleted vertices and edges. This result is obtained via the protrusion decomposition/replacement techniques introduced by Bodlaender et al.~\cite{BodlaenderFLPST09}.  

\begin{open}
Investigate the parameterized complexity of variants of \pname{DCF$(S)$} and \pname{EECG} for  graph classes. In particular, what can be said about planar graph when edge additions are allowed and the graph obtained by the modification should stay planar?
\end{open}

We conclude this section by discussing modification problems which deal with the parity constraints for degrees.  These problems are the most investigated degree constraint modification problems. Already in 1977 Boesch, Suffel and Tindell~\cite{BoeschST77} (see also~\cite{LesniakO86} and \cite{DornMNW13}) proved that \pname{Eulerian Completion} that asks about minimum number of edges that should be added to make the input graph Eulerian can be solved in polynomial time, and the same holds if multiple edges are allowed and 
for \pname{Even Graph Completion} where the aim is to obtain a graph with vertices of even degrees. Recall that a (directed) graph $G$ is \emph{Eulerian} if it contains a closed walk without repeating edges (arcs) that goes through every edge (arc). By the classical Euler theorem, a connected graph is Eulerian if and only if its vertices have even degrees. Respectively, a (weakly) connected directed graph is Eulerian if and only if for every vertex its in-degree is the same as its out-degree (see, e.g.,~\cite{Diestel12}). Following the same scheme as with the previous problems, we state the generalization of \pname{Eulerian Completion} for a set of modification operations $S$ as follows.

\defproblem{Connected Parity Constraint Editing $(S)$(CPCE$(S)$)}{A graph $G$, a parity function $f\colon V(G)\rightarrow\{0,1\}$ and 
a non-negative integer $k$.}{Is it possible to obtain a connected graph $G'$ from $G$  such that for every $v\in V(G')$, $d_{G'}(v)\equiv f(v)(\textrm{mod}~2) $, using at most $k$ modification operations from $S$?}

When we do not require connectivity, we refer to the problem as \pname{Parity Constraint Editing $(S)$} (\pname{PCE$(S)$}). 
 For directed graphs, we state the following problem.

\defproblem{Connected Degree Balance Editing $(S)$(CDBE$(S)$)}{A directed graph $G$, a  function $f\colon V(G)\rightarrow\mathbb{Z}$ and 
a non-negative integer $k$.}{Is it possible to obtain a weakly connected directed graph $G'$ from $G$  such that for every $v\in V(G')$, $d_{G'}^+(v)-d_{G'}^-(v)=f(v)$, using at most $k$ modification operations from $S$?}

Here $d_G^-(v)$ and $d_G^+(v)$ denote in- and out-degree of a vertex $v$ in a graph $G$.
Notice that if $f(v)=0$, then the question is equivalent to asking  whether we can obtain an Eulerian graph by at most $k$ operations.

Generalizing the results of Boesch, Suffel and Tindell~\cite{BoeschST77}, Dabrowski,  Golovach,  van 't Hof and Paulusma~\cite{DabrowskiGHP14} proved that \pname{CPCE$(S)$} and \pname{CDBE$(S)$} are polynomial if $S=\{\textrm{edge (arc) addition}\}$. Moreover, the problems remain polynomial if  $S=\{\textrm{edge (arc) addition},\textrm{edge (arc) deletion}\}$. It can be also observed that the same holds for \pname{PCE$(S)$}. If $\textrm{vertex deletion}\in S$, then CPCE$(S)$, \pname{PCE$(S)$} and \pname{CDBE$(S)$} are \NP-hard and \WOne-hard by the results of Cai and Yang~~\cite{cai2011parameterized}, Cygan et al.~\cite{cygan2014parameterized} and Dabrowski et al.~\cite{DabrowskiGHP14}. 
The most interesting from the parameterized complexity viewpoint case is the case $S=\{\textrm{edge (arc) deletion}\}$ for \pname{CPCE$(S)$} and \pname{CDBE$(S)$} (\pname{PCE$(S)$} is polynomial in this case as it was proven by  Cygan, Marx, Pilipczuk, Pilipczuk and Schlotter~\cite{cygan2014parameterized}).

Cygan et al.~\cite{cygan2014parameterized} observed that if $S=\{\textrm{edge (arc) deletion}\}$, then \pname{CPCE$(S)$} and \pname{CDBE$(S)$} are \NP-complete. They also proved that they can be solved in time $2^{\Oh(k\log k)}\cdot \poly(n)$ and complemented these results by proving that these problems have no polynomial kernel unless \ph. 
Their \FPT result is based on the following structural observation. If $G$ is an undirected graph and $T$ is the set of vertices for which degree constraints are broken, then the edges of a solution form a $T$-join, that is, they induce a forest that could be decomposed into edge disjoint paths that connect $|T|/2$ pairs of vertices of $T$. Hence, the task is to find a $T$-join of size at most $k$ such that the deletion of the edges of the join does not destroy connectivity.
Cygan et al.~\cite{cygan2014parameterized}
use a non-trivial application of the color coding technique to solve this problem.  Similar techniques work  also for directed graphs. Their results were improved by Goyal et al.~\cite{GoyalMPPS15}.
They showed that CPCE$(S)$ and CDBE$(S)$ for $S=\{\textrm{edge (arc) deletion}\}$ can be solved in time $2^{\Oh(k)}\cdot \poly(n)$. They use the same structural observations  as Cygan et al.~\cite{cygan2014parameterized}, but instead of color coding, they apply matroid representative sets techniques. In particular, for undirected graphs, they use the fact that the set of edges of a $T$-join is an independent set of the cographic (bond) matriod.  It should be noted that Cygan et al.~\cite{cygan2014parameterized} and Goyal et al.~\cite{GoyalMPPS15} gave their results for the \pname{Eulerian Edge Deletion} problem, that is for the special cases of  CPCE$(S)$ and CDBE$(S)$ where $f(v)=0$ (Goyal et al.~\cite{GoyalMPPS15} considered also \pname{Connected Odd Edge Deletion}), but the algorithm could be rewritten for CPCE$(S)$ and CDBE$(S)$ in a straightforward way.

Observe that in CPCE$(S)$ and CDBE$(S)$ the parameter $k$ upper bounds the number of modification operations. We can ask whether the modifications can be done by \emph{exactly} $k$ operations. In particular, Cai and Yang~\cite{cai2011parameterized} left open the following problem.

\begin{open}[\cite{cai2011parameterized}]
Is \pname{$(m-k)$-edge Eulerian Subgraph}, which asks whether a (directed) graph has an Eulerian subgraph with exactly $m-k$ edges (arcs), \FPT?
\end{open}

The same question can be asked for more general degree restriction given in  CPCE$(S)$ and CDBE$(S)$.

Notice that in  CDBE$(S)$ we require $G'$ to be weakly connected. It is natural to ask whether this condition could be strengthen.

\begin{open}
 Investigate the parameterized complexity of variant of CDBE$(S)$ where the graph $G'$ obtained by the modifications is required to be strongly connected.
\end{open}

A more special question was asked by Cygan et al.~\cite{cygan2014parameterized} (see also~\cite{cechlarova2010computing}).

\begin{open}[\cite{cechlarova2010computing,cygan2014parameterized}]
Is it \FPT to decide whether it is possible to delete at most $k$ arcs from a directed graph to obtain a graph where each
  strongly connected component is Eulerian?
\end{open}

This problem was considered by Crowston et al.~\cite{crowston2012parameterized} for tournaments. They proved that the problem  has a kernel with at most $4k\cdot (4k+2)$ vertices.

Recall that  Boesch, Suffel and Tindell~\cite{BoeschST77} proved that the \pname{Eulerian Completion}  problem can be solved in polynomial time, but the situation changes if we switch to the weighted variant of the problem. For directed graphs, \NP-hardness was proved by H{\"{o}}hn ,  Jacobs and Megow~\cite{HohnJM12} for special cases that occur in scheduling problems. It is also easy to see that the problem is \NP-hard for undirected graphs as well by a straightforward reduction from \pname{Eulerian Deletion}. The parameterized complexity of the following problem was considered by Dorn, Moser, Niedermeier and Weller~\cite{DornMNW13}.

\defproblem{Weighted Multigraph Eulerian Completion (WMEC)}{A directed multigraph $G$, a  weight finction $w\colon V(G)\times V(G)\rightarrow \mathbb{N}_0$, and a non-negative integer $k$.}{Is it possible to obtain an Eulerian multighraph $G'$ from $G$ by adding arcs of total weight at most $k$?}

Since \pname{WMEC} deals with multigraphs, the addition of parallel arcs is allowed. 
It can be noted that the classical \pname{Chinese Postman}
problem, where the aim is to find a shortest  closed walk that visits all arcs of a given directed graph, and the more general
\pname{Rural Postman}, where it is required to find a shortest walk that visits a given set of arcs, can be seen as special cases of \pname{WMEC}. 
 Dorn et al.~\cite{DornMNW13} showed that \pname{WMEC} can be solved in time $\Oh(4^k\cdot n^3)$. This results immediately implies the respective \FPT result for \pname{Rural Postman}.
They conjecture that similar result can be obtained for undirected graphs. 
They also leave open the question about the variant with arc deletion. Generalizing it, we obtain the following open problem. 

\begin{open}
 Investigate the parameterized complexity of weighted variants of CPCE$(S)$ and CDBE$(S)$ for graphs and muligraphs for $S\subseteq\{\textrm{edge (arc) deletion},\textrm{edge (arc) addition}\}$.
\end{open}

Another parameterization of \pname{WMEC} was considered by Sorge,  van Bevern, Niedermeier and Weller~\cite{SorgeBNW11} (see also~\cite{SorgeBNW12}). They proved that \pname{WMEC} is \FPT when parameterized by $b+c$, where $b=\sum_{v\in V(G)}|d_G^+(v)-d_G^-(v)|$ and $c$ in the number of weakly connected components of $G$. They complemented this result by showing that 
\pname{WMEC} has no polynomial kernel when parameterized by $b$, $c$, $k$ or $b_c$ unless \ph.

We conclude the section by the open problem stated in~\cite{DabrowskiGHP14}. We considered the degree constraint modification problems with parity restrictions. What can be said if we replace parity constraints by the more complicated ``modulo $d$ constraints'' for $d\geq 3$. 
It is observed in~\cite{DabrowskiGHP14} that this variant of CPCE$(S)$ is \NP-hard if $S=\{\textrm{edge deletion},\textrm{edge deletion}\}$ and $d=3$. Taking into account the \WOne-hardness of CPCE$(S)$ if $\textrm{vertex deletion}\in S$ (see~\cite{DabrowskiGHP14}), we ask the following.

\begin{open}[\cite{DabrowskiGHP14}]
 Investigate the parameterized complexity of the variants of CPCE$(S)$ for $S\subseteq\{\textrm{edge deletion},\textrm{edge addition}\}$, where a positive integer $d$ is given,
the parity function $f$ is replaced by a function $f\colon V(G)\rightarrow\{0,\ldots,d-1\}$ and where the aim is to obtain a connected graph $G'$ such that for every $v\in V(G')$, $d_{G'}(v)\equiv f(v)(\textrm{mod}~d)$?
\end{open}

Additionally, what can be said if we remove the connectivity restriction?

\subsection{Modification to satisfy degree sequence constraints}\label{sec:degseq}
In this section we consider  problems where the task is to modify a graph in order to satisfy constraints on degree sequences. Motivations for the problems considered here often come  from applications  like social networks.
The \emph{identity disclosure} is a specific type of privacy breach in social
networks.  It happens when an adversary is able to determine the identity of an
entity in a   network.  One can weaken this to the \emph{existence
  disclosure}, where one is able to identify whether an entity is present in a
social network or not.
\emph{Affiliation link disclosure} is the problem to determine whether an entity belongs to
a specific group in a social network.
As Zheleva and Getoor~\cite{zheleva2011social} say in their survey, 
\begin{quote}
  $k$-anonymity protection of data is met if the information for each person
  contained in the data cannot be distinguished from at least $k - 1$ other
  individuals in the data.
\end{quote}

%\marginpar{WARNING: during one page k is not anymore the number of edits}

In \emph{degree anonymization}, a graph is said to be $s$-degree-anonymous (or
simply $s$-anonymous when it is clear from context that we are talking about
degree anonymity) if for every vertex~$v$, there are at least $s-1$ vertices
with the same degree as~$v$. This leads to the modification problems where the aim is to achieve the desired level of anonymity by bounded number  of  operations.
We refer to the survey of Casas-Roma,  Herrera-Joancomart{\'{\i}} and  Torra~\cite{Casas-Roma2017} for the introduction to the  edge modification
techniques used in anonymization and focus on the parameterized complexity of the problems.
Following the style used in the previous section, we define the following problem for a set of modification operations $S$. 

\defproblem{Anonymization$(S)$}{A graph $G$, a positive integer $s$ and a non-negative integer $k$.}{Is it possible to obtain a $s$-anonymous graph $G'$ from $G$ using at most $k$ modification operations from $S$?}

Degree anonymization is perhaps one of few places where the operation of adding vertices is a natural operation as a ``dummy'' vertex can be crated in a social network.  Hence, the case $\textrm{vertex addition}\in S$ was investigated. 

Bazgan et al.~\cite{BazganBHNW16} obtained a number of hardness results. They proved that if $S=\{\textrm{edge deletion}\}$ or $S=\{\textrm{vertex deletion}\}$, then 
\pname{Anonymization$(S)$} is already \NP-hard for $s=2$ even for trees and the problem is \NP-hard if the maximum degree $\Delta$ of the input graph is 3 or 7 respectively, that is, the problem is \paraNP-hard for the respective parameterizations.
 They also sowed that 
\pname{Anonymization$(S)$} is \WOne or \WTwo-hard  when parameterized by $s+k$ if  $S=\{\textrm{edge deletion}\}$ or $S=\{\textrm{vertex deletion}\}$ respectively. 
Bazgan et al.~\cite{BazganBHNW16} also proved that \pname{$(\textrm{vertex deletion})$} has no polynomial kernel unless \ph{} when parameterized by $k+s+\Delta$. 
They obtained a number of inapproximabilty results. In particular, they initiated the investigation of the parameterized approximation/inapproximabilty for \pname{Anonymization$(S)$}.
Observe that we obtain a bicriteria optimization problem here. First, it is possible to maximize the anonymity level $s$ by performing at most $k$ modification operations, and the second option is to minimize the number of modification operations to obtain a $s$-anonymous graph. For the maximization of the anonymity level,
Bazgan et al.~\cite{BazganBHNW16} proved that the problem is not \FPT $n^{1/2-\varepsilon}$-approximable for every  $0<\varepsilon\leq 1/2$ when parameterized by $k$ even on trees unless $\FPT=\WTwo$ if $S=\{\textrm{vertex deletion}\}$, and it is not \FPT $n^{1-\varepsilon}$-approximable for every $1/2<\varepsilon\leq 1$ when parameterized by $k$  unless $\FPT=\WOne$ if $S=\{\textrm{edge deletion}\}$.  The following question is open.

\begin{open}[\cite{BazganBHNW16}]
  Are there ``reasonable'' (parameterized) approximation algorithms for the
  optimization variants of \pname{Anonymization$(S)$} parameterized by $s$ and $k$ if $S\subseteq\{\textrm{edge deletion},\textrm{edge addition}\}$?
\end{open}

From the positive side, Bazgan et al.~\cite{BazganBHNW16} considered a more general variant of the problem where non-negative integers $k_1,k_2,k_3,k_4$ are given and the question is whether it is possible to obtain a $k$-anonymous graph by at most $k_1$ vertex deletion, at most $k_2$ vertex additions, at most $k_3$ edge deletions and at most $k_4$ edge additions. 
It is proved that the problem is \FPT when parameterized by $k+\Delta$ for $k=k_1+k_2+k_3+k_4$. The result is obtained via the first-order logic machinery using the meta-theorem of Frick and Grohe~\cite{FrickG01}. Bazgan et al.~\cite{BazganBHNW16} also sketched a direct color coding algorithm for the problem. 

Bazgan et al.~\cite{BazganBHNW16} initiated investigation of \pname{Anonymization$(S)$} for graph classes and obtained a number of hardness results and distinguished some polynomial cases. This line of research should be definitely extended.

\begin{open}
 Investigate the parameterized complexity of  \pname{Anonymization$(S)$} for graph classes. In particular, what can be said about planar graphs?
\end{open}

Notice that here we can restrict only the input graphs or demand that both the input and the modified graph belong to a specific class.

Hartung et al.~\cite{hartung2015refined} considered the case $S=\{\textrm{edge addition}\}$. They proved that the problem is \WOne-hard even if $s=2$ when parameterized by $k$. The main result of the paper is that \linebreak \pname{Anonymization$(\{\textrm{edge addition}\})$} admits a kernel with $\Oh(\Delta^7)$ vertices implying that the problem is \FPT when parameterized by the maximum degree of the input graph. As the first step, they use the approach that is generic for similar problems.  If the set of vertices of the input graph of degree $0\leq d\leq \Delta$ is sufficiently large, then it is possible to select a block of such vertices of size that is bounded in $k$ and assume that for every added edge in a solution, if it has its end-vertex (both end-vertices) in the set of vertices of degree $d$, then this end-vertex (end-vertices) is in the selected block.  This observation leads to a kernel size that is polynomial in $\Delta$, $s$ and $k$. Hartung et al.~\cite{hartung2015refined} showed that it is possible to obtain a kernel whose size depend on $\Delta$ only by adjusting $s$ and showing that if $k$ is sufficiently large compared to $\Delta$, then the problem can be solved in polynomial time.

To conclude the part about anonymization,  Bredereck et al.~\cite{BredereckFHNNT15} considered  \pname{Anonymization$(S)$}
for the case $S=\{\textrm{vertex addition}\}$, and Talmon and Hartung~\cite{TalmonH17} investigated the case where the modification operations allowed are various types of contractions. 

The investigation of the modification problems with the aim to satisfy some general degree sequence properties was initiated by  Froese, Nichterlein and Niedermeier~\cite{FroeseNN16}. 
Recall that the degree sequence of an $n$-vertex graph $G$ is an $n$-tuple containing the degrees of the vertices.
Froese et al.~\cite{FroeseNN16} introduced the following problem for a tuple property $\Pi$.

\defproblem{$\Pi$-Degree Sequence Completion ($\Pi$-DSC)}{A graph $G$ and a non-negative integer $k$.}{Is it possible to obtain a  graph $G'$ with the degree sequence satisfying the property $\Pi$ from $G$ using at most $k$ edge additions?}

Notice that $\Pi$ is a tuple property. In particular, \pname{DCE$(\{\textrm{edge addition})\}$} is not a special case of \pname{$\Pi$-DSC}, but \pname{Anonymization($\{\textrm{edge addition}\}$)} is. 
They introduced the auxiliary \pname{$\Pi$-Decision} problem that asks whether an $n$-tuple $T=(d_1,\ldots,d_n)$ of non-negative integers satisfies $\Pi$ and proved, using the previous results about \pname{Anonymization($\{\textrm{edge addition}\}$)}~\cite{hartung2015refined}, that  if \pname{$\Pi$-Decision}  is \FPT when parameterized by $\Delta'=\max\{d_i\mid 1\leq i\leq n\}$, then \pname{$\Pi$-DSC} is \FPT when parameterized by $\Delta+k$. Recall now that  Bredereck et al.~\cite{BredereckFHNNT15} proved that 
\pname{Anonymization$(\{\textrm{edge addition}\})$} is \FPT when parameterized by $\Delta$ and has a kernel with $\Oh(\Delta^7)$ vertices. Generalizing this result, 
Froese et al.~\cite{FroeseNN16} defined the \pname{$\Pi$-Number Sequence Completion ($\Pi$-NSC)} problem that asks for a sequence $d_1,\ldots,d_n$ of non-negative integers and two non-negative integers $k$ and $\Delta'$, whether there are non-negative integers $x_1,\ldots,x_n$ such that the $n$-tuple $T=(d_1+x_1,\ldots,d_n+x_n)$ satisfies $\Pi$, $\sum_{i=1}^kx_i=k$ and $d_i+x_i\leq \Delta'$ for $i\in\{1,\dots,n\}$. They proved that if \pname{$\Pi$-NSC} is \FPT when parameterized by $\Delta'$, then \pname{$\Pi$-DSC} if \FPT when parameterized by $\Delta''$ where $\Delta''$ is the maximum degree of the output graph. It is also shown that if \pname{$\Pi$-NSC} can be solved in polynomial time and \pname{$\Pi$-DSC} has a polynomial in $k$ and $\Delta$ kernel, then \pname{$\Pi$-DSC}  has a polynomial kernel when parameterized by $\Delta''$. Froese et al.~\cite{FroeseNN16} were interested only in edge additions, but it is tempting to extend their results for other modification operations. 

\begin{open}
Investigate the (parameterized) complexity of the modification problems with the aim to satisfy some general degree sequence properties for wider sets of permitted  operations.
\end{open}

Some steps in this directions were done by  Golovach and Mertzios~\cite{GolovachM17}. They were interested in the case when the aim is to obtain a graph with the degree sequence $T=(d_1,\ldots,d_n)$ by at most $k$ modification operations from a set $S\subseteq\{\textrm{vertex deletion},\textrm{edge deletion}, \textrm{edge addition}\}$ and called the corresponding problem
\pname{Editing to a Graph with a Given Degree Sequence($S$)}.  They proved that for any non-empty $S$, the problem is $\WOne$ when parameterized by $k$. From the other side, it can be decided in time $2^{\Oh(k(\Delta'+k)^2)}\cdot \poly(n)$ whether a graph with the degree sequence $T$ can be obtained by at most $k_1$ vertex deletions, at most $k_2$ edge additions and at most $k_3$ edge additions where $k_1+k_2+k_3\leq k$ and $\Delta'=\max T$.  It also proved that the problem has a polynomial kernel when parameterized by $k+\Delta'$ if $S=\{\textrm{edge addition}\}$ and has no polynomial kernel unless \ph{} in all other cases.

Bredereck et al.~\cite{BredereckFKMNN19} extended some results of Froese et al.~\cite{FroeseNN16}, Golovach and Mertzios~\cite{GolovachM17}  and Hartung et al.~\cite{hartung2015refined}  for directed graphs.  We already mentioned 
\pname{Digraph Degree Constraint Completion} in the previous section but, they also considered  more general \pname{Digraph Degree Constraint Sequence Completion} that combines individual degree and degree sequence constraints. In this problem, we are given a directed graph, a degree list  function that assigns for each vertex a set of pairs of non-negative integers from $\{0,\ldots,r\}$ that specify the desired pairs of values of in- and out-degrees of vertices, and the degree sequence property $\Pi$, and the question is whether we can add at most $k$ arcs to obtain a directed graph with vertices whose pairs of in- and out-degree are from their lists and whose degree sequence satisfies $\Pi$. Working with directed graphs demands a great deal more efforts, but it proves that the behaviour of the problems for directed and undirected graphs is essentially the same. Again, it would be interesting to extend the set of considered operations.

\begin{open}
Investigate the (parameterized) complexity of the modification problems with the aim to satisfy some general degree sequence properties of directed graphs for wider sets of permitted modification operations.
\end{open}

The related \pname{DAG Realization} problem that asks whether there is a directed acyclic graph that realizes a given degree sequence was considered by Hartung and Nichterlein~\cite{HartungN15}. In particular, they showed that the problem is \NP-hard and proved that it is \FPT when parameterized by the maximum value in the input degree sequence.

\subsection{Modification to satisfy subgraph degree constraints}\label{sec:sub-deg}
In the above part of the section, we considered the problems where the modification aim is to make a graph to satisfy given degree constraints. In Section~\ref{sec:hereditary}, we considered the problems where the task is to obtain a graph that does not contain a given induced subgraph. Nevertheless, it is also possible to ask the question whether we can perform modifications to achieve the property that the obtained graph \emph{has} an induced subgraph with certain properties. In particular, the desired properties of a subgraph can include degree constraints. 

An induced subgraph $H$ of a graph $G$ is said to be a \emph{$k$-core} for a non-negative integer $k$ if the minimum degree $\delta(H)$ of $H$ is at least $k$. The introduction of this notion by 
Seidman~\cite{Seidman83} is motivated by the importance of $k$-cores in (social) networks. Intuitively, a $k$-core for a sufficiently large $k$ is a ``stable'' part of a network. 
Chitnis and Talmon asked in~\cite{ChitnisT18} whether it is possible to create a big $k$-core by edge additions. Formally,  the \pname{Edge $k$-Core} problem asks, given a graph $G$ and nonnegative integers $k$, $p$ and $b$, whether it is possible to add at most $b$ edges to $G$ in such a way that the obtained graph has a $k$-core with at least $p$ vertices.  Chitnis and Talmon proved that this problem is \NP-complete and analyzed its behavior with respect to the parameterizations by $k$, $p$, $b$ and the treewidth of the input graph. It is shown that \pname{Edge $k$-Core}  is \WOne-hard
 when parameterized by $k+p+k$ but can be solved in time $(k+tw)^{b+tw}\poly(n)$, where $tw$ is the treewidth of the input graph.

%!TEX root = survey.tex

\section{Miscellaneous problems}\label{sec:misc}
In this section, we consider several types of edge modification problems that do not fit into the framework of Sections~\ref{sec:hereditary}--\ref{sec:deg}.

\subsection{Diameter augmentation}
%We will start with the \pname{Diameter Augmentation} problem. 
Recall that 
the diameter of a graph $G$ is the longest shortest path between two vertices in a
graph, that is, if $d_G(u,v)$ is the distance in $G$ from $u$ to $v$ defined as the minimum number of edges (or the minimum sum of weights of edges, in the weighted case)  of a $(u,v)$-path, then
\[
\diam(G) = \max_{u,v \in V(G)}d_G(u,v) .
\]
Respectively, we obtain the following completion problem.

\defproblem{Diameter Augmentation}{A graph $G$ and non-negative integers $k$ and $d$.}{Is it possible to obtain a graph $G'$ with $\diam(G'')\leq d$ from $G$ by adding at most $k$ edges?}

The problem is known to be \NP-hard even if $d=2$ \cite{LiMcCSL92} as it was shown by Li, McCormick and Simchi-Levi, and it was proved by Gao, Hare and Nastos that the problem is \WOne-hard when parameterized by $k$ even if $d=2$. Frati et al.~\cite{frati2015augmenting} studied a
more general weighted optimization version of \pname{Diameter Augmentation}, where we have a weighted graph with a weight function
$w \colon V(G)\times V(G) \to \mathbb{N}$ and a cost function $c \colon V(G)\times V(G) \to\mathbb{N}$ and an integer bound $B$.  The goal is to add a set of edges $F$
such that $c(F) = \sum_{e \in F}c(e) \leq B$, and the diameter of $G+F$ is minimum. Frati et al.~\cite{frati2015augmenting} gave an \FPT $4$-approximation algorithm running in time $3^B \poly(n,B)$. They also established some inapproximability results. 

\pname{Diameter Augmentation} was actively investigated for graph classes, and the most famous in the parameterized framework and notoriously hard variant of the problem called \pname{Planar Diameter Augmentation} was introduced by Dejter and Fellows in 1993~\cite{DejterF93}.  In this variant of the problem, the input graph is planar, the value of $k$ is unbounded (it can be assumed that $k=3n-6$), and the graph obtained by adding edges should remain planar. Despite a lot of efforts, it is still unknown whether this problem can be solved in polynomial time or is \NP-hard, but the most interesting question is about the parameterized complexity of the problem. Already Dejter and Fellows~\cite{DejterF93} proved that \pname{Planar Diameter Augmentation} is \FPT when parameterized by $d$. This follows from the fact that for any $d$, the class of planar graph $\mathcal{C}_d$ containing all graphs that can be augmented to graphs of diameter at most $d$ is closed  under taking
minors. By the classical Robertson and Seymour theorem~\cite{RobertsonS04}, $\mathcal{C}_d$ can be characterized by a
finite set of forbidden minors. Together with the minor-checking algorithm of Robertson and Seymour~\cite{RobertsonS95b}, it implies that \pname{Planar Diameter Augmentation} is \FPT. 
Unfortunately, this algorithm is not uniform, because it depends on the set of forbidden minors for $\mathcal{C}_d$ that are distinct for different $d$ and, moreover, 
%Unfortunately, these argument do not give a constructive algorithm, because forbidden minors for $\mathcal{C}_d$ 
are unknown. This lead to the following long standing open problem.

\begin{open}[\cite{DejterF93}]
Give a uniform constructive  \FPT algorithm for \pname{Planar Diameter Augmentation}.%\todo{It should be explained why it is important}
\end{open}

In the last years, some partial results have been obtained.  Interestingly, it was unknown whether \pname{Planar Diameter Augmentation} can be solved by a constructive algorithm running in \XP time.
Lokshtanov, de Oliveira Oliveira, and Saurabh~\cite{LokshtanovO018} considered the \pname{Plane Diameter Augmentation} problem that differs from \pname{Planar Diameter Augmentation}
by the assumption that we are given a plane embedding of the input graph and new edges should be inserted within the faces of the embedding. They constructed an algorithm running in $n^{\Oh(d)}$ time. For the version of \pname{Plane Diameter Augmentation}, where the augmented graph should be $h$-outerplanar, an algorithm with runnning time $f(d)n^{\Oh(h)}$ was given. This extends the result of  Cohen,  Gon{\c{c}}alves, Kim,  Paul, Sau,  Thilikos and Weller~\cite{Cohen0KPSTW15} who proved that  \pname{Outerplanar Diameter Augmentation}, is polynomial. 
For the variant of  \pname{Plane Diameter Augmentation} where the budget parameter $k$ is a part of the input, Golovach,  Requil{\'{e}} and Thilikos~\cite{GolovachRT15}  proved that the problem is \NP-hard and \FPT when parameterized by $k+d$. They also considered the variant where each face of the input graph is bounded by at most $f$ edges and proved that \pname{Plane Diameter Augmentation} is \FPT when parameterized by $d+f$.

\subsection{Local edge modifications}
In the previous sections, we were dealing with edge modification problems where the only constraint on the set of modified edges itself was its cardinality. Nevertheless, there are problems when the set of modified edges should satisfy some additional, usually local, combinatorial  property. In this subsection, we consider such problems.

\emph{Seidel’s switching} is a graph operation which makes a given vertex adjacent to precisely those vertices to which it was non-adjacent before, while keeping the rest of the graph unchanged. 
In~\cite{KratochvilNZ92}, Kratochv\'{\i}l, Ne\v{s}et\v{r}il, and Z\'{y}ka initiated the study of the \pname{Switching to $\mathcal{C}$} problem, whose task is to decide whether a graph can be modified to belong to a given graph class $\mathcal{C}$ by a series of Seidel’s switching. There are various algorithmic and hardness results for the problem, but since we are interested in Parameterized Complexity, we only mention the results of  Jel{\'{\i}}nkov\'{a},  Such\'{y}, Hlinen{\'{y}}, and Kratochv\'{\i}l~\cite{JelinkovaSHK11}. In particular, they proved that if $\mathcal{C}$ is the class of graphs of minimum (maximum, respectively) degree at least (at most, respectively) $d$ or the class of $d$-regular graphs, then  the problem is \FPT when parameterized by $d$.

If Seidel’s switching complements adjacencies of a vertex,  the \emph{local complementation} introduced by Kotzig~\cite{Kotzig68}  complements the edges between the neighbors of a vertex. 
More formally, the graph $G'$ is obtained from $G$ by the local complementation with respect to a vertex $v$ if $G'=G-E(G(N_G(u)))+E(\overline{G(N_G(u))]})$. The study of this operation is mainly motivated by its importance for vertex minors and rank-width (see, e.g.,~\cite{Oum05}) but, similarly to   \pname{Switching to $\mathcal{C}$}, we can define the \pname{Local Complementation to $\mathcal{C}$} problem. The investigation of the parameterized complexity of this problem was initiated by Cattan\'{e}o and Perdrix in~\cite{CattaneoP15}, where they proved that the problem is \WOne-hard  if $\mathcal{C}$ is the class of graphs of minimum degree at most $d$ when parameterized by $d$.

Fomin et al. in~\cite{FominGST18} considered complementations with respect to vertex subsets. For a set $S\subseteq V(G)$, the \emph{partial complement} of $G$ with respect to $S$ is the graph $G'$ obtained by taking the complement of $G[S]$ in $G$, that is, $G'=G-E(G[S])+E(\overline{G[S]})$. For a graph class $\mathcal{C}$, they defined 
 the \pname{Partial Complement to $\mathcal{C}$} problem that asks whether there is a partial complement of a graph $G$ belonging to $\mathcal{C}$. Among the obtained results, they proved that  \pname{Partial Complement to $\mathcal{C}$} is \FPT when parameterized by $w$ for some subclasses $\mathcal{C}$ of the graphs of clique-width at most $w$.
 
\begin{open}[\cite{FominGST18}] 
 What is the complexity of \pname{Partial Complement to $\mathcal{C}$} when $\mathcal{G}$ is 
\begin{itemize}
\item the class of chordal graphs, 
\item the class of interval graphs,
\item the class of graphs excluding a path $P_5$ as an induced subgraph,  
\item the class of graphs with minimum degree $\geq r$ for some constant $r$?\end{itemize}
\end{open}

Fomin, Golovach and Thilikos in~\cite{FominGT18,FominGT19} introduced problems where the structure of the modified edges is defined by a given pattern graph $H$. 

In~\cite{FominGT18}, they defined the notion of graph superposition. Let $G$ and $H$ be graphs such that $|V(G)|\geq |V(H)|$ and let $\varphi\colon V(H)\rightarrow V(G)$ be an injective mapping.  The graph $G'$ is 
the \emph{superposition of $G$ and $H$ (with respect to $\varphi$)} if $V(G')=V(G)$ and two vertices $u,v\in V(G')$ are adjacent in this graph if and only $uv\in E(G)$ or $u,v\in \varphi(V(H))$ and $\varphi^{-1}(u)\varphi^{-1}(v)\in E(H)$. Informally, we select $|V(H)|$ vertices in $G$ and ``glue'' a copy of $H$ into $G$ using these vertices. Fomin at al.~\cite{FominGT18} considered the \pname{Structural Connectivity and $2$-Connectivity Augmentation} problems that ask, given graphs $G$ and $H$, whether there is a superposition of $G$ and $H$ such that the obtained graph is connected and $2$-connected respectively. They showed a computational complexity dichotomy for the problem depending on the properties of the graph class $\mathcal{C}$ containing $H$. If the vertex cover number of graphs in $\mathcal{C}$ is at most $t$, then \pname{Structural Connectivity and $2$-Connectivity Augmentation} can be solved in polynomial time, that is, they are in \XP when parameterized by $t$, and the problems are \NP-hard if $\mathcal{C}$ contains graphs with arbitrarily large vertex cover number. 

In~\cite{FominGT19}, Fomin et al. proposed a very general edge modification model. The  allowed changes are defined through  {\em replacement actions}. 
Let $\mathcal{L}$ be a mapping that assigns to every  labeled  $k$-vertex graph $H$  a list  $L(H)$ of labeled $k$-vertex  graphs.  
 Then the replacement action selects a subset of $k$ vertices $S$ in the graph $G$  and replaces  the subgraph $G[S]$ induced by $S$ by a graph $F$ from the list $L(G[S]))$.  More precisely, the action selects a $k$-sized vertex subset $S$ of $G$  labeled by numbers $\{1, \dots , k\}$ and, given that $H$ is the  labeled $k$-vertex graph  obtained from $G[S]$, we select a labelled $k$-vertex graph $F$ from $L(H)$ and replace $H$ by $F$. Thus, the vertex set of the new graph $G'$ is $V(G)$ and it has the same adjacencies as in  $G$ except pairs of vertices from $S$. In the transformed graph, vertices $u,v\in S$ labeled by $i,j\in\{1,\dots, k\}$ are adjacent  in $G'$ if and only if $\{i,j\}$ is an edge of $F$.  
Using replacement actions we can express various modification problems. 
For example, we can express the deletion of at most $\ell$ edges as a family of actions that map graphs with at most $2\ell$ vertices into graphs that can be obtained from them by at most $\ell$ edge deletions.  Similarly, we can express edge additions. Fomin et al. considered the \pname{$\mathcal{L}$-Replacement to a Planar Graph} problem, 
whose task is to decide,  given a graph $G$ and a positive integer $k$, whether there is an action that makes $G$ planar. They proved that this problem is \FPT when parameterized by $k$ and got a number of related results where it is required to obtain a planar graph with some specific properties.

It can be seen from our brief description that, up to now, we have just a scattered set of parameterized complexity results for the aforementioned problems. We believe that these problems are natural and their systematic study for various parameterizations may lead to interesting findings.

%\todo{What is known about parameterized complexity of the increasing diameter problem?}

\subsection{Flip distance}

Here we briefly discuss the geometric \pname{Flip Distance} problem which, strictly speaking, is not defined as a graph modification problem but is closely related to our subject. 

Let $\mathcal{T}$ be a triangulation of a set of points $\mathcal{P}$ on the Euclidean plane. Let $ABC$ and $BCD$ be triangles of $\mathcal{T}$ such that $ABCD$ is a convex quadrilateral. The \emph{flip} operation for $ABC$ and $BCD$ replaces these triangles by $ABD$ and $ADC$, that is, the diagonal $BC$ in the quadrilateral $ABCD$ is replaced by $AD$. The \emph{flip} distance between two triangulations $\mathcal{T}_1$ and $\mathcal{T}_2$ of $\mathcal{P}$ is the minimum number of flips needed to transform $\mathcal{T}_1$ into $\mathcal{T}_2$. The \pname{Flip Distance} problem asks, given two triangulations $\mathcal{T}_1$ and $\mathcal{T}_2$ of a set of points $\mathcal{P}$ and a non-negative integer $k$, whether the flip distance between $\mathcal{T}_1$ and $\mathcal{T}_2$ is at most $k$. Note that this problem can be considered as an edge modification problem on triangulated plane graphs. We refer to the survey of Bose and
Hurtado~\cite{BoseH09} for the discussion of the relations between geometric and graph variants. 
 
Lubiw and  Pathak proved in~\cite{LubiwP15} that \pname{Flip Distance} is \NP-complete. Cleary and St. John initiated the study of the parameterized complexity of the problem. They considered the case when $\mathcal{P}$ defines a convex polygon and gave a kernel with $5k$ points using the relation between the flip distance and the so named \emph{rotation distance} between two rooted binary trees. The kernel size for convex polygons was improved to $2k$ by  Lucas in~\cite{Lucas10}. The first \FPT algorithm for the general case running in $\Oh(n+k\cdot c^k)$ time for $c\approx 2\cdot 14^{11}$ was given by  Kanj,  Sedgwick and Xia in~\cite{KanjSX17}. The running time was recently improved by Li, Feng,  Meng and Wang~\cite{LiFMW17}.

\begin{open}[\cite{KanjSX17,LiFMW17}]
Does \pname{Flip Distance} admit a polynomial kernel when parameterized by $k$?
\end{open}

\subsection{Strong triadic closure and related problems}
In the classical setting for graph editing problems, the task is to delete and/or add some edges to satisfy a certain property. There are closely related variants where the aim is to label edges of a graph to achieve a given property of labeled graphs. Considering all problems of this type is far beyond the scope of the survey and here we mention only a few of them that are related to our subject. 

The notion of \emph{triadic closure} was introduced in social network theory (see the book of Easley and Kleinberg~\cite{EasleyK10} for details). In terms of graphs, this property is stated as follows. Let $G$ be a graph, whose edges are labeled \emph{strong} and \emph{weak}. It is said that $G$ satisfies the \emph{strong triadic closure} property if for every two distinct strong edges $uv$ and $uw$ with a common end-vertex, $vw\in E(G)$. Informally, this means that if there are strong connections between $v$ and $u$ and between $u$ and $w$, then there is a connection (either strong or weak) between $v$ and $w$. The task of the \pname{Strong Triadic Closure} problem is, given a graph $G$ and a non-negative integer $k$, to decide whether there is a strong/weak labeling of the edges of $G$ with at most $k$ weak edges such that the labeled graph satisfies the strong triadic closure property. Observe that this problem is closely related to \pname{Cluster Deletion} or \pname{$P_3$-Free Deletion}, because for every induced path on three vertices at least one of its edges should be labeled weak.   
  
\pname{Strong Triadic Closure} is known to be \NP-complete~\cite{SintosT14} and the parameterized complexity of the problem was considered in~\cite{GolovachHKLP18,DBLP:conf/wg/GruttemeierK18,SintosT14}. In particular, Sintos and Tsaparas~\cite{SintosT14} observed that the problem is \FPT when parameterized by $k$ by a reduction to \pname{Vertex Cover}. Golovach et al.~\cite{GolovachHKLP18} and Gr{\"{u}}ttemeier and Komusiewicz~\cite{DBLP:conf/wg/GruttemeierK18} observed that it admits a polynomial kernel for this parameterization. For the dual parameterization by $\ell=|E(G)|-k$, that is, by the number of strong edges, \pname{Strong Triadic Closure} is \FPT but does not admit a polynomial kernel unless \ph~\cite{GolovachHKLP18,DBLP:conf/wg/GruttemeierK18}. Notice that the kernelization lower bound holds for \pname{Cluster Deletion} as well.

It was observed in~\cite{GolovachHKLP18} that if $M$ is a matching of a graph $G$, the edges of $M$ are strong and the remaining edges are weak, then the labeled graph  satisfies the strong triadic closure property. This means that the maximum size of a matching $\mu(G)$ gives a lower bound for the maximum number of strong edges. This lead to the following open problem.

 \begin{open}[\cite{GolovachHKLP18}]
Is \pname{Strong Triadic Closure} \FPT when parameterized by $h=|E(G)|-k-\mu(G)$, that is, by the number of strong edges above the maximum matching size?
\end{open}

Golovach et al.~\cite{GolovachHKLP18} proved that the problem is \FPT on graph of maximum degree at most four. Notice that the  question for the same parameterization is also open for \pname{Cluster Deletion}. 

Golovach et al.~\cite{GolovachHKLP18} also considered the more general variant called  \pname{Strong $F$-Closure} that is related to \pname{$F$-Free Deletion}. Here, $F$ is a fixed graph and the task is to label the edges of an input graph $G$ in such a way that if the subgraph of $G$ composed by strong edges contains a copy of $F$ as an induced subgraph, then there is a weak edge with both end-vertices in this copy.   Bulteau et al. in~\cite{BulteauGKS19} introduced another generalization, where there are $c$ strong labels (or colors) and the constraint is that  if  $uv$ and $uw$ are distinct edges with a common end-vertex and the same strong label, then $uv\in E(G)$. In~\cite{BulteauGKS19,GolovachHKLP18}, the authors obtained various results that generalize the aforementioned results for \pname{Strong Triadic Closure}.

Gr{\"{u}}ttemeier et al. considered in~\cite{GruttemeierKSS19} the \pname{Bicolored $P_3$-Deletion} problem: given a graph $G$, whose edges are partitioned into two sets $E_r$ and $E_b$ of \emph{red} and \emph{blue} edges respectively, and a non-negative integer $k$, the task is to decide whether it is possible to delete at most $k$ edges in such a way that the obtained graph has no bicolored induced $P_3$. It was proved that \pname{Bicolored $P_3$-Deletion} can be solved in time $\Oh(1.85^k n^5)$ and has a polynomial kernel when parameterized by $k$ and the maximum degree of the input graph $\Delta$.

\subsection{Beyond forbidden subgraphs}
In Section~\ref{sec:hereditary}, we considered editing problems whose task is to obtain a graph belonging to a given hereditary graph class, that is, a graph class defined by a family of forbidden induced subgraphs.  Here we survey some variations and generalizations of these problems.

Besides forbidding induced subgraphs, it is possible to forbid other structures. In particular, there is a plethora of results for graph classes defined by families of forbidden minors or topological minors. However, these problems have been mainly investigated for vertex deletions and the results about edge deletions are corollaries. Therefore, we do not consider them in this survey. The situation is different if we forbid containment of some graphs as \emph{immersions}.   A graph $H$ is an \emph{immersion} of $G$ if there is an injective mapping of the vertices of $H$ to the vertices of $G$ and a mapping of the edges of $H$ to pairwise edge-disjoint paths of $G$ such that for every two adjacent vertices $u$ and $v$ of $H$, the edge $uv$ is mapped to a path of $G$ whose end-vertices are the images of $u$ and $v$. For a family of graphs $\mathcal{F}$, a graph $G$ is \emph{$\mathcal{F}$-immersion free} if $H$ is not an immersion of $G$ for every $H\in\mathcal{F}$. 
In~\cite{GiannopoulouPRT17}, Giannopoulou et al. initiated the study of the \pname{$\mathcal{F}$-Immersion Deletion} problem. Given a (finite) family of graphs $\mathcal{F}$, the task is to decide whether a graph $G$ can be made $\mathcal{F}$-immersion free by at most $k$ edge deletions. Giannopoulou et al.~\cite{GiannopoulouPRT17} proved that if $\mathcal{F}$ consists of connected graphs and at least one graph in the family is planar, then \pname{$\mathcal{F}$-Immersion Deletion} admits a linear kernel when parameterized by $k$ and can be solved in time $2^{\Oh(k)}\poly(n)$.

Fomin, Golovach and Thilikos considered in~\cite{FominGT2019} a generalization of another type in which the property that a graph $G$ does not contain an induced subgraph isomorphic to $H$ is \emph{local}. The most general way to express local properties is via the \emph{first-order logic} (FOL) formulas on graphs.  
Recall that the syntax of FOL-formulas on graphs includes the logical connectives 
$\vee$, $\wedge$, $\neg$, variables for vertices, 
the quantifiers $\forall$, $\exists$ that are applied to these variables, and the adjacency and equality predicates.
An FOL-formula $\phi$ is  in {\em prenex normal form} if it is written as $\phi={\tt Q}_{1}x_{1}{\tt Q}_{2}x_{2}\cdots{\tt Q}_{t}x_{t} \chi$  where 
each ${\tt Q}_i\in\{\forall,\exists\}$ is a quantifier,  $x_i$ is a variable, and $\chi$ is a quantifier-free part.  Let $\varphi$ be a FOL-formula. The task of the \pname{Edge Deletion (Completion, Editing) to $\varphi$} problem is, given a graph $G$ and a non-negative integer $k$, decide whether there is $S\subseteq E(G)$ ($S\subseteq \binom{V(G)}{2}$, $S\subseteq \binom{V(G)}{2}$, respectively) of size at most $k$ such that $G-S\models \varphi$ ($G+F\models \varphi$, $G\bigtriangleup F\models \varphi$ respectively).  Fomin et al.~\cite{FominGT2019} characterized the  complexity of  \pname{Edge Deletion (Completion, Editing) to $\varphi$} (and the vertex deletion analog) with respect to the prefix structure of $\varphi$. More precisely, they obtained the following parameterized complexity dichotomy depending on the quantifier alternations in the prefix. If $\varphi$ can be written in the form $\exists x_1\ldots \exists x_s\forall y_1\ldots \forall y_t\psi$ (we assume that either forall or existential quantification part may be empty), where $\psi$ is a quantifier-free part, then \pname{Edge Deletion (Completion, Editing) to $\varphi$} can be solved in time $|\varphi|^{\Oh(k)}\cdot n^{\Oh(|\varphi|)}$, that is, the problem is \FPT when parameterized by $k$. If we allow at least two quantifier alternations  or one alternation but $\forall$ occurs first, then there is $\varphi$ with the corresponding structure of the prefix for which the problem is \wtwo-hard. Notice that the property that $G$ has no induced subgraph isomorphic to $H$ can be expressed in FOL. Hence, these result, indeed, generalize the results of Cai~\cite{cai1996fixedparameter}. For kernelization,  Fomin et al.~\cite{FominGT2019} established a similar dichotomy: if $\varphi=\exists x_1\ldots \exists x_s\psi$, then \pname{Edge Deletion (Completion, Editing) to $\varphi$} admits a trivial kernel when parameterized by $k$, and for every other prefix structure, there is a formula such that the problem has no polynomial kernel unless \ph. 

\medskip
\noindent\textbf{Acknowledgement}. We thank Marcin Pilipczuk, William Lochet, and 
Dekel Tsur  for helpful comments. 

\bibliographystyle{abbrv}
 %\bibliography{bibliography,book_pc,book_kernels_fvf,deg-mis,chris}
% \bibliography{bibliography,book_kernels_fvf,deg-mis,chris}
\bibliography{survey_edge_modification}

\end{document}